\documentclass[12pt]{article}
\usepackage{epsf,amsfonts,latexsym}
\newsymbol\ltimes 226E
\newsymbol\rtimes 226F
\epsfverbosetrue
\def\goth{\mathfrak}
\def\double{\mathbb}
\def\ccal{\cal}

\def\cc{{\double C}}

\def\rr{{\double R}}
\def\zz{{\double Z}}

\def\llll{{\double L}}

\def\aa{{\cal A}}
\def\ccc{{\cal C}}
\def\dd{{\cal D}}
\def\gg{{\goth g}}
\def\hh{{\cal H}}
\def\hhh{{{\double H}}}

\def\mm{{{\ccal M}}}

\def\aa{{\cal A}}
\def\dd{{\cal D}}
\def\hh{{\cal H}}

\def\lll{{\cal L}}
\def\sss{{\cal S}}
\def\jj{{\cal J}}
\def\t{{\rm tr}\,}

\def\ddd{{\,\hbox{$\partial\!\!\!/$}}}

\def\dee{\hbox{\rm D}}
\def\de{\hbox{\rm d}}

\def\pa{{\partial}}
\def\Box{\,\hbox{$\sqcap\!\!\!\!\sqcup$}}

\def\lb{\left[}
\def\rb{\right]}

\def\ot{\otimes}
\def\op{\oplus}

\def\bb{\begin{eqnarray}}
\def\ee{\end{eqnarray}}
\def\eee{\nonumber\end{eqnarray}}
\def\pp{\pmatrix}
\def\qq{\quad}

\begin{document}

\hsize 17truecm
\vsize 24truecm
\font\twelve=cmbx10 at 13pt
\font\eightrm=cmr8
\baselineskip 18pt

\begin{titlepage}

\centerline{\twelve CENTRE DE PHYSIQUE TH\'EORIQUE}
\centerline{\twelve CNRS - Luminy, Case 907}
\centerline{\twelve 13288 Marseille Cedex 9}
\vskip 3truecm

\centerline{\twelve FORCES FROM CONNES' GEOMETRY}

\bigskip

\begin{center} {\bf Thomas SCH\"UCKER}
\footnote{\, and Universit\'e de Provence\\
schucker@cpt.univ-mrs.fr } \\

\end{center}

\vskip 2truecm
\leftskip=1cm
\rightskip=1cm
\centerline{\bf Abstract}

\medskip
\noindent
We try to give a pedagogical introduction to Connes' derivation
 of the
standard model of electro-magnetic, weak and strong forces from
gravity.

\vskip 2truecm\noindent
Lectures given at the  Autumn School ``Topology
and Geometry in Physics''\\
of the Graduiertenkolleg `Physical Systems with Many Degrees of
Freedom'\\ Universit\"at Heidelberg\\
 September 2001, Rot an der Rot,
Germany\\ Editors: Eike Bick
\& Frank Steffen\\
Lecture Notes in Physics 659, Springer, 2005

\vskip 1truecm
 PACS-92: 11.15 Gauge field theories\\
\indent MSC-91: 81T13 Yang-Mills and other gauge theories

\vskip 1truecm

\vskip 1truecm
\noindent CPT-01/P.4264\\
\noindent hep-th/0111236

\vskip1truecm

 \end{titlepage}

\tableofcontents \vfil\eject

\section{Introduction}

Still today one of the major summits in physics is the
understanding of the spectrum of the hydrogen atom. The
phenomenological formula by Balmer and Rydberg was a
remarkable pre-summit on the way up. The true summit was
reached by deriving this formula from quantum mechanics. We
would like to compare the standard model of electro-magnetic,
weak, and strong forces with the Balmer-Rydberg formula \cite{cls}
and review the present status of Connes' derivation of this model
from noncommutative geometry, see table 1. This geometry extends
Riemannian geometry, and Connes' derivation is a natural extension
of another major summit in physics: Einstein's derivation of general
relativity from Riemannian geometry. Indeed, Connes' derivation
unifies gravity with the other three forces.
 \begin{table}[h]
\begin{center}
\begin{tabular}{ll}
atoms&particles and forces\\[1ex]
Balmer-Rydberg formula \qq\qq&standard model\\[1ex]
quantum
mechanics&noncommutative geometry
\end{tabular}
\end{center}
\caption{An analogy}
\end{table}

Let us briefly recall four nested, analytic geometries and their
impact on our understanding of forces and time, see table 2. {\it
Euclidean geometry} is underlying Newton's mechanics as space of
positions. Forces are described by vectors living in the same space
and the Euclidean scalar product is needed to define work and
potential energy. Time is not part of geometry, it is absolute. This
point of view is abandoned in special relativity unifying space
and time into {\it Minkowskian geometry}. This new point of view
allows to derive the magnetic field from the electric field as a
pseudo force associated to a Lorentz boost. Although time has
become relative, one can still imagine a grid of synchronized
clocks, i.e. a universal time. The next generalization is
{\it Riemannian geometry} = curved spacetime. Here gravity can be
viewed as the pseudo force associated to a uniformly accelerated
coordinate transformation. At the same time, universal time loses
all meaning and we must content ourselves with proper time. With
today's precision in time measurement, this complication of life
becomes a bare necessity, e.g. the global positioning system (GPS).

\begin{table}[h]
\begin{center}
\begin{tabular}{lll}
geometry&force & time\\[3ex]
Euclidean &$E=\int\vec F\cdot\de \vec x$&absolute\\[1ex]
Minkowskian&$\vec E,\epsilon _0\Rightarrow\vec B,\mu
_0=\,\frac{1}{\epsilon _0c^2}\, $&universal\\[1ex]
Riemannian&Coriolis $\leftrightarrow$ gravity&proper, $\tau$
\\[1ex]
noncommutative\qq\qq&gravity $\Rightarrow$ YMH, $\lambda
={\textstyle\frac{1}{3}} g_2^2$\qq\qq&$\Delta \tau \sim 10^{-40}$
 s
\end{tabular}
\end{center}
\caption{Four nested analytic geometries}
\end{table}

 Our last
generalization is to Connes'  {\it noncommutative geometry} = curved
space(time) with uncertainty. It allows to understand some
Yang-Mills and some Higgs forces as pseudo forces associated to
transformations that extend the two coordinate transformations
above to the new geometry without points. Also, proper time comes
with an uncertainty. This uncertainty of some hundred Planck
times might be accessible to experiments through gravitational
wave detectors within the next ten years \cite{ac}.

\subsection*{Prerequisites}

On the physical side, the reader is supposed to be
acquainted with general relativity, e.g. \cite{gr}, Dirac spinors at
the level of e.g. the first few chapters in
\cite{bd}  and
Yang-Mills theory with spontaneous symmetry
break-down, for example the standard model, e.g. \cite{or}. I am not
ashamed to adhere to the minimax principle: a
maximum of pleasure with a minimum of effort. The
effort is to do a calculation, the pleasure is when its
result coincides with an experiment result.
Consequently our mathematical treatment is as
low-tech as possible. We do need {\it local }
differential and Riemannian geometry at the level of
e.g. the first few chapters in
\cite{gs}. Local means that our spaces or manifolds can
be thought of as open subsets of
$\rr^4$. Nevertheless, we sometimes use compact
spaces like the torus: only to simplify some integrals.
We do need some group theory, e.g. \cite{group}, mostly
matrix groups and their representations. We also need
a few basic facts on associative algebras. Most of
them are recalled as we go along and can be found for
instance in \cite{algebra}. For the reader's convenience, a
few simple definitions from groups and algebras are
collected in the appendix. And, of course, we need some
chapters of noncommutative geometry which are developped in
the text.  For a more detailed presentation still with particular care
for the physicist see \cite{jogi, costarica}.

\section{Gravity from Riemannian geometry}

In this section, we briefly review Einstein's derivation of general
relativity from Riemannian geometry. His derivation is in two
strokes, kinematics and dynamics.

\subsection{First stroke: kinematics}

Consider flat space(time) $M$ in inertial or Cartesian coordinates
$\tilde x^{\tilde \lambda } $.
Take as matter  a free, classical point particle. Its dynamics,
Newton's free equation, fixes the trajectory $\tilde x^{\tilde
\lambda }(p) $:
\bb {\frac{\de^2\tilde x^{\tilde \lambda}}{\de p^2}}\,  =0.\ee
After a general coordinate transformation, $x^\lambda =\sigma
^\lambda (\tilde x)$, Newton's equation reads
\bb {\,\frac{\de^2 x^{
\lambda }}{\de p^2}\,}+ \hbox{$ \Gamma
^{
\lambda }$}_ {
\mu  \nu }( g) {\,\frac{\de x^{
\mu  }}{\de p}\,} {\,\frac{\de x^{ \nu
}}{\de p}\,}
 =0.\label{1}\ee
Pseudo forces have appeared. They are coded in the
Levi-Civita connection
\bb\hbox{$ \Gamma
^{
\lambda }$}_ {
\mu  \nu }( g)={\textstyle\frac{1}{2}}
 g^{ \lambda \kappa}\left[
\,\frac{\pa}{\pa x^{ \mu}}\,
 g_{ \kappa
\nu}+\,\frac{\pa}{\pa x^{ \nu}}\,
 g_{ \kappa \mu}
-\,\frac{\pa}{\pa x^{ \kappa}}\,
 g_{ \mu \nu}\right],\ee
where $g_{ \mu \nu}$ is obtained by `fluctuating'
 the flat metric $\tilde \eta_{\tilde \mu \tilde \nu }={\rm
diag}(1,-1,-1,-1,)
$ with the Jacobian of the coordinate transformation $\sigma  $:
\bb g_{ \mu \nu}(x)={\jj( x)^{-1\tilde \mu }}_{ \mu}\,
\eta_{\tilde \mu \tilde \nu }\,{\jj( x)^{-1\tilde \nu }}_{ \nu},\qq
{\jj( \tilde x)^{\mu }}_{\tilde \mu}:=
\pa
 \sigma ^{ \mu }(\tilde x)/\pa \ \tilde x^{ \tilde \mu }.\ee
For the coordinates of the rotating disk, the pseudo forces are
precisely the centrifugal and Coriolis forces. Einstein takes
uniformly accelerated coordinates, $ct=c\tilde t,\ z=\tilde
z+{\textstyle\frac{1}{2}} {\textstyle\frac{g}{c^2}} (c\tilde t)^2$
with $g=  9.81\ \rm m/s^2$. Then the geodesic equation (\ref{1})
reduces to $\de^2z/\de t^2=-g$. So far this gravity is still a pseudo
force which means that the curvature of its Levi-Civita
connection vanishes. This constraint is relaxed by the
equivalence principle: pseudo forces and true gravitational forces
are coded together in a not necessarily flat connection $\Gamma $,
that derives from a potential, the not necessarily flat metric $g$.
The kinematical variable to describe gravity is therefore the
Riemannian metric. By construction the dynamics of matter, the
geodesic equation, is now covariant under general coordinate
transformations.

\subsection{Second stroke: dynamics}

Now that we know the kinematics of gravity let us see how Einstein
obtains its dynamics, i.e. differential equations for the metric
tensor
$g_{\mu \nu }$. Of course Einstein wants these equations to be
covariant under general coordinate transformations and he wants
the energy-momentum tensor $T_{\mu \nu }$ to be the source of
gravity. From Riemannian geometry he knew that there is no
covariant, first order differential operator for the metric. But
there are second order ones:\\
{\bf Theorem:} The most general tensor of degree 2 that can be
constructed from the metric tensor $g_{\mu \nu }(x)$ with at most
two partial derivatives is
 \bb\alpha R_{\mu \nu }+\beta R g_{\mu
\nu }+\Lambda  g_{\mu \nu },\qq \alpha ,\ \beta ,\
\Lambda \in\rr.\label{tensor}.\ee
 Here are our conventions for the curvature tensors:
\bb {\rm Riemann\ tensor:}&& {R^\lambda}_{\mu\nu\kappa}=
\pa_\nu {\Gamma^\lambda}_{\mu\kappa}
-\pa_\kappa {\Gamma^\lambda}_{\mu\nu}
+{\Gamma^\eta}_{\mu\kappa}{\Gamma^\lambda}_
{\nu\eta}
-{\Gamma^\eta}_{\mu\nu}{\Gamma^\lambda}_
{\kappa\eta},\\
{\rm Ricci\ tensor:}&&
R_{\mu\kappa}=
{R^\lambda}_{\mu\lambda\kappa},\\
{\rm curvature\ scalar:}&&
R=R_{\mu\nu}g^{\mu\nu}.\ee

 The miracle is that the tensor (\ref{tensor}) is symmetric just as
the energy-momentum tensor. However, the latter is
covariantly conserved, $\dee^\mu T_{\mu \nu }=0$,
while the former one is conserved if and
only if $\beta =-{\textstyle\frac{1}{2}}
\alpha $. Consequently, Einstein puts his equation
\bb  R_{\mu \nu }-{\textstyle\frac{1}{2}}  R g_{\mu
\nu }-\Lambda _c g_{\mu \nu }={\textstyle\frac{8\pi G}{c^4}} \,
T_{\mu
\nu }.\ee
He chooses a vanishing cosmological constant, $\Lambda_c =0$. Then
for small static mass density $T_{00}$, his equation reproduces
Newton's universal law of gravity with $G$ the Newton constant.
However for not so small masses there are corrections to Newton's
law like  precession of perihelia. Also Einstein's theory applies to
massless matter and produces the curvature of light. Einstein's
equation has an agreeable formal property, it derives via the
Euler-Lagrange variational principle
 from an action, the famous Einstein-Hilbert action:
\bb S_{EH}[g]=\frac{-1}{16\pi G}\,
\int_MR\,\de V\ -\,\frac{2\Lambda _c}{16\pi G}\,
\int_M\de V,\ee
with the invariant volume element $\de V:=|\det
g_{\cdot\cdot}|^{1/2}\,\de^ 4x.$

 General relativity has a precise geometric origin: the
left-hand side of Einstein's equation is a sum of some
$80\,000$ terms in first and second partial derivatives of $g_{\mu
\nu }$ and its matrix inverse $g^{\mu \nu }$. All of these terms
are completely fixed by the requirement of covariance under
general coordinate transformations.  General relativity is verified
experimentally to an extraordinary accuracy, even more, it has
become a cornerstone of today's technology. Indeed length
measurements had to be abandoned in favour of proper time
measurements, e.g. the GPS. Nevertheless, the theory still leaves a
few questions unanswered:
\begin{itemize}\item
Einstein's equation is nonlinear and therefore does not allow point
masses as source, in contrast to Maxwell's equation that does allow
point charges as source. From this point of view it is not satisfying to
consider point-like matter.
\item
The gravitational force is coded in the connection $\Gamma $.
Nevertheless we have accepted its potential, the metric $g$, as
kinematical variable.
\item
The equivalence principle states that locally, i.e. on the
trajectory of  a point-like particle, one cannot distinguish gravity
from a pseudo force. In other words, there is always a coordinate
system, `the freely falling lift', in which gravity is absent. This is
not true for electro-magnetism and we would like to derive this force
(as well as the weak and strong forces) as a pseudo force coming
from a geometric transformation.
\item
So far general relativity has resisted  all attempts to reconcile it
with quantum mechanics.
\end{itemize}

\section{Slot machines and the standard model}

Today we have a very precise phenomenological description of
electro-magnetic, weak, and strong forces. This description, the
standard model, works on a perturbative quantum  level and, as
classical gravity, it derives from an action principle. Let us
introduce this action by analogy with the Balmer-Rydberg
formula.

\begin{figure}[h]
\label{slot1}
\epsfxsize=4cm
\hspace{5.5cm}
\epsfbox{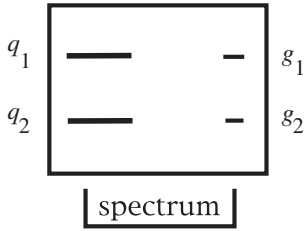}
\caption{A slot machine for atomic spectra}
\end{figure}

One of the new features of atomic physics was the appearance of
discrete frequencies and the measurement of atomic spectra
became a highly developed art. It was natural to label the discrete
frequencies $\nu $ by natural numbers
$n$. To fit the spectrum of a given atom, say hydrogen, let us try
the ansatz
\bb\nu =g_1n_1^{q_1}+g_2n_2^{q_2}.\label{ansatz}\ee
We view this ansatz as a slot machine. You input two bills, the
integers $q_1$, $q_2$ and two coins, the two real numbers
$g_1$, $g_2$, and compare the output with the measured
spectrum. (See Figure 1.) If
you are rich enough, you play and replay on the slot machine
until you win. The winner is the Balmer-Rydberg formula, i.e.,
$q_1=q_2=-2$ and $g_1=-g_2=\ 3.289\ 10^{15}$ Hz, which is the
famous
Rydberg constant $R$. Then came quantum mechanics. It explained
why the spectrum of the hydrogen atom was discrete in the first
place and derived the exponents and the Rydberg constant,
\bb R=\,\frac{m_e}{4\pi \hbar^3}\,\frac{e^4}{(4\pi \epsilon
_0)^2}\, ,\ee
 from a noncommutativity,
$[x,p]=i\hbar 1$.

To cut short its long and complicated history we introduce the
standard model as the winner of a particular slot machine. This
machine, which has become popular under the names Yang,
Mills and Higgs, has
 four slots for four bills. Once you have decided which
bills you choose and entered them, a certain number of
small slots will open for coins. Their number depends
on the choice of bills. You make your choice of coins,
feed them in, and the machine starts working. It
produces as output a Lagrange density. From this density,
perturbative quantum field theory allows you to compute a
complete particle phenomenology: the particle spectrum with the
particles' quantum numbers, cross sections, life times, and
branching ratios. (See Figure 2.) You compare the
phenomenology to experiment to find out whether your input
wins or loses.

\begin{figure}[h]
\label{slot2}
\epsfxsize=6cm
\hspace{4.7cm}
\epsfbox{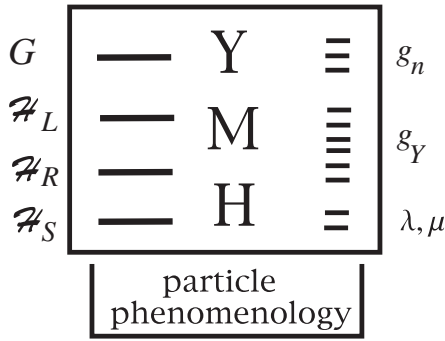}
\caption{The Yang-Mills-Higgs slot machine}
\end{figure}

\subsection{Input}

The first bill is a finite dimensional, real, compact Lie
group $G$. The gauge bosons, spin 1, will live in its
adjoint representation whose Hilbert space is the
complexification of the Lie algebra $\gg$ (cf Appendix).

The remaining bills are three unitary
representations of $G$, $\rho_L,\ \rho_R,\ \rho_S$,
defined on the complex Hilbert spaces, $\hh_L,\
\hh_R,\
\hh_S$. They classify the left- and right-handed
fermions, spin ${\textstyle\frac{1}{2}}$, and the
scalars, spin 0. The group $G$ is chosen compact to
ensure that the unitary representations are finite
dimensional, we want a finite number of `elementary particles'
according to the credo of particle physics that particles are
orthonormal basis vectors of the Hilbert spaces which carry the
representations. More generally, we might also admit multi-valued
representations, `spin representations', which would open the
debate on charge quantization. More on this later.

The coins are numbers, coupling constants, more
precisely coefficients of invariant polynomials. We
need an invariant scalar product on $\gg$. The set of
all these scalar products is a cone and the gauge
couplings are particular coordinates of this cone. If
the group is simple, say $G=SU(n)$, then the most
general, invariant scalar product is
\bb (X,X')={\textstyle\frac{2}{g_n^2}}\t [X^*X'],\qq
X,X'\in su(n).\ee
If $G=U(1)$, we have
\bb (Y,Y')={\textstyle\frac{1}{g_1^2}}\bar YY',\qq
Y,Y'\in u(1).\ee
We denote by $\bar\cdot$ the complex conjugate and by $\cdot^*$
the Hermitean conjugate. Mind the different normalizations, they
are conventional. The $g_n$ are positive numbers, {\it the
gauge couplings.} For every simple factor of $G$ there
is one gauge coupling.

Then we need the Higgs potential $V(\varphi)$. It is
an invariant, fourth order, stable polynomial on
$\hh_S\owns\varphi$. Invariant means $V(\rho _S(u)\varphi
)=V(\varphi )$ for all $u\in G$.  Stable means
bounded from below. For $G=U(2)$ and the Higgs scalar in the
fundamental or defining representation,
$\varphi\in\hh_S=\cc^2$, $\rho_S(u)=u$, we have
\bb V(\varphi)=\lambda\,(\varphi^*\varphi)^2-
{\textstyle\frac{1}{2}}\mu^2\,\varphi^*\varphi.\ee
The coefficients of the Higgs potential are the Higgs
couplings, $\lambda$ must be positive for stability. We
say that the potential breaks $G$ spontaneously if no
minimum of the potential is a trivial orbit under $G$. In our
example, if $\mu$ is positive, the minima of
$V(\varphi)$ lie on the 3-sphere $|\varphi|=v:=
{\textstyle\frac{1}{2}}\mu/\sqrt\lambda$. $v$ is
called vacuum expectation value and $U(2)$ is said to
break down spontaneously to its little group
\bb U(1)\owns \pp{1&0\cr 0&e^{i\alpha }}.\ee
 The little group leaves invariant any
given point of the minimum, e.g. $\varphi=(v,0)^T$. On the other
hand, if $\mu$ is purely imaginary, then the
minimum of the potential is the origin, no
spontaneous symmetry breaking and the little group is all of $G$.

Finally, we need the Yukawa couplings $g_Y$. They
are the coefficients of the most general, real, trilinear
invariant on $\hh_L^*\,\ot\,\hh_R\,\ot\,(
\hh_S\op\hh_S^*)$. For every 1-dimensional
invariant subspace in the reduction of this tensor
representation, we have one complex Yukawa
coupling. For example $G=U(2)$, $\hh_L=\cc^2$, $\rho _L(u)\psi _L
=(\det u)^{q_L}u\,\psi _L,$ $\hh_R=\cc$, $\rho _R(u)\psi _R
=(\det u)^{q_R}\psi _R,$ $\hh_S=\cc^2$, $\rho _S(u)\varphi
=(\det u)^{q_S}u\,\varphi $. If $-q_L+q_R+q_S\not=0$ there is no
Yukawa coupling, otherwise there is one: $(\psi _L,\psi _R,\varphi
)={\rm Re} (g_Y\,\psi _L^*\psi _R\varphi )$.

If the symmetry is broken
spontaneously, gauge and Higgs bosons acquire masses
related to gauge and Higgs couplings, fermions
acquire masses equal to the `vacuum expectation value' $v$ times
the Yukawa couplings.

As explained in Jan-Willem van Holten's and Jean Zinn-Justin's
lectures at this School
\cite{van,zinn}, one must require for
consistency of the quantum theory that the  fermionic representations
be free of Yang-Mills
anomalies,
\bb\t ((\tilde \rho _L(X))^3)-\t ((\tilde \rho _R(X))^3)=0,\qq
{\rm for\ all}\ X\in\gg.\ee
We denote by $\tilde \rho $ the Lie algebra representation of the
group representation $\rho $. Sometimes one also wants  the mixed
Yang-Mills-gravitational anomalies to vanish:
\bb\t \tilde \rho _L(X)-\t \tilde \rho _R(X)=0,\qq
{\rm for\ all}\ X\in\gg.\ee

\subsection{Rules} \label{rules}

It is time to open the slot machine and to see how it
works. Its mechanism has five pieces:

\medskip
\noindent
{\bf The Yang-Mills action:}
 The actor in this piece is
$A=A_\mu \de x^\mu $, called connection, gauge potential, gauge
boson or Yang-Mills field. It is a 1-form on spacetime $M\owns x$
with values in the Lie algebra $\gg$,
$ A\in\Omega^1(M,\gg).$
We define its curvature or field strength,
\bb F:=\de A+{\textstyle\frac{1}{2}}[A,A]={\textstyle\frac{1}{2}}
F_{\mu \nu }\de x^\mu \de x^\nu \ \in
\Omega^2(M,\gg),\ee
and the Yang-Mills action,
\bb
S_{YM}[A]=-{\textstyle\frac{1}{2}}\int_M(F,*F)
=\,\frac{-1}{2g_n^2}\,\int_M\t F^*_{\mu \nu }F^{\mu \nu }
\de V .\ee
The gauge group $^MG$ is the infinite dimensional group of
differentiable functions $g:M\rightarrow G$ with pointwise
multiplication. $\cdot^*$ is the Hermitean conjugate of matrices,
$*\cdot$ is the Hodge star of differential forms. The space of all
connections carries an affine representation (cf Appendix)
$\rho_V$ of the gauge group:
\bb\rho_V(g)A=gAg^{-1}+g\de
g^{-1}.\label{inhom}\ee
Restricted to $x$-independent (`rigid')
gauge transformation, the representation is linear,
the adjoint one. The field strength transforms
homogeneously even under $x$-dependent (`local') gauge
transformations,
$g:M\rightarrow G$ differentiable,
\bb \rho_V(g)F=gFg^{-1},\ee
and, as the scalar product $(\cdot,\cdot)$ is invariant,
the Yang-Mills action is gauge invariant,
\bb S_{YM}[\rho_V(g)A]=S_{YM}[A] \qq{\rm for\ all}
\ g\in \ ^MG.\ee
Note that a mass term for the gauge bosons,
\bb{\textstyle\frac{1}{2}}\int_M m^2_A(A,*A)
=\,\frac{1}{g_n^2}\,\int_Mm^2_A\t A_\mu ^*A^\mu \de V ,\ee
is not gauge invariant because of the inhomogeneous
term in the transformation law of a connection
(\ref{inhom}). Gauge invariance forces the gauge
bosons to be massless.

In the Abelian case $G=U(1)$,  the Yang-Mills Lagrangian is
nothing but Maxwell's Lagrangian, the gauge boson $A$ is the
photon and its coupling constant $g$ is
$e/\sqrt{\epsilon _0}$.  Note however, that the Lie algebra of
$U(1)$ is $i\rr$ and the vector potential is purely
imaginary, while conventionally, in Maxwell's theory it is chosen
real. Its quantum
 version is $QED$, quantum electro-dynamics.
For $G=SU(3)$ and $\hh_L=\hh_R=\cc^3$ we have
today's theory of strong interaction, quantum
chromo-dynamics, $QCD$.

\medskip
\noindent
{\bf The Dirac action:}
 Schr\"odinger's action is
non-relativistic. Dirac generalized it to be Lorentz
invariant, e.g. \cite{bd}. The price to be paid is
twofold. His generalization only works for
spin ${\textstyle\frac{1}{2}}$ particles and requires
that for every such particle there must be an
antiparticle with same mass and opposite charges.
Therefore, Dirac's wave function $\psi(x)$ takes
values in $\cc^4$, spin up, spin down, particle,
antiparticle. Antiparticles have been discovered and
Dirac's theory was celebrated. Here it is in short for
(flat) Minkowski space of signature $+---$,
$\eta _{\mu \nu }=\eta ^{\mu \nu }={\rm diag }(+1,-1,-1,-1)$.
Define the four Dirac matrices,
\bb \gamma ^0=\pp{0&-1_2\cr -1_2&0},\qq \gamma
^j=
\pp{0&\sigma _j\cr -\sigma _j&0},\ee
for $j=1,2,3$ with the three Pauli matrices,
\bb\sigma
_1=\pp{0&1\cr 1&0},\qq\sigma _2=\pp{0&-i\cr
i&0},\qq\sigma _3=\pp{1&0\cr 0&-1}.\ee
They satisfy the anticommutation relations,
\bb\gamma^\mu\gamma^\nu+
\gamma^\nu\gamma^\mu=2\eta^{\mu\nu}1_4.\ee
In even spacetime dimensions, the chirality,
\bb \gamma_5:=-{\textstyle\frac{i}{4!}}\epsilon_{
\mu\nu\rho\sigma}\gamma^\mu\gamma^\nu
\gamma^\rho\gamma^\sigma=
- i\gamma^0\gamma^1\gamma^2\gamma^3=\pp{
-1_2&0\cr 0&1_2}\ee
is a natural operator and it paves the way to an
understanding of parity violation in weak interactions.
The chirality is a unitary matrix of unit square, which
anticommutes with all four Dirac matrices.
$(1-\gamma_5)/2$ projects a Dirac spinor onto its left-handed
part,
$(1+\gamma_5)/2$ projects onto the right-handed part. The
two parts are called Weyl spinors. A massless left-handed
(right-handed) spinor, has its spin parallel (anti-parallel) to its
direction of propagation.
 The chirality maps a left-handed spinor
to a right-handed spinor. A space reflection or parity
transformation changes the sign of the velocity vector and
leaves the spin vector unchanged. It therefore has the same
effect on Weyl spinors as the chirality operator.
 Similarly, there is the charge
conjugation, an anti-unitary operator (cf Appendix) of unit
square, that applied on a particle $\psi$ produces its
antiparticle
\bb J={\textstyle\frac{1}{i}}\gamma ^2\circ\,{\rm
complex\ conjugation}=\pp{0&0&0&-1\cr  0&0&1&0\cr
0&1&0&0\cr -1&0&0&0}\,\circ\,{\rm c\ c},\ee
i.e. $J\psi ={\textstyle\frac{1}{i}} \gamma ^2\
\bar\psi $. Attention, here and for the last time $\bar\psi $
stands for the complex conjugate of $\psi $. In a few lines we will
adopt a different more popular convention. The charge
conjugation commutes with all four Dirac matrices times $i$. In flat
spacetime, the free Dirac operator is simply defined by,
\bb \ddd:=i\hbar\gamma^\mu\pa_\mu.\ee
It is sometimes referred to as square root of the wave
operator because $\ddd^2=-\Box$.
 The coupling of the
Dirac spinor to the gauge potential $A=A_\mu\de
x^\mu$ is done via the covariant derivative, and called
minimal coupling. In order to break parity, we write
left- and right-handed parts independently:
\bb S_D[A,\psi_L,\psi_R]&=&
\int_M\bar\psi_L\left[\ddd+i\hbar
\gamma^\mu\tilde\rho_L(A_\mu)\right]
\,\frac{1-\gamma_5}{2}\,\psi_L\,\de V\cr
&&+\int_M\bar\psi_R\left[\ddd+i\hbar
\gamma^\mu\tilde\rho_R(A_\mu)\right]
\,\frac{1+\gamma_5}{2}\,\psi_R\,\de V.
\label{diracaction}\ee
The new actors in this piece are $\psi_L$ and $\psi_R$,
two multiplets of Dirac spinors or fermions, that is with values
in $\hh_L$ and
$\hh_R$. We use the notations,
$\bar\psi:=\psi^*\gamma^0$, where $\cdot^*$ denotes the
Hermitean conjugate with respect to the four spinor components
and the dual with respect to the scalar product in the (internal)
Hilbert space
$\hh_L$ or $\hh_R$. The $\gamma^0$ is needed for
energy reasons and for invariance of the
pseudo--scalar product of spinors under lifted
Lorentz transformations. The
$\gamma^0$ is absent if spacetime is Euclidean. Then
we have a genuine scalar product and the square
integrable  spinors form a Hilbert space
$\lll^2(\sss)=\lll^2(\rr^4)\ot\cc^4$, the infinite dimensional
brother of the internal one. The Dirac operator is then self
adjoint in this Hilbert space. We denote by
$\tilde\rho_L$ the Lie algebra
representation in $\hh_L$.
 The covariant derivative,
$\dee_\mu:=\pa_\mu+\tilde\rho_L(A_\mu)$, deserves
its name,
\bb
\left[\pa_\mu+\tilde\rho_L(\rho_V(g)A_\mu)\right]
(\rho_L(g)\psi_L)=\rho_L(g)
\left[\pa_\mu+\tilde\rho_L(A_\mu)\right]\psi_L,\ee
for all gauge transformations $g\in\, ^M\!G$. This
ensures that the Dirac action (\ref{diracaction}) is gauge invariant.

If parity is conserved, $\hh_L=\hh_R$, we may add a mass term
\bb -c\int_M\bar\psi_R\, m_\psi
\,\frac{1-\gamma_5}{2}\,\psi_L\,\de V\ -\
c\int_M\bar\psi_L\, m_\psi
\,\frac{1+\gamma_5}{2}\,\psi_R\,\de V
\ =\ -c\int_M\bar\psi\, m_\psi
\,\psi\,\de V\ee
 to the Dirac action. It gives identical masses to all
members of the multiplet.
The fermion masses are gauge invariant if all
fermions in $\hh_L=\hh_R$ have the same mass.  For instance
 $QED$ preserves parity,
$\hh_L=\hh_R=\cc$, the representation being characterized by
the electric charge, $-1$ for both the left- and right handed
electron.
Remember that gauge invariance forces gauge bosons
to be massless. For fermions, it is parity {\it non-}invariance
that forces them to be massless.

 Let us conclude by reviewing
briefly why the Dirac equation is the Lorentz
invariant generalization of the Schr\"odinger
equation. Take the free Schr\"odinger equation on
(flat)
$\rr^4$. It is a linear differential equation with
constant coefficients,
\bb \left(\,\frac{2m}{i\hbar}\,\frac{\pa}{\pa
t}\,-\Delta\right)\psi=0.\label{deB}\ee
We compute its polynomial following Fourier and de
Broglie,
\bb -\,\frac{2m}{\hbar}\,\omega+k^2=
-\,\frac{2m}{\hbar
^2}\,\left[E-\,\frac{p^2}{2m}\,\right].\ee
Energy conservation in Newtonian mechanics is
equivalent to the vanishing of the polynomial.
Likewise, the polynomial of the free, massive Dirac
equation $(\ddd -cm_\psi)\psi=0$ is
\bb {\textstyle\frac{\hbar}{c}}\,\omega\gamma^0
+\hbar\,k_j\gamma^j-c\,m1.\ee
Putting it to zero implies energy conservation in
special relativity,
\bb ({\textstyle\frac{\hbar}{c}})^2\,\omega^2
-\hbar^2\,\vec k^2-c^2\,m^2=0.\ee
In this sense, Dirac's equation generalizes Schr\"odinger's to
special relativity. To see that Dirac's equation is really Lorentz
invariant we must lift the Lorentz transformations to the space
of spinors. We will come back to this lift.

So far we have seen the two noble pieces by Yang-Mills
and Dirac.
The remaining three pieces are cheap copies of the two noble ones
with the gauge boson $A$ replaced by a scalar
$\varphi$. We need these three pieces to cure only
one problem, give masses to some gauge bosons and to
some fermions. These masses are forbidden by gauge
invariance and parity violation.  To simplify the
notation we will work from now on in units with
$c=\hbar=1$.

\medskip
\noindent
{\bf The Klein-Gordon action:}
The Yang-Mills action contains the kinetic term for
the gauge boson. This is simply the quadratic term,
$(\de A,\de A)$, which by Euler-Lagrange produces
linear field equations. We copy this for our new actor,
a multiplet of scalar fields or Higgs bosons,
\bb \varphi\in\Omega^0(M,\hh_S), \ee
by writing the Klein-Gordon action,
\bb S_{KG}[A,\varphi]={\textstyle\frac{1}{2}}
\int_M (\dee \varphi)^**\dee \varphi
=  {\textstyle\frac{1}{2}} \int_M(\dee_\mu \varphi )^*
\dee^\mu \varphi\, \de V,\ee
with the covariant derivative here defined with
respect to the scalar representation,
\bb\dee\varphi:=\de
\varphi+\tilde\rho_S(A)\varphi.\ee
Again we need this minimal coupling
$\varphi^*A\varphi$ for gauge invariance.

\medskip
 \noindent
{\bf The Higgs potential:}
The non-Abelian Yang-Mills action contains
interaction terms for the gauge bosons, an
invariant, fourth order polynomial, $2(\de
A,[A,A])+([A,A],[A,A])$. We mimic these interactions
for scalar bosons by adding the integrated Higgs
potential $\int_M*V(\varphi)$ to the action.

\medskip
\noindent
{\bf The Yukawa terms:}
We also mimic the (minimal) coupling of the gauge
boson to the fermions $\psi^*A\psi$ by writing all
possible trilinear invariants,
\bb S_Y[\psi_L,\psi_R,\varphi]:={\rm Re}
\int_M*\left(
\sum_{j=1}^ng_{Yj}\left(\psi_L^*,\psi_R,\varphi
\right)_j+\sum_{j=n+1}^mg_{Yj}\left(\psi_L^*,\psi_R,
\varphi^*\right)_j\right)
.\ee
In the standard model, there are 27 complex Yukawa
couplings, $m=27$.

\medskip

The Yang-Mills and Dirac actions, contain three
types of couplings, a trilinear self coupling $AAA$, a
quadrilinear self coupling $AAAA$ and the trilinear
minimal coupling $\psi^*A\psi$. The gauge self
couplings are absent if the group $G$ is Abelian, the
photon has no electric charge, Maxwell's
equations are linear.
 The beauty of gauge
invariance is that if $G$ is simple, all these couplings
are fixed in terms of one positive number, the gauge
coupling $g$. To see this, take an orthonormal basis
$T_b,\ b=1,2,...\dim G$ of the complexification $\gg^\cc$ of the
Lie algebra with respect to the invariant scalar
product and an orthonormal basis $F_k,\
k=1,2,...\dim\hh_L$, of the fermionic Hilbert space, say
$\hh_L$, and expand the actors,
\bb A =: A_\mu^b T_b \de x^\mu,\qq
\psi=:\psi^kF_k.\ee
Insert these expressions into the Yang-Mills and Dirac
actions, then you get the following interaction terms, see
Figure 3,
\bb g\,
\pa_\rho A_\mu^aA_\nu^bA_\sigma^c\,f_{abc}\,
\epsilon^{\rho\mu\nu\sigma},\qq
g^2\,A_\mu^aA_\nu^bA_\rho^cA_\sigma^d\,{f_{ab}}^e
{f_{ecd}}\,
\epsilon^{\rho\mu\nu\sigma},\qq
g\,\psi^{k*}A^b_\mu\gamma^\mu\psi_\ell\,
{{t_b}_k}^\ell,\ee
with the structure constants ${f_{ab}}^e$,
\bb [T_a,T_b]=:{f_{ab}}^eT_e.\ee
The indices of the structure constants are raised and
lowered with the matrix of the invariant scalar
product in the basis $T_b$, that is the identity matrix.
The ${{t_b}_k}^\ell$ is the matrix  of the operator
$\tilde\rho_L(T_b)$ with respect to the basis $F_k$.
The difference between the noble and the cheap actions
is that the Higgs couplings, $\lambda$ and $\mu$ in
the standard model, and the Yukawa couplings
$g_{Yj}$ are arbitrary, are neither connected among
themselves nor connected to the gauge couplings
$g_i$.

\begin{figure}[h]
\epsfxsize=16cm
\hspace{.25cm}
\epsfbox{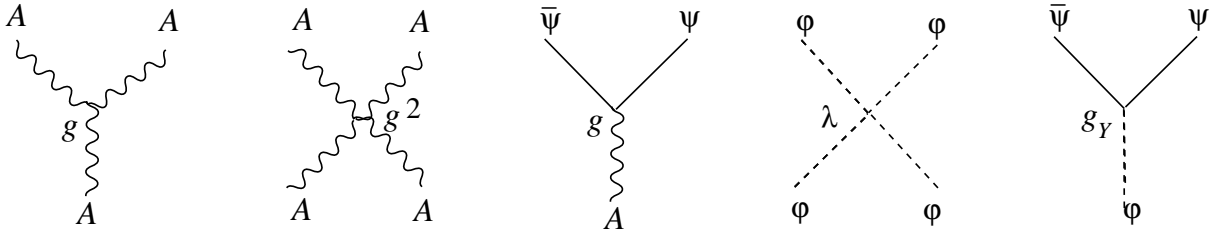}
\caption{Tri- and quadrilinear gauge couplings,
 minimal gauge coupling to fermions, Higgs selfcoupling and
Yukawa coupling }
\label{couplings}
\end{figure}

\subsection{The winner} \label{winner}

Physicists have spent some thirty years and billions of
Swiss Francs playing on the slot machine by
Yang, Mills and Higgs. There is a winner, the standard
model of electro-weak and strong forces. Its
bills are
\bb G&=&SU(2)\times U(1)\times
SU(3)/(\zz_2\times\zz_3),\label{smgr}\\ \cr
\hh_L &=& \bigoplus_1^3\lb
(2,{\textstyle\frac{1}{6}},3)\op
(2,-{\textstyle\frac{1}{2}},1)
\rb  ,\label{hl}\\
 \hh_R& = &\bigoplus_1^3\lb
(1,{\textstyle\frac{2}{3}},3)\oplus
(1,-{\textstyle\frac{1}{3}},3)\op (1,-1,1)
\rb,\label{hr} \\
 \hh_S &= &(2,-{\textstyle\frac{1}{2}},1)\label{hs},
\ee
where $(n_2, y, n_3)$
denotes the tensor product of an $n_2$ dimensional
representation of $SU(2)$, an $n_3$ dimensional
representation of $SU(3)$ and the one dimensional
representation of $U(1)$ with hypercharge $y$:
$\rho(\exp (i\theta)) = \exp (iy\theta) $. For
historical reasons the hypercharge is an integer
multiple of ${\textstyle\frac{1}{6}}$.
This is irrelevant: only the product of the
hypercharge with its gauge coupling is measurable and we do not
need multi-valued representations, which are characterized by
non-integer, rational hypercharges.
 In the direct sum, we recognize the three
generations of fermions, the quarks are $SU(3)$
colour triplets, the leptons colour singlets. The basis
of the fermion representation space is
\bb \pp{u\cr d}_L,\ \pp{c\cr s}_L,\ \pp{t\cr b}_L,\
\pp{\nu_e\cr e}_L,\ \pp{\nu_\mu\cr\mu}_L,\
\pp{\nu_\tau\cr\tau}_L\eee
\bb\matrix{u_R,\cr d_R,}\qq \matrix{c_R,\cr s_R,}\qq
\matrix{t_R,\cr b_R,}\qq  e_R,\qq \mu_R,\qq
\tau_R\eee
The
parentheses indicate isospin doublets.

The eight gauge bosons associated to
$su(3)$ are called gluons. Attention, the $U(1)$ is not the one of electric
charge, it is called hypercharge, the electric charge
is a linear combination of hypercharge and weak
isospin, parameterized by the weak mixing angle
$\theta_w$ to be introduced below. This mixing is
necessary to give electric charges to the $W$ bosons.
The $W^+$ and $W^-$ are pure isospin states, while the
$Z^0$ and the photon are (orthogonal) mixtures of the
third isospin generator and hypercharge.

 Because of the high
degree of reducibility in the bills, there are many
coins, among them 27 complex Yukawa couplings. Not
all Yukawa couplings have a physical meaning and we only remain
with 18 physically significant, positive numbers
\cite{data}, three gauge couplings at energies corresponding to
the $Z$ mass,
\bb g_1=0.3574\pm 0.0001,&g_2=0.6518\pm 0.0003,&
g_3=1.218\pm 0.01,\label{gaugecoup}\ee
two Higgs couplings, $\lambda$ and $\mu$, and
13 positive parameters from the Yukawa couplings.
The Higgs couplings are related to the boson masses:
\bb m_W&=&{\textstyle\frac{1}{2}}g_2\,v
\,=\,80.419\pm 0.056\ {\rm GeV},\\
m_Z&=&{\textstyle\frac{1}{2}}\sqrt{g_1^2+g_2^2}\ v
=m_W/\cos\theta_w
\,=\,91.1882\,\pm\,0.0022\ {\rm GeV},\\
m_H&=&2\sqrt 2\sqrt\lambda\,v\,>\,98\ {\rm GeV},\ee
with the vacuum expectation value
$v:={\textstyle\frac{1}{2}}\mu/\sqrt\lambda$ and the
weak mixing angle $\theta_w$ defined by
\bb\sin^2\theta_w:=g_2^{-2}/(g_2^{-2}+g_1^{-2})=
0.23117\,\pm\,0.00016.\ee
 For
the standard model, there is a one--to--one
correspondence between the physically relevant part
of the Yukawa couplings and the fermion masses and
mixings,
\bb m_e=0.510998902\pm 0.000000021\ {\rm MeV},&
m_u=3\pm 2\ {\rm MeV},&m_d=6\pm 3\ {\rm MeV},
\cr
m_\mu=0.105658357\pm 0.000000005\ {\rm GeV},&
m_c=1.25\pm 0.1\ {\rm GeV},&
m_s=0.125\pm 0.05\ {\rm GeV},\cr
m_\tau=1.77703 \pm 0.00003\ {\rm GeV},&
m_t=174.3\pm 5.1\ {\rm GeV},&
m_b=4.2\pm 0.2\ {\rm GeV}.\eee
For simplicity, we take massless neutrinos. Then  mixing only
occurs for quarks and is given by
a unitary matrix, the Cabibbo-Kobayashi-Maskawa
matrix
\bb C_{KM}:=\pp{V_{ud}&V_{us}&V_{ub}\cr
V_{cd}&V_{cs}&V_{cb}\cr  V_{td}&V_{ts}&V_{tb}}.\ee
For physical purposes it can be parameterized by
three angles $\theta_{12}$,
$\theta_{23}$, $\theta_{13}$ and
one $CP$ violating phase $\delta$:
\bb C_{KM}=\pp{
c_{12}c_{13}&s_{12}c_{13}&s_{13}e^{-i\delta}\cr
-s_{12}c_{23}-c_{12}s_{23}s_{13}e^{i\delta}&
c_{12}c_{23}-s_{12}s_{23}s_{13}e^{i\delta}&
s_{23}c_{13}\cr
s_{12}s_{23}-c_{12}c_{23}s_{13}e^{i\delta}&
-c_{12}s_{23}-s_{12}c_{23}s_{13}e^{i\delta}&
c_{23}c_{13}},\ee
with $c_{kl}:=\cos \theta_{kl}$,
$s_{kl}:=\sin \theta_{kl}$.
The
absolute values of the matrix elements in $C_{KM}$ are:
\bb \pp{
0.9750\pm 0.0008&0.223\pm 0.004&0.004\pm
0.002\cr
0.222\pm 0.003&0.9742\pm 0.0008&0.040\pm 0.003\cr
0.009\pm 0.005&0.039\pm 0.004&0.9992\pm 0.0003}.
\ee
The physical meaning of the quark mixings is the
following: when a sufficiently energetic $W^+$ decays
into a $u$ quark, this
$u$ quark is produced together with a
$\bar d$ quark with probability $|V_{ud}|^2$, together
with a
$\bar s$ quark with probability $|V_{us}|^2$, together
with a
$\bar b$ quark with probability $|V_{ub}|^2$. The
fermion masses and mixings together are an entity,
the fermionic mass matrix or the matrix of Yukawa
couplings multiplied by the vacuum expectation
value.

Let us note
six intriguing properties of the standard model.
\begin{itemize}\item
The gluons couple in the same way to left- and
right-handed fermions, the gluon coupling is
vectorial, the strong interaction does not break parity.
\item
The fermionic mass matrix commutes with $SU(3)$, the three
colours of a given quark have the same mass.
\item
 The scalar is a colour singlet, the
$SU(3)$ part of $G$ does not suffer spontaneous symmetry break
down, the gluons remain massless.
\item
The $SU(2)$ couples only to left-handed fermions, its
coupling is chiral, the weak interaction breaks parity
maximally.
\item
The scalar is an isospin doublet, the $SU(2)$ part
suffers spontaneous symmetry break down, the $W^\pm$ and the
$Z^0$ are massive.
\item
The remaining colourless and neutral gauge boson, the photon, is
massless and couples vectorially. This is certainly the most ad-hoc
feature of the standard model. Indeed the photon is a linear
combination of isospin, which couples only to left-handed
fermions, and of a $U(1)$ generator, which may couple to both
chiralities. Therefore  only the
careful fine tuning of the hypercharges in the three input
representations (\ref{hl}-\ref{hs}) can save parity conservation
and gauge invariance of electro-magnetism,
\bb
 y_{u_R}=y_{q_L}-y_{\ell_L}\qq
y_{d_R}=y_{q_L}+y_{\ell_L},\qq
y_{e_R}=2y_{\ell_L},\qq
y_\varphi =y_{\ell_L} ,
\label{4cond}\ee
The subscripts label the multiplets, $qL$ for the left-handed
quarks,
$\ell L$ for the left-handed leptons,
$uR$ for the right-handed up-quarks and
so forth and $\varphi $ for the scalar.
\end{itemize}
Nevertheless the phenomenological success of the standard model
is phenomenal: with only a handful of parameters, it reproduces
correctly some millions of experimental numbers. Most of these
numbers are measured with an accuracy of a few percent and
they can be reproduced by classical field theory, no $\hbar$
needed. However, the experimental precision has become so good
that quantum corrections cannot be ignored anymore. At this point
it is important to note that the fermionic representations of the
standard model are free of Yang-Mills (and mixed) anomalies. Today
the standard model stands uncontradicted.

Let us come back to our analogy between the Balmer-Rydberg
formula and the standard model. One might object that the ansatz
for the spectrum, equation (\ref{ansatz}), is completely ad hoc,
while the class of all (anomaly free) Yang-Mills-Higgs models is
distinguished by perturbative renormalizability. This is true, but
this property was proved \cite{renorm} only years after
the electro-weak part of the standard model was published
\cite{gsw}.

By placing the hydrogen atom in an electric or magnetic field, we
know experimentally that every frequency `state'
$n$, $n=1,2,3,...$, comes with $n$ irreducible unitary
representations of the rotation group $SO(3)$. These representations
are labelled by
$\ell$,
$\ell=0,1,2,...n-1$, of dimensions $2\ell+1$. An orthonormal basis of
each representation
$\ell$ is labelled by another integer $m$, $m=-\ell,-\ell+1,...\ell$.
This experimental fact has motivated the credo that particles are
orthonormal basis vectors of unitary representations of compact
groups. This credo is also behind the standard model. While $SO(3)$
has a clear geometric interpretation, we are still looking for
such an interpretation of $SU(2)\times U(1)\times
SU(3)/[\zz_2\times\zz_3].$

We close this subsection with Iliopoulos' joke \cite{joke} from 1976:
\vfil\eject
\medskip
\noindent{\bf Do-it-yourself kit for gauge models:}
\begin{enumerate}
\item[\bf 1)]
Choose a gauge group $G$.
\item[\bf 2)]
Choose the fields of the ``elementary particles'' you want to
introduce, and their representations. Do not forget to include
enough fields to allow for the Higgs mechanism.
\item[\bf 3)]
Write the most general renormalizable Lagrangian invariant
under $G$. At this stage gauge invariance is still exact and all
vector bosons are massless.
\item[\bf 4)]
Choose the parameters of the Higgs scalars so that spontaneous
symmetry breaking occurs. In practice, this often means to choose
a negative value [positive in our notations] for the parameter $\mu
^2$.
\item[\bf 5)]
Translate the scalars and rewrite the Lagrangian in terms of the
translated fields. Choose a suitable gauge and quantize the theory.
\item[\bf 6)]
Look at the properties of the resulting model. If it resembles
physics, even remotely, publish it.
\item[\bf 7)]
GO TO \bf 1.
\end{enumerate}
Meanwhile his joke has become experimental reality.

\subsection{Wick rotation}

Euclidean signature is technically easier to handle than
Minkowskian. What is more, in Connes' geometry it will be vital
that the spinors form a Hilbert space with a true scalar product
and that the Dirac action takes the form of a scalar product. We
therefore put together the Einstein-Hilbert and Yang-Mills-Higgs
actions  with emphasis on the relative signs and indicate the
changes necessary to pass from Minkowskian to Euclidean
signature.

In 1983 the meter disappeared as fundamental unit of
science and technology. The conceptual revolution of
general relativity, the abandon of length in favour of
time, had made its way up to the domain of
technology. Said differently, general relativity is not
really  geo-metry, but chrono-metry. Hence our choice of
Minkowskian signature is $+---$.

With this choice
the combined
Lagrangian reads,
\bb
&&\{
-\,{\textstyle\frac{2\Lambda _c}{16\pi G}}
\,-\,{\textstyle\frac{1}{16\pi G}}\,R
\,-\,
{\textstyle\frac{1}{2g^2}}\t (
F_{\mu\nu}^{*}F^{\mu\nu})\,+\,
{\textstyle\frac{1}{g^2}}m_A^2\t (
A_{\mu}^{*}A^{\mu})\cr
&&+\,{\textstyle\frac{1}{2}}\,(\dee_\mu\varphi)^*
\dee^\mu\varphi
\,-\,{\textstyle\frac{1}{2}}\,m_\varphi^2|\varphi|^2
\,+\,{\textstyle\frac{1}{2}}\,\mu^2|\varphi|^2
\,-\,\lambda |\varphi|^4\cr
&&+\,
\psi^*\gamma^{0}\,[i\gamma^\mu\dee_\mu
\,-\,m_\psi 1_4]\, \psi\}
\ |\!\det g_{\cdot\cdot}|^{1/2}.\ee
This Lagrangian is real if we suppose that all fields
vanish at infinity.  The relative coefficients between
kinetic terms and mass terms are chosen as to
reproduce the correct energy momentum relations
from the free field equations using Fourier transform
and the de Broglie relations as explained after
equation (\ref{deB}). With the chiral decomposition
\bb\psi_L&=&{\textstyle\frac{1-\gamma_5}{2}}\,\psi,
\qq
\psi_R\ =\ {\textstyle\frac{1+\gamma_5}{2}}\,\psi,
\label{project}\ee
the Dirac Lagrangian reads
\bb &&\psi^*\gamma^0\,[i\gamma^\mu\dee_\mu
\,-\,m_\psi 1_4]\,
\psi\cr
&&\qq=\psi_L^*\gamma^0\,i\gamma^\mu\dee_\mu\,
\psi_L
\,+\,
\psi_R^*\gamma^0\,i\gamma^\mu\dee_\mu\,\psi_R
\,-\,m_\psi\psi_L^*\gamma^0\psi_R
\,-\,m_\psi\psi_R^*\gamma^0 \psi_L.\ee
The  relativistic energy momentum
relations are quadratic in the masses. Therefore the
sign of the fermion mass
$m_\psi$ is conventional and merely reflects the
choice: who is particle and who is antiparticle. We can
even adopt one choice for the left-handed fermions
and the opposite choice for the right-handed
fermions. Formally this can be seen by the change of
field variable (chiral transformation):
\bb \psi:=\exp(i\alpha\gamma_5)\,\psi'.\ee
It leaves invariant the kinetic term and the mass term
transforms as,
\bb -m_\psi{\psi'}^*\gamma^0[\cos (2\alpha)\,1_4
+i\sin (2\alpha)\,\gamma_5]{\psi'}. \ee
With
$\alpha=-\pi/4$ the Dirac Lagrangian becomes:
\bb
&&{\psi'}^*\gamma^0[\,i\gamma^\mu\dee_\mu\,+i
m_\psi\gamma_5]{\psi}'\cr
&&\qq =
{\psi'}_L^*\gamma^0\,i\gamma^\mu\dee_\mu\,
{\psi}'_L
\,+\,
{\psi'}_R^*\gamma^0\,i\gamma^\mu\dee_\mu\,
{\psi'}_R
\,+\,m_\psi {\psi'}_L^*\gamma^0i
\gamma_5{\psi'}_R
\,+\,m_\psi {\psi'}_R^*\gamma^0i
\gamma_5 {\psi'}_L\cr
&&\qq =
{\psi'}_L^*\gamma^0\,i\gamma^\mu\dee_\mu\,
{\psi}'_L
\,+\,
{\psi'}_R^*\gamma^0\,i\gamma^\mu\dee_\mu\,
{\psi'}_R
\,+\,im_\psi {\psi'}_L^*\gamma^0{\psi'}_R
\,-\,im_\psi {\psi'}_R^*\gamma^0 {\psi'}_L.\ee
We have seen that gauge invariance forbids
massive gauge bosons, $m_A=0$, and that parity
violation forbids massive fermions, $m_\psi=0$. This
is fixed by spontaneous symmetry breaking, where we
take the scalar mass term with wrong sign,
$m_\varphi=0,\
\mu>0$. The shift of the scalar then induces masses
for the gauge bosons, the fermions and the physical
scalars. These masses are calculable in terms of the
gauge, Yukawa, and Higgs couplings.

The other relative signs in the combined Lagrangian
are fixed by the requirement that the energy density
of the non-gravitational part $T_{00}$ be positive
(up to a cosmological constant) and that gravity in the
Newtonian limit be attractive. In particular this
implies that the Higgs potential must be bounded from
below,
$\lambda>0$. The sign of the Einstein-Hilbert action
may also be obtained from an
asymptotically flat space of weak curvature, where we
can define gravitational energy density. Then
the requirement is that the kinetic terms
of all physical bosons, spin 0, 1, and 2, be of the same
sign. Take the metric of the form
\bb g_{\mu\nu}=\eta_{\mu\nu}+h_{\mu\nu},\ee
$h_{\mu\nu}$ small. Then the Einstein-Hilbert
Lagrangian becomes \cite{gilles},
\bb -\,{\textstyle\frac{1}{16\pi G}}\,R\,
|\!\det g_{\cdot\cdot}|^{1/2}&=&
{\textstyle\frac{1}{16\pi G}}\{
{\textstyle\frac{1}{4}}\pa_\mu h_{\alpha\beta}
\pa^\mu h^{\alpha\beta}\,-\,
{\textstyle\frac{1}{8}}\pa_\mu{h_\alpha}^\alpha
\pa^\mu{h_\beta}^\beta\cr
&&-\,[\pa_\nu{h_\mu}^\nu
-{\textstyle\frac{1}{2}}\pa_\mu{h_\nu}^\nu]
[\pa_{\nu'}{h^\mu}^{\nu'}
-{\textstyle\frac{1}{2}}\pa^\mu{h_{\nu'}}^{\nu'}
]\,+\,O(h^3)\}.\ee
Here indices are raised with $\eta^{\cdot\cdot}$. After
an appropriate choice of coordinates, `harmonic
coordinates', the bracket $\left[\pa_\nu{h_\mu}^\nu
-{\textstyle\frac{1}{2}}\pa_\mu{h_\nu}^\nu
\right]$ vanishes and only two independent
components of $h_{\mu\nu}$ remain,
$h_{11}=-h_{22}$ and $h_{12}$.  They represent
the two physical states of the graviton,
helicity $\pm 2$. Their kinetic terms are both positive,
e.g.:
\bb +{\textstyle\frac{1}{16\pi
G}}{\textstyle\frac{1}{4}}
\pa_\mu h_{12}\pa^\mu h_{12}.\ee
Likewise, by an appropriate gauge transformation, we
can achieve $\pa_\mu A^\mu=0$, `Lorentz gauge',
and remain with only two `transverse' components
$A_1,\ A_2$ of helicity $\pm 1$. They have positive
kinetic terms, e.g.:
\bb+{\textstyle\frac{1}{2g^2}}\t (\pa_\mu
A_1^*\pa^\mu A_1).\ee
Finally, the kinetic term of the scalar is positive:
\bb +{\textstyle\frac{1}{2}}\pa_\mu\varphi^*
\pa^\mu\varphi.\ee

An old recipe from quantum field theory, `Wick
rotation', amounts to replacing spacetime by a
Riemannian manifold with Euclidean signature. Then
certain calculations become feasible or easier. One of
the reasons for this is that Euclidean quantum field
theory resembles statistical mechanics, the imaginary
time playing formally the role of the inverse
temperature. Only at the end of the calculation the
result is `rotated back' to real time. In some cases, this
recipe can be justified rigorously. The precise
formulation of the recipe is that the $n$-point
functions computed from the Euclidean Lagrangian
be the analytic continuations in the complex time
plane of the Minkowskian
$n$-point functions. We shall indicate a hand
waving formulation of the recipe, that
is sufficient for our purpose: In a first stroke we pass to the signature
$-+++$. In a second stroke we replace $t$ by $it$ and
replace all Minkowskian scalar products by the
corresponding Euclidean ones.

The first stroke amounts simply to replacing the
metric by its negative. This leaves invariant the
Christoffel symbols, the Riemann and Ricci tensors,
but reverses the sign of the curvature scalar.
Likewise, in the other terms of the Lagrangian we get
a minus sign for every contraction of indices, e.g.:
$\pa_\mu\varphi^*\pa^\mu\varphi=
\pa_\mu\varphi^*\pa_{\mu'}\varphi g^{\mu\mu'}$
becomes $\pa_\mu\varphi^*\pa_{\mu'}\varphi
(-g^{\mu\mu'})=-\pa_\mu\varphi^*\pa^\mu\varphi$.
After multiplication by a conventional overall minus
sign the combined Lagrangian reads now,
\bb
&&\{\,{\textstyle\frac{2\Lambda _c}{16\pi
G}}\,-\,{\textstyle\frac{1}{16\pi G}}\,R
\,+\,
{\textstyle\frac{1}{2g^2}}\t (
F_{\mu\nu}^{*}F^{\mu\nu})\,
\,+\,{\textstyle\frac{1}{g^2}}m_A^2\t (A^*_\mu A^\mu )
\cr && +\,
{\textstyle\frac{1}{2}}\,(\dee_\mu\varphi)^*
\dee^\mu\varphi
\,+\,{\textstyle\frac{1}{2}}\,m_\varphi ^2|\varphi|^2
\,-\,{\textstyle\frac{1}{2}}\,\mu^2|\varphi|^2
\,+\,\lambda |\varphi|^4\cr
&&\qq+\,
\psi^*\gamma^0[\,i\gamma^\mu\dee_\mu
\,+\,m_\psi 1_4\,]\psi\,\}
\ |\!\det g_{\cdot\cdot}|^{1/2}.\label{-+++}\ee

To pass to the Euclidean signature, we multiply time,
energy and mass by $i$. This amounts to
$\eta^{\mu\nu}=\delta^{\mu\nu}$ in the scalar
product. In order to have the Euclidean
anticommutation relations,
\bb   \gamma ^\mu \gamma ^\nu +\gamma ^\nu
\gamma ^\mu =  2\delta ^{\mu \nu }1_4,\ee
we change the Dirac matrices to the Euclidean ones,
\bb \gamma ^0=\pp{0&-1_2\cr -1_2&0},\qq \gamma
^j= {\textstyle\frac{1}{i}}
\pp{0&\sigma _j\cr -\sigma _j&0},\ee
All four are now self adjoint.
For the chirality we take
\bb \gamma_5:=
\gamma^0\gamma^1\gamma^2\gamma^3=\pp{
-1_2&0\cr 0&1_2}.\ee
 The Minkowskian scalar product for
spinors has a
$\gamma^0$. This $\gamma^0$ is needed for the
correct physical interpretation of the energy of
antiparticles and for invariance under lifted Lorentz
transformations,
$Spin(1,3)$. In the Euclidean, there is no physical interpretation
and we can only retain the requirement of a
$Spin(4)$ invariant scalar product. This scalar
product has no $\gamma^0$. But then we have a
problem if we want to write the Dirac Lagrangian in
terms of chiral spinors as above. For instance, for a purely
left-handed neutrino, $\psi _R=0$ and
$\psi_L^*\,i\gamma^\mu\dee_\mu\, \psi_L$ vanishes
identically because $\gamma_5$ anticommutes with
the four $\gamma^\mu$. The standard trick of
Euclidean field theoreticians \cite{zinn} is fermion doubling,
$\psi_L$ and $\psi_R$ are treated as two
{\it independent}, four component spinors. They are
not chiral projections of one four component spinor
as in the Minkowskian, equation (\ref{project}). The
spurious degrees of freedom in the Euclidean are kept
all the way through the calculation. They are
projected out only after the Wick rotation back to
Minkowskian, by imposing $\gamma_5\psi_L=-\psi_L,
\gamma_5\psi_R=\psi_R$.

 In noncommutative
geometry the Dirac operator must be self adjoint,
which is not the case for the Euclidean Dirac operator
$i\gamma^\mu\dee_\mu+im_\psi 1_4$ we get from
the  Lagrangian (\ref{-+++}) after multiplication of
the mass by $i$. We therefore prefer the primed
spinor variables $\psi'$ producing the self adjoint
Euclidean Dirac  operator
$i\gamma^\mu\dee_\mu+m_\psi\gamma_5$.
Dropping the prime, the combined Lagrangian in the
Euclidean then reads:
\bb
&&\{\,{\textstyle\frac{2\Lambda _c}{16\pi
G}}\,-\,{\textstyle\frac{1}{16\pi G}}\,R
\,+\,
{\textstyle\frac{1}{2g^2}}\t (
F_{\mu\nu}^{*}F^{\mu\nu})\,
\,+\,{\textstyle\frac{1}{g^2}}m_A^2\t (A^*_\mu A^\mu )
\cr && +\,
{\textstyle\frac{1}{2}}\,(\dee_\mu\varphi)^*
\dee^\mu\varphi
\,+\,{\textstyle\frac{1}{2}}\,m_\varphi ^2|\varphi|^2
\,-\,{\textstyle\frac{1}{2}}\,\mu^2|\varphi|^2
\,+\,\lambda |\varphi|^4\cr
&&+\,
{\psi}_L^*\,i\gamma^\mu\dee_\mu\,
{\psi}_L
\,+\,
{\psi}_R^*\,i\gamma^\mu\dee_\mu\,
{\psi}_R
\,+\,m_\psi {\psi}_L^*\gamma_5{\psi}_R
\,+\,m_\psi {\psi}_R^*\gamma_5 {\psi}_L
\}
\,(\det g_{\cdot\cdot})^{1/2}.\label{++++}\ee

\section{Connes' noncommutative geometry}

Connes equips Riemannian spaces with an uncertainty principle.
As in quantum mechanics, this uncertainty principle is derived from
noncommutativity.

\subsection{Motivation: quantum mechanics}

Consider the classical harmonic oscillator. Its phase space is
$\rr^2$ with points labelled by position $x$ and momentum $p$. A
classical observable is a differentiable function on phase
space such as the total energy $p^2/(2m)\,+\,kx^2$.
Observables can be added and multiplied, they form the algebra
$\ccc^\infty(\rr^2)$, which is associative and commutative. To pass
to quantum mechanics, this algebra is rendered noncommutative
by means of the following noncommutation relation for the
generators
$x$ and $p$,
\bb [x,p]=i\hbar 1.\label{heisen}\ee
Let us call $\aa$ the resulting algebra `of quantum observables'. It
is still associative, has an involution $\cdot^*$ (the adjoint or Hermitean
conjugation)  and a unit 1. Let us briefly recall the defining properties
of an involution: it is a linear map from the {\it real} algebra into itself
that reverses the product, $(ab)^*=b^*a^*$, respects the unit,
$1^*=1$, and is such that $a^{**}=a$.
 \begin{figure}[h]
\hspace{5.5cm}
\setlength{\unitlength}{1.0cm}
\begin{picture}(10,6.5)(0.5,0)
\put(3,2){\framebox(1,1)}
\put(0,0.5){\vector(1,0){6}}
\put(0.5,5.3){\parbox[b]{2cm}{$p$}}
\put(0.5,0){\vector(0,1){5}}
\put(6.2,0.4){\parbox{1cm}{$x$}}
\put(4.15,2.7){\parbox{1cm}{$\hbar/2$}}
\put(3.4,2.6){\circle*{0.1}}
\end{picture}
\caption{The first example of noncommutative geometry}
\end{figure}
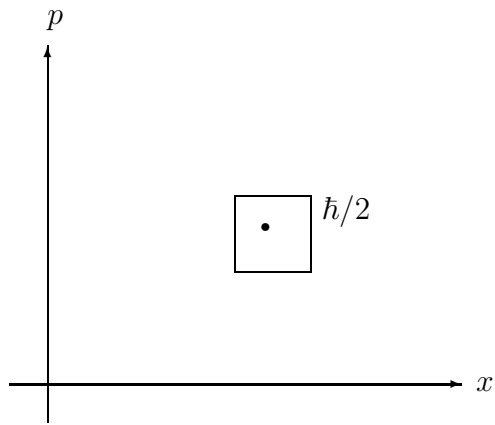

 Of course, there is no space anymore of which
$\aa$ is the algebra of functions. Nevertheless, we talk about such a
`quantum phase space' as a space that has no points or a space with
an uncertainty relation. Indeed, the noncommutation relation
(\ref{heisen}) implies Heisenberg's uncertainty relation
\bb \Delta x\Delta p \geq \hbar /2\ee
and tells us that points in phase space lose all meaning, we can
only resolve cells in phase space of volume $\hbar/2$, see Figure 4.
To define the uncertainty $\Delta a$ for an observable $a\in\aa$,
we need a faithful representation of the algebra on a Hilbert
space,  i.e. an injective homomorphism
$\rho :\aa\rightarrow {\rm End}(\hh)$ (cf Appendix).
For the harmonic oscillator, this Hilbert space is
$\hh=\lll^2(\rr)$. Its elements are the wave functions $\psi (x)$,
 square integrable functions on configuration space. Finally,
the dynamics is defined by a self adjoint observable $H=H^*\in\aa$
via Schr\"odinger's equation
\bb \left( i\hbar \,\frac{\pa}{\pa t}\, -\,\rho (H)\right) \psi
(t,x)=0.\ee
Usually the representation is not written explicitly. Since it is
faithful, no confusion should arise from this abuse.  Here time is
considered an external parameter, in particular, time is not
considered an observable. This is different in the special
relativistic setting where Schr\"odinger's equation is replaced by
Dirac's equation,
\bb\ddd\psi =0.\ee
Now the wave function $\psi $ is the four-component spinor
consisting of left- and right-handed, particle and antiparticle
wave functions.  The Dirac operator is not in
$\aa$ anymore, but
$\ddd\in{\rm End}(\hh)$. The Dirac operator is only formally self
adjoint because there is no {\it positive definite} scalar product,
whereas in Euclidean spacetime it is truly self adjoint,
$\ddd^*=\ddd.$

Connes' geometries are described by these three purely algebraic
items, $(\aa,\hh,\ddd)$, with $\aa$ a real, associative, possibly
noncommutative involution algebra with unit, faithfully
represented on a complex Hilbert space $\hh$, and $\ddd$ is a
self adjoint operator on $\hh$.

\subsection{The calibrating example: Riemannian spin geometry}

Connes'  geometry \cite{book}
 does to spacetime what quantum mechanics does
to phase space. Of course, the first thing we have to learn is how to
reconstruct the Riemannian geometry from the algebraic data
$(\aa,\hh,\ddd)$ in the case where the algebra is commutative. We
start the easy way and construct the triple $(\aa,\hh,\ddd)$
given a four dimensional, compact, Euclidean spacetime $M$. As
before $\aa=\ccc^\infty(M)$ is the real algebra of complex valued
differentiable functions on spacetime and $\hh=\lll^2(\sss)$ is the
 Hilbert space of complex,
square integrable spinors $\psi $ on
$M$. Locally, in any coordinate neighborhood,  we write the
spinor as a
 column vector, $\psi (x)\in\cc^4,\
x\in M$. The scalar product of two
spinors is defined by
\bb (\psi ,\psi ')=\int_M \psi ^*(x)\psi '(x)\,\de V,\ee
with the invariant volume form $\de V:=|\det
g_{\cdot\cdot}|^{1/2}\,\de^ 4x$ defined with the metric
tensor,
\bb g_{\mu \nu }=g\left( \,\frac{\pa}{\pa x^\mu }\,,
\,\frac{\pa}{\pa x^\nu }\,\right),
\ee
 that is the matrix of the Riemannian metric
$g$ with respect to the coordinates $x^\mu $,
$\mu =0,1,2,3.$ Note -- and this is important -- that with
Euclidean signature the Dirac action is simply a scalar product,
$S_D=(\psi ,\ddd\psi )$. The representation is defined by pointwise
multiplication,
$(\rho(a)\,\psi )(x):=a(x)\psi (x),$ $a\in \aa$.
For a start, it is sufficient to know the Dirac operator on a flat
manifold $M$ and with respect to inertial or Cartesian coordinates
$\tilde x^{\tilde \mu} $ such that $\tilde g_{\tilde \mu \tilde \nu
}={\delta ^{\tilde \mu}}_
{\tilde \nu }$. Then we use Dirac's original definition,
\bb\dd=\ddd=i\gamma ^{\tilde \mu} \pa/\pa
\tilde x^{\tilde \mu},\ee
 with the self adjoint $\gamma $-matrices
\bb \gamma ^0=\pp{0&-1_2\cr -1_2&0},\qq \gamma
^j= {\textstyle\frac{1}{i}}
\pp{0&\sigma _j\cr -\sigma _j&0},\label{diracmat}\ee
with the Pauli matrices
\bb \sigma
_1=\pp{0&1\cr 1&0},\qq\sigma _2=\pp{0&-i\cr
i&0},\qq\sigma _3=\pp{1&0\cr 0&-1}. \label{pauli}\ee
 We will construct the general curved Dirac operator
later.

When the dimension
of the manifold is even like in our case, the representation $\rho $
is reducible. Its Hilbert space decomposes into left- and
right-handed spaces,
\bb\hh=\hh_L\op\hh_R,\qq\hh_L=\,\frac{1-\chi }{2}\,  \hh,\qq
\hh_R=\,\frac{1+\chi }{2}\,  \hh.\ee
Again we make use of the unitary chirality operator,
\bb \chi =\gamma_5:=
\gamma^0\gamma^1\gamma^2\gamma^3=\pp{
-1_2&0\cr 0&1_2}.\ee
We will also need the charge conjugation or real structure,
the anti-unitary operator:
\bb J=C:=\gamma ^0\gamma ^2\circ\,{\rm complex\
conjugation}=\pp{0&-1&0&0\cr  1&0&0&0\cr
0&0&0&1\cr 0&0&-1&0}\,\circ\,{\rm c\ c},\ee
that permutes particles and antiparticles.

The five items $(\aa,\hh,\dd,J,\chi )$ form what Connes calls an
even, real spectral triple \cite{tresch}.\\
$\aa$ is a real, associative involution algebra with unit,
represented faithfully by bounded operators on the Hilbert space
$\hh$.\\
$\dd$ is an unbounded self adjoint operator on $\hh$.\\
$J$ is an anti-unitary operator, \\
$\chi $ a unitary one.

They enjoy the following properties:
\begin{itemize}\item
$J^2=-1$ in four dimensions (\,$J^2=1$ in zero dimensions).
\item
$[\rho(a),J\rho(\tilde a)J^{-1}]=0$
for all $a,\tilde a\in\aa$.
\item
$\dd J=J\dd$, particles and antiparticles have the same dynamics.
\item
$[\dd,\rho(a)] $ is bounded
for all $a\in\aa$ and $[[\dd,\rho(a)],J\rho(\tilde
a)J^{-1}]=0$ for all $a,\tilde a\in\aa$. This property is
called first order condition because in the calibrating
example it states that the genuine Dirac operator is a
first order differential operator.
\item
$\chi^2=1$ and $[\chi ,\rho (a)]=0$ for all $a\in\aa$. These
properties allow the decomposition $\hh=\hh_L\op\hh_R$.
\item
 $J\chi=\chi J$.
\item
$\dd\chi=-\chi\dd$, chirality does not change under time
evolution.
\item
   There are three more
properties, that we do not spell out, orientability, which relates the
chirality to the volume form, Poincar\'e duality and regularity,
which states that our functions $a\in\aa$ are differentiable.
\end{itemize}
Connes promotes these properties to the axioms defining an even,
real spectral triple. These axioms are justified by his \\
{\bf Reconstruction theorem} (Connes 1996 \cite{grav}): Consider an
(even) spectral triple $(\aa,\hh,\dd,J,(\chi ))$ whose algebra $\aa$ is
commutative. Then here exists a compact, Riemannian spin manifold
$M$ (of even dimensions), whose spectral triple $(\ccc^\infty (M),
\lll^2(\sss),\ddd,C,(\gamma _5))$ coincides with $(\aa,\hh,\dd,J,(\chi ))$.

For details on this theorem and noncommutative geometry in
general, I warmly recommend the Costa Rica book
\cite{costarica}. Let us try to get a feeling of the {\it local}
information contained in this theorem. Besides describing the
dynamics of the spinor field $\psi $, the Dirac operator $\ddd$
encodes the dimension of spacetime, its Riemannian metric, its
differential forms and its integration, that is all the tools that we
need to define a Yang-Mills-Higgs model. In Minkowskian
signature, the square of the Dirac operator is the wave operator,
which in 1+2 dimensions governs the dynamics of a drum. The
deep question: `Can you hear the shape of a drum?' has been raised.
This question concerns a global property of spacetime, the
boundary. Can you reconstruct it from the spectrum of the wave
operator?
\begin{description}\item[The dimension of spacetime]
is a local property.
It can be retrieved from the asymptotic behaviour of the spectrum
of the Dirac operator for large eigenvalues. Since
$M$ is compact, the spectrum is discrete. Let us order the
eigenvalues, $...\lambda _{n-1}\leq\lambda _n\leq\lambda
_{n+1}...$ Then Weyl's spectral theorem states that the eigenvalues
grow asymptotically as
$n^{1/{\rm dim}M}$. To explore a local property of spacetime we
only need the high energy part of the spectrum. This is in nice
agreement with our intuition from quantum mechanics and
motivates the name `spectral triple'.
\item[The metric] can be
reconstructed from the commutative spectral triple by Connes'
distance formula (\ref{dist}) below. In the commutative case a point
$x\in M$ is reconstructed as the pure state. The general definition of a
pure state of course does not use the commutativity. A state $\delta  $ of
the algebra $\aa$ is a linear form on $\aa$, that is normalized, $\delta
(1)=1$, and positive, $\delta  (a^*a)\geq 0$ for all $a\in\aa$. A state
is pure if it cannot be written as a linear combination of two states.
For the calibrating example, there is a one-to-one correspondence
between points $x\in M$ and pure states $\delta  _x$
 defined by the Dirac distribution,
$\delta  _x(a):=a(x)=\int_M\delta _x(y) a(y)\de^4 y$. The geodesic
distance between two points $x$ and $y$ is reconstructed from the
triple as:
\bb {\rm sup}\left\{ |\delta _x(a)-\delta _y(a)|;\
a\in\ccc^\infty(M)\ {\rm such \ that}\ ||[\ddd,\rho (a)]||\leq
1\right\} .\label{dist}\ee
For the calibrating example, $[\ddd,\rho
(a)]$ is a bounded operator. Indeed, $[\ddd,\rho (a)]\psi =i\gamma ^\mu
\pa_\mu (a\psi )-i a\gamma ^\mu  \pa_\mu \psi =i\gamma ^\mu(
\pa_\mu a)\psi $, and $\pa_\mu a$ is bounded as a differentiable
function on a compact space.

 For a general spectral triple this
operator is bounded by axiom. In any case,  the operator norm
$||[\ddd,\rho (a)]||$ in the distance formula is finite.

Consider the circle, $M=S^1$, of circumference $2\pi $ with Dirac
operator
$\ddd=i\,\de/\de x$. A function $a\in \ccc^\infty(S^1)$ is
represented faithfully on a wavefunction $\psi \in \lll^2(S^1)$ by
pointwise multiplication,
$(\rho (a)\psi )(x)=a(x)\psi (x)$. The commutator $[\ddd,\rho
(a)]=i\rho (a')$ is familiar from quantum mechanics. Its operator
norm is $ ||[\ddd,\rho (a)]||: =\sup_\psi |[\ddd,\rho
(a)]\psi |/|\psi |=\sup_x|a'(x)|$, with $|\psi |^2=\int_0^{2\pi
}\bar\psi (x)\psi (x)\,\de x$. Therefore, the distance between two
points $x$ and $y$ on the circle is
\bb\sup_a\{|a(x)-a(y)|;\,\sup_x|a'(x)|\leq 1\}=|x-y|.\ee

Note that Connes' distance
formula continues to make sense for non-connected manifolds, like
discrete spaces of dimension zero, i.e. collections of points.
\item[Differential forms,] for example of degree one like
$\de a$ for a function $a\in\aa$, are reconstructed as
$(-i)[\ddd,\rho (a)]$. This is again motivated from quantum
mechanics. Indeed in a 1+0 dimensional spacetime $\de a$ is just
the time derivative of the `observable' $a$ and is associated with the
commutator of the Hamilton operator with $a$.
\end{description}
Motivated from quantum mechanics, we define a noncommutative
geometry by a real spectral triple with noncommutative algebra
$\aa$.

\subsection{Spin groups}

Let us go back to quantum mechanics of spin and recall how a
space rotation acts on a spin ${\textstyle\frac{1}{2}} $ particle. For
this we need group homomorphisms between the rotation group
$SO(3)$ and the probability preserving unitary group $SU(2)$. We
construct first the group homomorphism
\bb p:SU(2)&\longrightarrow&SO(3)\cr
U&\longmapsto & p(U).\eee
 With the help of the auxiliary function
\bb f:\rr^3&\longrightarrow&su(2)\cr \cr
\vec x=\pp{x^1\cr x^2\cr x^3}&\longmapsto&-
{\textstyle\frac{1}{2}} ix^j\sigma _j \, ,\eee
we define the rotation $p(U)$ by
\bb p(U)\vec x:=f^{-1}(Uf(\vec x)U^{-1}).\ee
The conjugation by the unitary $U$ will play an important role
and we give it a special name, $i_U(w):=UwU^{-1}$, $i$ for inner.
Since $i_{(-U)}=i_U$, the projection $p$ is two to one,
Ker$(p)=\{\pm1\}$. Therefore the spin lift
\bb L:SO(3)&\longrightarrow&SU(2)\cr
R=\exp(\omega) &\longmapsto & \exp({\textstyle\frac{1}{8}}
\omega ^{jk}[\sigma_j,\sigma _k]) \label{spi1} \ee
is double-valued.
It is a local group homomorphism and satisfies $p(L(R))=R$.
Its double-valuedness is
accessible to quantum mechanical experiments: neutrons
have to be rotated through an angle of $720^{\circ}$  before
interference patterns repeat \cite{neu}.

 \begin{figure}[h]
\setlength{\unitlength}{1.0cm}
\begin{picture}(20,3)(-1.1,0)
\put(0,2.4){\parbox{2cm}{${\rm Aut}_\hh(\aa)$}}
\put(1.8,2.4){\parbox{1cm}{$\hookleftarrow$}}
\put(2.5,2.4){\parbox{4cm}{${\rm Diff}(M)\ltimes
\,^MSpin(1,3)$}}
\put(6.7,2.4){\parbox{1cm}{$\hookleftarrow$}}
\put(7.4,2.4){\parbox{4cm}{$SO(1,3)\times
Spin(1,3)$}}
\put(11.5,2.4){\parbox{1cm}{$\hookleftarrow$}}
\put(12.2,2.4){\parbox{3cm}{$SO(3)\times
SU(2)$}}

 \put(0.1,0){\parbox{6mm}{${\rm Aut}(\aa)$}}
\put(1.8,0){\parbox{6mm}{$\hookleftarrow$}}
\put(3.6,0){\parbox{6mm}{${\rm Diff}(M)$}}
\put(6.7,0){\parbox{6mm}{$\hookleftarrow$}}
\put(8.5,0){\parbox{6mm}{$SO(1,3)$}}
\put(11.5,0){\parbox{6mm}{$\hookleftarrow$}}
\put(13.1,0){\parbox{6mm}{$SO(3)$}}

\put(0.3,1.4){\parbox{2cm}{\hskip -2mm p\
\parbox{6mm}{
\setlength{\unitlength}{0.4mm}
\begin{picture}(20,10)
\put(0,15){\vector(0,-1){30}}
\put(15,-15){\vector(-1,4){8}}
\put(15,-15){\vector(0,1){33}}
\end{picture}} L }}

\put(3.7,1.4){\parbox{2cm}{\hskip -2mm p\
\parbox{6mm}{
\setlength{\unitlength}{0.4mm}
\begin{picture}(20,10)
\put(0,15){\vector(0,-1){30}}
\put(15,-15){\vector(-1,4){8}}
\put(15,-15){\vector(0,1){33}}
\end{picture}} L }}

\put(8.8,1.4){\parbox{2cm}{\hskip -2mm p\
\parbox{6mm}{
\setlength{\unitlength}{0.4mm}
\begin{picture}(20,10)
\put(0,15){\vector(0,-1){30}}
\put(15,-15){\vector(-1,4){8}}
\put(15,-15){\vector(0,1){33}}
\end{picture}} L }}

\put(13.1,1.4){\parbox{2cm}{\hskip -2mm p\
\parbox{6mm}{
\setlength{\unitlength}{0.4mm}
\begin{picture}(20,10)
\put(0,15){\vector(0,-1){30}}
\put(15,-15){\vector(-1,4){8}}
\put(15,-15){\vector(0,1){33}}
\end{picture}} L }}
\end{picture}

\caption{The nested spin lifts of Connes, Cartan, Dirac,
and Pauli}
\end{figure}

The lift $L$ was generalized by Dirac to the special relativistic
setting, e.g. \cite{bd}, and by E. Cartan \cite{cartan} to the general
relativistic setting. Connes \cite{bris} generalizes it to noncommutative
geometry, see Figure 5. The transformations we need to lift are
Lorentz transformations in special relativity, and general coordinate
transformations in general relativity, i.e. our calibrating example. The
latter transformations are the local elements of the diffeomorphism
group Diff$(M)$. In the setting of noncommutative geometry, this group
is the group of algebra automorphisms Aut$(\aa)$. Indeed, in the
calibrating example we have Aut$(\aa)$=Diff$(M)$. In order to
generalize the spin group to spectral triples, Connes defines the
receptacle of the group of `lifted automorphisms',
\bb{\rm Aut}_\hh(\aa):=\{U\in {\rm End}(\hh),\
UU^* =U^*U=1,\ UJ=J U,\ U\chi =\chi U,\
i_U\in{\rm Aut}(\rho (\aa))\}.\ee
The first three properties
say that a lifted automorphism $U$ preserves probability,
charge conjugation, and chirality. The
fourth, called {\it covariance property}, allows to
define the projection
$p:\ {\rm Aut}_\hh(\aa)\longrightarrow {\rm
Aut}(\aa)$ by
\bb p(U)=\rho ^{-1}i_U\rho \ee
We will see that the covariance property will protect the locality of
field theory. For the calibrating example of a four dimensional
spacetime, a local calculation, i.e. in a coordinate patch, that
we still denote by $M$, yields the semi-direct product (cf
Appendix) of diffeomorphisms with local or gauged spin
transformations,
${\rm Aut}_{\lll^2(\sss)}(\ccc^\infty (M))={\rm
Diff}(M)\ltimes \,^MSpin(4)$. We say receptacle because
already in six dimensions,
${\rm Aut}_{\lll^2(\sss)}(\ccc^\infty (M))$ is larger
than ${\rm Diff}(M)\ltimes \,^MSpin(6)$.
However we can use the lift $L$ with
$p(L(\sigma ))=\sigma   $, $\sigma  \in$Aut$(\aa)$ to correctly
identify the spin group in any dimension of $M$. Indeed we will
see that the spin group is the image of the spin lift
$L($Aut$(\aa))$, in general a proper subgroup of the receptacle
${\rm Aut}_\hh(\aa)$.

Let $\sigma  $ be a diffeomorphism close to the identity. We
interpret
$\sigma  $ as coordinate transformation, all our
calculations will be local, $M$ standing for one chart,
on which the coordinate systems $\tilde x^{\tilde\mu } $ and $
x^{\mu }=(\sigma   (\tilde x))^{\mu }$ are
defined. We will work out the local expression of a lift
of $\sigma $ to the Hilbert space of spinors. This lift
$U=L(\sigma )$ will depend on the metric and on the
initial coordinate system $\tilde x^{\tilde\mu }$.

In a first step, we construct a group homomorphism
$\Lambda:{\rm Diff}(M)\rightarrow {\rm
Diff}(M)\ltimes\,^MSO(4)$ into the group of local `Lorentz'
transformations, i.e. the group of differentiable functions from
spacetime into
$SO(4)$ with pointwise multiplication. Let
${(\tilde e^{-1}(\tilde x))^{\tilde \mu} }_a={(\tilde
g^{-1/2}(\tilde x))^{\tilde \mu} }_a$ be the inverse of the square
root of the positive matrix $\tilde g$ of the metric with respect to the
initial coordinate system
$\tilde x^{\tilde \mu} $. Then the four vector fields $\tilde e_a$,
$a=0,1,2,3$, defined by
\bb \tilde e_a:= {(\tilde e^{-1})^{\tilde \mu} }_a\,\frac{\pa}{\pa
\tilde x^{\tilde \mu }}\,
\ee
give
 an orthonormal frame of the tangent bundle.
This frame defines a complete gauge fixing of the
Lorentz gauge group $^MSO(4)$
because it is the only orthonormal frame to have
symmetric coefficients ${(\tilde e^{-1})^{\tilde \mu} }_a$ with
respect to the coordinate system $\tilde x^{\tilde \mu} $. We call
this gauge the symmetric gauge for the coordinates
$\tilde x^{\tilde \mu} .$ Now let us perform a local change of
coordinates,
$ x =\sigma  (\tilde x)$. The holonomic frame with
respect to the new coordinates is related to the former
holonomic one by the inverse Jacobian matrix of
$\sigma  $
\bb \,\frac{\pa}{\pa x^{ \mu }}\,
=\,\frac{\pa \tilde x^{\tilde \mu }}{\pa x^{\mu }}\,
\,\frac{\pa}{\pa\tilde  x^{\tilde  \mu }}\,=\hbox{${\left(
\jj^{-1}\right) ^{\tilde \mu }}$}_{\mu }\,\frac{\pa}{\pa\tilde
x^{\tilde \mu }},&& {\left(
\jj^{-1}( x)\right) ^{\tilde \mu }}_{\mu
}=\,\frac{\pa \tilde x^{\tilde \mu }}{\pa x^{
\mu }}\,.\ee
The matrix $g$ of the metric  with respect
to the new coordinates reads,
\bb  g_{\mu \nu }( x):=
g\!\!\left.\left(\,\frac{\pa}{\pa x^{ \mu
}}\,,\,\frac{\pa}{\pa x^{ \nu }}\,
\right)\right|_{ x} =\left( \jj^{-1T}( x)
\tilde g(\sigma  ^{-1}( x))\jj^{-1}( x)
\right) _{\mu \nu },\ee
 and the symmetric
gauge for the new coordinates $ x$ is the new
orthonormal frame
\bb  e_b=\hbox{$ e^{-1\mu
}$}_b\,\frac{\pa  }{\pa x^{ \mu
}}\,=\hbox{$ g^{-1/2\,\mu
}$}_b{\jj^{-1\,\tilde \mu }}_{ \mu
}\,\frac{\pa}{\pa \tilde x^{\tilde \mu }}\, ={\left(
\jj^{-1}\sqrt{\jj \tilde g^{-1}\jj^T}\right) ^{\tilde \mu}
}_b\,\frac{\pa}{\pa\tilde  x^{\tilde
\mu}}\,.\ee
New and old orthonormal frames are
related by a Lorentz transformation $\Lambda $,
$ e_b={\Lambda ^{-1\,a}}_b\tilde e_a$, with
\bb\left.\Lambda(\sigma )\right|_{\tilde x}
=\left.\sqrt{\jj^{-1T}\tilde g\jj^{-1}}\right|_{\sigma
(\tilde x)}\left.\jj
\right|_{\tilde x}\left.\sqrt{\tilde g^{-1}}\right|_{\tilde
x}=\sqrt{ g}\jj\sqrt{\tilde g^{-1}}.\ee

If $M$ is flat and $\tilde x^{\tilde \mu} $ are `inertial' coordinates,
i.e. $\tilde g_{\tilde \mu \tilde \nu }={\delta ^{\tilde
\mu}}_{\tilde \nu} $, and
$\sigma  $ is a local isometry then
$\jj(\tilde x)\in SO(4)$ for all
$\tilde x$ and
$\Lambda(\sigma  ) =\jj$. In special relativity,
therefore, the symmetric gauge ties together Lorentz
transformations in spacetime with Lorentz
transformations in the tangent spaces.

In general, if the coordinate transformation $\sigma
$ is close to the identity, so is its Lorentz transformation
$\Lambda(\sigma  )$ and it can be lifted to the spin
group,
\bb S: SO(4)&\longrightarrow& Spin(4)\cr
\Lambda =\exp\omega &\longmapsto&\exp \left[
{\textstyle\frac{1}{4}} \omega _{ab}
\gamma ^{ab}\right] \label{spin}\ee with $\omega
=-\omega ^T\,\in so(4)$ and $\gamma
^{ab}:={\textstyle\frac{1}{2}} [\gamma ^a,\gamma ^b]$. With our
choice (\ref{diracmat}) for the $\gamma $ matrices, we
have
\bb \gamma ^{0j}=i\pp{-\sigma _j&0\cr 0&\sigma _j},\qq
\gamma ^{jk}=i\epsilon ^{jk\ell}
\pp{\sigma _\ell&0\cr 0&\sigma _\ell},\qq j,k=1,2,3,\qq
\epsilon ^{123}=1.\ee
We can write the local
expression \cite{lift} of the lift
 $L:{\rm Diff}(M)\rightarrow {\rm Diff}(M)\ltimes \,^MSpin(4)$,
\bb \left( L(\sigma  )\psi \right) (
x)=\left.S\left(\Lambda (\sigma
)\right)\right|_{\sigma  ^{-1}( x)}\psi
({\sigma ^{-1}( x)})\label{spi2}.\ee
$L$ is a double-valued group homomorphism. For
any $\sigma  $ close to the identity,
$L(\sigma  )$ is unitary, commutes with charge
conjugation and chirality, satisfies the covariance
property, and $p(L(\sigma  ))=\sigma  $. Therefore,
we have locally
\bb L({\rm Diff}(M))\ \subset\  {\rm
Diff}(M)\ltimes\,^MSpin(4)={\rm Aut}_{\lll^2(\sss)}(\ccc^\infty
(M)).\ee
The symmetric
gauge is a complete gauge fixing and this reduction follows
Einstein's spirit in the sense that the only arbitrary choice is the
one of the initial coordinate system
$\tilde x^{\tilde \mu} $ as will be illustrated in the next section.
Our computations are deliberately local. The global
picture can be found in reference \cite{bourg}.

\section{The spectral action}

\subsection{Repeating Einstein's derivation in the commutative
case}

We are ready to parallel Einstein's derivation of general relativity
in Connes' language of spectral triples. The associative algebra
$\ccc^\infty(M)$ is commutative, but this property will never be
used. As a by-product, the lift
$L$ will reconcile Einstein's and Cartan's formulations of general
relativity and it will  yield a self contained introduction to Dirac's
equation in a gravitational field accessible to particle physicists.
For a comparison of Einstein's and Cartan's  formulations of general
relativity see for
example
\cite{gs}.

\subsubsection{First stroke: kinematics}

Instead of a point-particle, Connes takes as matter a field, the free,
massless Dirac particle
$\psi (\tilde x)$ in the flat spacetime of special relativity. In
inertial coordinates $\tilde x^{\tilde \mu} $, its dynamics is given
by the Dirac equation,
\bb \tilde \ddd\psi =i{\delta ^{\tilde \mu }}_a\gamma
^a{\,\frac{\pa}{\pa\tilde  x^{\tilde \mu }}}\, \psi =0.\ee
We have written ${\delta ^{\tilde \mu }}_a\gamma
^a$ instead of $\gamma^{\tilde \mu}$ to stress that the
$\gamma$ matrices are $\tilde x$-independent. This Dirac
equation is covariant under Lorentz transformations.
Indeed if
$\sigma  $ is a local isometry then
\bb L(\sigma  )\tilde \ddd L(\sigma  )^{-1}=\ddd=
i{\delta ^{\mu} }_a\gamma
^a{\frac{\pa}{\pa  x^{\mu} }}.\ee
To prove this special relativistic covariance, one needs
the identity
$S(\Lambda )\gamma ^a S(\Lambda )^{-1}={\Lambda
^{-1\,a}}_b\gamma ^b$ for Lorentz transformations
$\Lambda \in SO(4)$ close to the identity. Take a
general coordinate transformation
$\sigma  $ close to the identity. Now comes a long, but
straight-forward calculation. It is a useful exercise requiring only
matrix multiplication and standard calculus, Leibniz and chain
rules. Its result is the Dirac operator in curved coordinates,
\bb L(\sigma  )\tilde \ddd L(\sigma  )^{-1}=\ddd=
i\hbox{$ e^{-1\,\mu}$}_a\gamma^a\left[
\,\frac{\pa}{\pa x^{ \mu}}\,+s(\omega
_{\mu})\right],\label{2}\ee
where
$ e^{-1}=\sqrt{\jj \jj^T}$ is a symmetric matrix,
\bb s: so(4)&\longrightarrow& spin(4)\cr
\omega &\longmapsto&
{\textstyle\frac{1}{4}} \omega _{ab}
\gamma ^{ab} \ee
is the Lie algebra isomorphism corresponding to the
lift (\ref{spin}) and
\bb
\omega_{\mu}(x)=\left.\Lambda\right|_{\sigma ^{-1}(x)}
\pa_{\mu}
\left.\Lambda^{-1}\right|_{x}.\ee
The `spin connection' $ \omega$ is the gauge transform of
the Levi-Civita connection $ \Gamma$, the latter is expressed
with respect to the holonomic frame $\pa_{\mu}$, the former is
written with respect to the orthonormal frame $ e_a=\hbox{$
e^{-1\,\mu}$}_a\pa_{\mu}$. The gauge transformation passing
between them is $e\in\,^MGL_4$,
\bb \omega =e\Gamma e^{-1}+e\de e^{-1}.\ee

 We recover
 the well known explicit expression
\bb  {\omega^a}_{b \mu}( e)=
{\textstyle\frac{1}{2}}\left[( \pa_{
\beta}\hbox{$ e^a$}_{ \mu})-
( \pa_{
\mu}\hbox{$ e^a$}_{ \beta})+
\hbox{$ e^m$}_{ \mu}(
\pa_{\beta  }
\hbox{$ e^m$}_{
\alpha})\hbox{$ e^{-1\,\alpha}$}_a
\right]\hbox{$ e^{-1\,
\beta}$}_b\ -\ [a \leftrightarrow b]\ee
 of the spin
connection in terms of the first derivatives of
$\hbox{$ e^a$}_{ \mu}= {\sqrt{
g}^a}_{
\mu}.$ Again the spin connection has zero curvature
and the equivalence principle relaxes this
constraint. But now equation (\ref{2}) has an
advantage over its analogue (\ref{1}). Thanks to
Connes' distance formula (\ref{dist}), the metric can
be read explicitly in (\ref{2}) from the matrix of
functions
$\hbox{$ e^{-1\,\mu}$}_a$, while in
(\ref{1}) first derivatives of the metric are
present. We are used to this nuance from
electro-magnetism, where the classical particle feels
the force while the quantum particle feels the
potential. In Einstein's approach, the zero connection
fluctuates, in Connes' approach, the flat metric
fluctuates. This means that the constraint $
e^{-1}=\sqrt{\jj
\jj^T}$ is relaxed and $ e^{-1} $ now is an
arbitrary symmetric matrix depending smoothly
on $ x$.

Let us mention two experiments with neutrons confirming the
`minimal coupling' of the Dirac operator to curved coordinates,
equation (\ref{2}). The first takes place in flat spacetime. The
neutron interferometer is mounted on a loud speaker and shaken
periodically \cite{bonse}. The resulting pseudo forces coded in the
spin connection do  shift  the interference patterns observed. The
second experiment takes place in a true gravitational field in
which the neutron interferometer is placed \cite{cow}. Here shifts
of the interference patterns are observed that do depend on the
gravitational {\it potential}, ${e^a}_\mu $ in equation (\ref{2}).

\subsubsection{Second stroke: dynamics}

 The second stroke, the covariant dynamics
for the new class of Dirac operators $ \ddd$ is
due to Chamseddine \& Connes \cite{cc}. It is the
celebrated spectral action.
 The beauty of their
approach to general relativity is that it works
precisely because the Dirac operator $ \ddd$
plays two roles simultaneously, it defines the
dynamics of matter and the kinematics of gravity. For
a discussion of the transformation passing from the metric to the
Dirac operator I recommend the article \cite{lr} by Landi
\& Rovelli.

The starting point of Chamseddine \& Connes is the
simple remark that the spectrum of the Dirac operator
is invariant under diffeomorphisms interpreted as
general coordinate transformations. From
$ \ddd\chi=-\chi \ddd$ we know that the
spectrum of
$ \ddd$ is even. Indeed, for every eigenvector $\psi $ of $\ddd$ with
eigenvalue $E$, $\chi \psi $ is eigenvector with eigenvalue $-E$. We
may therefore consider only the spectrum of the positive operator $
\ddd^2/\Lambda^2$ where we have divided by a fixed
arbitrary energy scale to make the spectrum
dimensionless. If it was not divergent the trace $\t
 \ddd^2/\Lambda^2$ would be a general
relativistic action functional. To make it convergent,
take a differentiable function
$f:\rr_+\rightarrow\rr_+$ of sufficiently fast
decrease such that the action
\bb S_{CC}:=\t f( \ddd^2/\Lambda^2)\ee
converges. It is still a diffeomorphism invariant
action. The following theorem, also known as heat kernel expansion, is
a local version of an index theorem \cite{heat}, that as explained in
Jean Zinn-Justin's lectures \cite{zinn} is intimately related to Feynman
graphs with one fermionic loop.\\
 {\bf Theorem:} Asymptotically for high
energies, the spectral action is
\bb S_{CC}=
\int_M
[{\textstyle\frac{2\Lambda_c}{16\pi G}}-{\textstyle\frac{1}{16\pi
G}}R +a(5\,R^2-8\,{\rm Ricci}^2-7\,{\rm
Riemann}^2)]\,\de V \,+\,
O(\Lambda^{-2}), \label{heat}\ee
where the cosmological constant
is $\Lambda_c=
{\textstyle\frac{6f_0}{f_2}}\Lambda^2$, Newton's constant is
$G={\textstyle\frac{3\pi }{f_2}}\Lambda^{-2}$ and
$a={\textstyle\frac{f_4}{5760\pi^2}}$. On the right-hand
side of the theorem we have omitted surface terms, that is
terms that do not contribute to the Euler-Lagrange equations.
 The
Chamseddine-Connes action is universal in the sense
that the `cut off' function $f$ only enters through its
first three `moments', $f_0:=\int_0^\infty uf(u)\de u$,
$f_2:=\int_0^\infty f(u)\de u$ and $f_4=f(0)$.

If we take for $f$ a differentiable approximation of the
characteristic function of the unit interval, $f_0=1/2$,
$f_2=f_4=1$, then the spectral action just counts the number of
eigenvalues of the Dirac operator whose absolute values are below
the `cut off' $\Lambda $. In four dimensions, the minimax example
 is  the flat
 4-torus with all circumferences measuring $2\pi$.
 Denote by $\psi_B(x)$, $B=1, 2, 3, 4$,
the four components of the spinor. The Dirac
operator is
\bb \ddd = \pmatrix{
0&0&-i\pa_0+\pa_3&\pa_1-i\pa_2\cr
0&0&\pa_1+i\pa_2&-i\pa_0-\pa_3\cr
-i\pa_0-\pa_3&-\pa_1+i\pa_2&0&0\cr
-\pa_1-i\pa_2&-i\pa_0+\pa_3&0&0
 }
.\ee
After a Fourier transform
\bb \psi_B(x)\ =:\ \sum_{j_0,...,j_3\in\zz}
\hat\psi
_B(j_0,...,j_3)\exp(-ij_\mu x^\mu),\quad B=1,2,3,4\ee
the eigenvalue equation $\ddd\psi=\lambda\psi$ reads
\bb \pmatrix{
0&0&-j_0-ij_3&-ij_1-j_2\cr
0&0&-ij_1+j_2&-j_0+ij_3\cr
-j_0+ij_3&ij_1+j_2&0&0\cr
ij_1-j_2&-j_0-ij_3&0&0
}
\pmatrix{\hat\psi_1\cr \hat\psi_2\cr \hat\psi_3\cr
\hat\psi_4}\ =\
\lambda
\pmatrix{\hat\psi_1\cr \hat\psi_2\cr \hat\psi_3\cr
\hat\psi_4}. \ee
Its characteristic equation is
$ \lb \lambda^2-(j_0^2+j_1^2+j_2^2+j_3^2)\rb^2=0$
and for fixed $j_\mu$, each eigenvalue
$ \lambda=\pm\sqrt{j_0^2+j_1^2+j_2^2+j_3^2}$
has multiplicity two. Therefore asymptotically for large
$\Lambda$ there are
$ 4B_4\Lambda^4$ eigenvalues (counted with their
multiplicity) whose absolute values are smaller than
$\Lambda$.
$ B_4=\pi^2/2$
denotes the volume of the unit ball
in $\rr^4$. En passant, we check Weyl's spectral theorem.
Let us arrange the absolute values of the
eigenvalues in an increasing
sequence and number them by naturals $n$, taking due account of
their multiplicities. For large $n$, we have
\bb |\lambda_n|\approx
\left({n\over{2\pi^2}}\right)^{1/4}.\ee
The  exponent is
indeed the inverse dimension.
 To check the heat kernel expansion,
we compute the right-hand side of equation (\ref{heat}):
\bb S_{CC}=\int_M\,\frac{\Lambda _c}{8\pi G}\,\de V =(2\pi
)^4\,{\textstyle\frac{f_0}{4\pi^2}}\Lambda^4
=2\pi ^2\Lambda ^4,\ee
which agrees with the asymptotic count of eigenvalues, $
4B_4\Lambda^4$. This example was the flat torus. Curvature
 will modify the spectrum and this modification can be used
to measure the curvature =
gravitational field,  exactly as the Zeemann or Stark effect
measures the electro-magnetic field by observing how it
modifies the spectral lines of an atom.

In the spectral action, we find the Einstein-Hilbert action, which is
linear in curvature. In addition, the spectral action contains terms
quadratic in the curvature. These terms can safely be
neglected in weak gravitational fields like in our solar system.
In homogeneous, isotropic cosmologies, these terms are a
surface term and do not modify Einstein's equation. Nevertheless
the quadratic terms render the (Euclidean) Chamseddine-Connes
action positive. Therefore this action has minima. For instance, the
4-sphere with a radius of
the order of
 the Planck length $\sqrt G$ is a minimum, a `ground state'.
This minimum breaks the diffeomorphism group
spontaneously \cite{bris} down to the isometry group $SO(5)$.
The little group is the isometry group, consisting of those lifted
automorphisms that commute with the Dirac operator
$ \ddd$. Let us anticipate that the spontaneous
symmetry breaking via the Higgs mechanism will be a
mirage of this gravitational break down.
Physically this ground state seems to regularize the initial
cosmological singularity with its ultra strong gravitational field in
the same way in which quantum mechanics regularizes the Coulomb
singularity of the hydrogen atom.

We close this subsection with a technical remark. We noticed
that the matrix $\hbox{$
e^{-1\,\mu}$}_a$ in equation (\ref{2}) is
symmetric. A general, not necessarily symmetric
matrix
$\hbox{$\hat e^{-1\,\mu}$}_a$ can be obtained
from a general Lorentz transformation
$\Lambda\in\,^MSO(4)$:
\bb \hbox{$ e^{-1\,\mu}$}_a {\Lambda^a}_b
=\hbox{$\hat e^{-1\,\mu}$}_b,\ee
which is nothing but the polar decomposition of the
matrix $\hat e^{-1}$. These transformations are the gauge
transformations of general relativity in Cartan's formulation.
They are invisible in Einstein's formulation because of the
complete (symmetric) gauge fixing coming from the initial
coordinate system $\tilde x^{\tilde \mu }$.

\subsection{Almost commutative geometry}

We are eager to see the spectral action in a noncommutative
example. Technically the simplest noncommutative examples are
almost commutative. To construct the latter, we need a natural
property of spectral triples, commutative or not: The tensor
product of two even spectral triples is an even spectral triple. If
both are commutative, i.e. describing two manifolds, then their
tensor product simply describes the direct product of the two
manifolds.

Let $(\aa_i,\hh_i,\dd_i,
J_i,\chi _i)$, $i=1,2$ be two even, real spectral triples
of even dimensions
$d_1$ and
$d_2$. Their tensor product is the triple
$(\aa_t,\hh_t,\dd_t,J_t,\chi_t)$ of dimension $d_1+d_2$ defined by
\bb &&\aa_t=\aa_1\ot\aa_2,\qq
\hh_t=\hh_1\ot\hh_2,\cr
&&\dd_t=\dd_1\ot 1_2\ +\ \chi _1\ot\dd_2,\cr
&&J_t=J_1\ot J_2,\qq\chi _t=\chi _1\ot\chi _2.\eee
The other obvious choice for the Dirac operator,
$\dd_1\ot\chi _2\ +\ 1_1\ot\dd_2$, is unitarily
equivalent to the first one. By definition, an almost commutative
geometry is a tensor product of two spectral triples, the first triple
is a 4-dimensional spacetime, the calibrating example,
\bb\left(
\ccc^\infty(M),\lll^2(\sss),\ddd,C,\gamma_5\right),
\ee
and the second is 0-dimensional. In accordance with Weyl's spectral
theorem, a 0-dimensional spectral triple has a finite dimensional
algebra and a finite dimensional Hilbert space. We will  label
the second triple by the subscript $\cdot_f$ (for finite) rather
than by $\cdot_2$. The origin of the word almost commutative is
clear: we have a tensor product of an infinite dimensional
commutative algebra with a finite dimensional, possibly
noncommutative algebra.

This tensor product is, in fact, already familiar to you from the
quantum mechanics of spin, whose Hilbert space is the infinite
dimensional Hilbert space of square integrable functions on
configuration space tensorized with the 2-dimensional Hilbert
space
$\cc^2$ on which acts the noncommutative algebra of spin
observables. It is the algebra $\hhh$ of quaternions, $2\times 2$
complex matrices of the form $\pp{x&-\bar y\cr y&\bar x}\
x,y\in\cc$. A basis of
$\hhh$ is given by $\{1_2,i\sigma _1,i\sigma
_2,i\sigma _3\}$, the identity matrix and the three Pauli
matrices (\ref{pauli}) times $i$.  The group of unitaries of $\hhh$
is
$SU(2)$, the spin cover of the rotation group, the group of
automorphisms of $\hhh$ is $SU(2)/\zz_2$, the rotation
group.

 A commutative 0-dimensional or finite spectral triple is just a
collection of points, for examples see \cite{ikm}. The simplest
example is the two-point space,
\bb\aa_f= \cc_L\op\cc_R\owns
(a_L,a_R),\qq
\hh_f=\cc^4,\qq
\rho_f(a_L,a_R)=\pp{a_L&0&0&0\cr
0&a_R&0&0\cr
0&0&\bar a_R&0\cr
0&0&0&\bar a_R},\ee
\bb\dd_f =\pp{0&m&0&0\cr \bar m &0&0&0\cr
0&0&0&\bar m\cr0&0&m&0
},\ m\in\cc,\qq
 J_f=\pp{0&1_2\cr
1_2&0}\circ {\rm c\ c},\qq
\chi _f=\pp{-1&0&0&0\cr 0&1&0&0\cr 0&0&-1&0\cr
0&0&0&1}.\ee
The algebra has two points = pure states, $\delta _L$ and $\delta
_R$, $\delta _L(a_L,a_R)=a_L$.  By Connes' formula (\ref{dist}), the
distance between the two points is $1/|m|$. On the other hand
$\dd_t=\ddd\ot1_4\ +\
\gamma_5\ot\dd_f$ is precisely the free massive
Euclidean Dirac operator. It describes one Dirac spinor of mass
$|m|$ together with its antiparticle.
 The tensor product of the calibrating example
and the two point space is the two-sheeted universe,
two identical spacetimes at constant distance. It was the first
example in noncommutative geometry to exhibit spontaneous
symmetry breaking
\cite{dkm, ccl}.

One of the major advantages of the algebraic
description of space in terms of a spectral triple, commutative or
not, is that continuous and discrete spaces are included in the same
picture. We can view almost commutative geometries
as Kaluza-Klein models \cite{kk} whose fifth dimension is
discrete. Therefore we will also call the finite spectral
triple `internal space'. In noncommutative geometry, 1-forms are
naturally defined on discrete spaces where they play the role of
connections. In almost commutative geometry, these discrete,
internal connections will turn out to be the Higgs scalars
responsible for spontaneous symmetry breaking.

Almost commutative geometry is an ideal play-ground for the
physicist with low culture in mathematics that I am.
Indeed Connes' reconstruction theorem immediately reduces the
infinite dimensional, commutative part to Riemannian geometry
and we are left with the internal space, which is accessible to
anybody mastering matrix multiplication. In particular, we
can easily make precise the last three axioms of spectral triples:
orientability, Poincar\'e duality and regularity. In the finite
dimensional case -- let us drop the $\cdot_f$ from now on --
orientability means that the chirality can be written as a finite
sum,
\bb \chi = \sum_j \rho (a_j)J\rho (\tilde a_j)J^{-1},\qq a_j,\tilde
a_j\in\aa.\ee
The Poincar\'e duality says that the
intersection form
\bb \cap_{ij}:=\t
\left[\chi\,\rho(p_i)\,J\rho(p_j)J^{-1}
\right]\ee
 must be non-degenerate, where the $p_j$ are a set of
minimal projectors of $\aa$. Finally, there is the regularity
condition. In the calibrating example, it ensures that the algebra
elements, the functions on spacetime $M$, are not only continuous
but differentiable. This condition is of course
empty for  finite spectral triples.

 Let us come back to our finite, commutative
example.  The two-point space is orientable, $\chi =\rho
(-1,1)J\rho (-1,1)J^{-1}$. It also
satisfies Poincar\'e duality, there are two minimal
projectors, $p_1=(1,0)$, $p_2=(0,1)$, and the intersection form is
$\cap=\pp{0&-1\cr -1&2}$.

\subsection{The minimax example}

It is time for a noncommutative internal space, a mild
variation of the two point space:
\bb\aa= \hhh\op\cc\owns
(a,b),\qq
\hh=\cc^6,\qq
\rho(a,b)=\pp
{a&0&0&0\cr
0&\bar b&0&0\cr
0&0& b1_2&0\cr
0&0&0&b},\ee
\bb\tilde \dd =\pp{0&\mm&0&0\cr \mm^* &0&0&0\cr
0&0&0&\bar \mm\cr0&0&\bar\mm^*&0
},\qq \mm=\pp{0\cr m},\qq m\in\cc,\ee\bb
 J=\pp{0&1_3\cr
1_3&0}\circ {\rm c\ c},\qq
\chi =\pp{-1_2&0&0&0\cr 0&1&0&0\cr 0&0&-1_2&0\cr
0&0&0&1}.\ee
The unit is $(1_2,1)$ and the involution is $(a,b)^*=(a^*,\bar b),$
where $a^*$ is the Hermitean conjugate of the quaternion $a$.
The  Hilbert space now contains one massless,
left-handed Weyl spinor and one Dirac spinor of mass $|m|$ and
$\mm$ is the fermionic mass matrix. We
denote the canonical basis of $\cc^6$ symbolically by $(\nu
,e)_L,e_R,(\nu^c ,e^c)_L,e^c_R$. The spectral triple still describes
two points,
$\delta _L(a,b)={\textstyle\frac{1}{2}}
\t a$ and
$\delta _R(a,b)=b$ separated by a distance $1/|m|$. There are still
two minimal projectors, $p_1=(1_2,0)$, $p_2=(0,1)$ and the
intersection form $\cap=\pp{0&-2\cr -2&2}$ is invertible.

Our next task is to lift the automorphisms to the Hilbert space and
fluctuate the `flat' metric $\tilde \dd$. All automorphisms of the
quaternions are inner, the complex numbers considered as
2-dimensional real algebra only have one non-trivial
automorphism, the complex conjugation. It is disconnected from
the identity and we may neglect it. Then
\bb {\rm Aut}(\aa)=SU(2)/\zz_2\owns \sigma _{\pm u},\qq
\sigma _{\pm u}(a,b)=(uau^{-1},b).\ee
The receptacle group, subgroup of $U(6)$ is readily calculated,
\bb {\rm Aut}_\hh(\aa)=U(2)\times U(1)\owns U=
\pp{U_2&0&0&0\cr 0&U_1&0&0\cr 0&0&\bar U_2&0\cr 0&0&0&\bar
U_1},\qq U_2\in U(2),\ U_1\in U(1).\ee
The covariance property is fulfilled, $i_U\rho (a,b)=\rho
(i_{U_2}a,b)$ and the projection, $p(U)=\\ \pm (\det
U_2)^{-1/2}\,U_2$, has kernel $\zz_2$. The lift,
\bb L(\pm u)=\rho (\pm u,1)J\rho (\pm u,1)J^{-1}=
\pp{\pm u&0&0&0\cr 0&1&0&0\cr 0&0&\pm \bar u&0\cr 0&0&0&1},
\label{mmlift}\ee
is double-valued. The spin group is the image of the lift, $L({\rm
Aut}(\aa))=SU(2)$, a proper subgroup of the receptacle ${\rm
Aut}_\hh(\aa)=U(2)\times U(1)$.
The fluctuated Dirac operator is
\bb\dd:=L(\pm u)\tilde \dd L(\pm u)^{-1}=
\pp{0&\pm u\mm&0&0\cr (\pm u\mm)^*&0&0&0\cr
0&0&0&\overline{\pm u\mm}\cr 0&0&\overline{(\pm u\mm)^*}
&0}.\ee
An absolutely remarkable property of the fluctuated Dirac
operator in internal space is that it can be written as the flat Dirac
operator plus a 1-form:
\bb \dd=\tilde \dd+\rho (\pm u,1)\,[\dd,\rho (\pm u^{-1},1)]+
J\,\rho (\pm u,1)\,[\dd,\rho (\pm
u^{-1},1)]\,J^{-1}.\label{mmcov}\ee
The anti-Hermitean 1-form
\bb (-i)\rho (\pm u,1)\,[\dd,\rho (\pm u^{-1},1)] =(-i)
\pp{0&h&0&0\cr h^*&0&0&0\cr 0&0&0&0\cr 0&0&0&0},\qq
h:=\pm u\mm-\mm\ee
 is the
internal connection. The fluctuated Dirac operator is the
covariant one with respect to this connection. Of course, this
connection is flat, its field strength = curvature 2-form vanishes, a
constraint that is relaxed by the equivalence principle. The result
can be stated without going into the details of the reconstruction
of 2-forms from the spectral triple:
$h$ becomes a general complex doublet, not necessarily of the
form $\pm u\mm-\mm$.

Now we are ready to tensorize the spectral triple of spacetime with
the internal one and compute the spectral action. The algebra
$\aa_t=\ccc^\infty(M)\ot \aa$ describes a two-sheeted universe.
Let us call again its sheets `left' and `right'. The Hilbert space
$\hh_t=\lll^2(\sss)\ot\hh$ describes the neutrino and the electron
as genuine fields, that is spacetime dependent. The Dirac operator
$\tilde \dd_t=\tilde
\ddd\ot 1_6
\,+\,\gamma _5\ot\tilde \dd$ is the flat, free, massive Dirac
operator and it is impatient to fluctuate.

The automorphism group close to the identity,
\bb {\rm Aut}(\aa_t)=[{\rm Diff}(M)\ltimes
\,^MSU(2)/\zz_2]\,\times\,{\rm Diff}(M)\ \owns ((\sigma _L,
\sigma _{\pm u}),\sigma _R),\ee
now contains two independent coordinate transformations $\sigma
_L$ and $\sigma _R$ on each sheet and a {\it gauged}, that is
spacetime dependent, internal transformation $\sigma _{\pm u}$.
The gauge transformations are inner, they act by conjugation
$i_{\pm u}$. The receptacle group is
\bb{\rm Aut}_{\hh_t}(\aa_t)={\rm Diff}(M)\ltimes
\,^M(Spin(4)\times U(2)\times U(1)).\ee
It only contains one coordinate transformation, a point on the left
sheet travels together with its right shadow. Indeed the covariance
property forbids to lift an automorphism with $\sigma _L\not=
\sigma _R$. Since the mass term multiplies left- and right-handed
electron fields, the covariance property saves the locality of field
theory, which postulates that only fields at the same spacetime
point can be multiplied. We have seen examples where the
receptacle has more elements than the automorphism group, e.g.
six-dimensional spacetime or the present internal space. Now we
have an example of automorphisms that do not fit into the
receptacle. In any case the spin group is the image of the
combined, now 4-valued lift $L_t(\sigma ,\sigma _{\pm u})$,
\bb L_t({\rm Aut}(\aa_t))={\rm Diff}(M)\ltimes
\,^M(Spin(4)\times SU(2)).\ee
The fluctuating Dirac operator is
\bb \dd_t=L_t(\sigma ,\sigma _{\pm u})\tilde \dd_t L_t(\sigma
,\sigma _{\pm u})^{-1}=
\pp{\ddd_L&\gamma _5\varphi   &0&0\cr
\gamma _5\varphi   ^*&\ddd_R&0&0\cr
0&0&C\ddd_LC^{-1}&\gamma _5\bar\varphi   \cr
0&0&\gamma _5\bar\varphi  ^*&C\ddd_RC^{-1}},\ee
with
\bb e^{-1}=\sqrt{\jj \jj^T},
&& \ddd_L=i\hbox{$
e^{-1\,\mu}$}_a\gamma^a[
\pa _{ \mu}+s(\omega (e)_{\mu})+A_\mu ],
\\ A_\mu =-\pm u\,\pa_\mu (\pm u^{-1}), &&
\ddd_R=i\hbox{$ e^{-1\,\mu}$}_a\gamma^a[
\pa _{ \mu}+s(\omega (e)_{\mu}) ], \\ \varphi =\pm
u\mm.&&\ee
Note that the sign ambiguity in $\pm u$ drops out from the
$su(2)$-valued 1-form
$A=A_\mu \de x^\mu$ on spacetime. This is not the case for the
ambiguity in the `Higgs' doublet
$\varphi $ yet, but this ambiguity does drop out from the spectral
action. The variable
$\varphi $ is the homogeneous variable corresponding to the
affine variable $h=\varphi -\mm$ in the connection 1-form on
internal space. The fluctuating Dirac operator $\dd_t$ is still flat.
This constraint has now three parts, $e^{-1}=\sqrt{\jj(\sigma )
\jj(\sigma )^T},\ A=-u\de (u^{-1}),$ and $\varphi =\pm
u\mm$. According to the equivalence principle, we will take $e$ to
be any symmetric, invertible matrix depending differentiably on
spacetime, $A$ to be any $su(2)$-valued 1-form on spacetime and
$\varphi $ any complex doublet depending differentiably on
spacetime. This defines the new kinematics. The dynamics of the
spinors = matter is given by the fluctuating Dirac operator
$\dd_t$, which is covariant with respect to i.e. minimally coupled
to gravity, the gauge bosons and the Higgs boson. This dynamics is
equivalently given by the Dirac action $(\psi, \dd_t \psi )$ and
this action delivers the awkward Yukawa couplings for free.
The Higgs boson
$\varphi $ enjoys two geometric
interpretations, first as connection in the discrete direction. The
second derives from Connes' distance formula:
$1/|\varphi (x) |$ is the -- now
$x$-dependent -- distance between the two sheets. The calculation
behind the second interpretation makes explicit use of
the Kaluza-Klein nature of almost commutative geometries
\cite{mw}.

As in pure gravity, the dynamics of the new kinematics derives
from the Chamseddine-Connes action,
\bb S_{CC}[e,A,\varphi ]&=&\t
f(\dd_t^2/\Lambda ^2)\cr &=&\int_M
[\ {\textstyle\frac{2\Lambda_c}{16\pi G}} -{\textstyle\frac{1}{16\pi
G}}R +a(5\,R^2-8\,{\rm Ricci}^2-7\,{\rm
Riemann}^2)\cr
&&\qq\qq + {\textstyle\frac{1}{2g_2^2}} \t F^*_{\mu \nu} F^{\mu
\nu } +{\textstyle\frac{1}{2}} (\dee_\mu \varphi )^*\dee^\mu
\varphi
\cr &&\qq\qq + \lambda |\varphi |^4-{\textstyle\frac{1}{2}} \mu
^2|\varphi |^2+{\textstyle\frac{1}{12}} |\varphi |^2R\
 ]\,\de V \,+\,
O(\Lambda^{-2}), \label{mmsp}\ee
where the coupling constants are
 \bb \Lambda_c=
{\frac{6f_0}{f_2}}\,\Lambda^2,\qq
G={\frac{\pi}{2f_2}}\,\Lambda^{-2},\qq
a={\frac{f_4}{960\pi^2}},\qq
g^2_2={\frac{6\pi ^2}{f_4}},\qq
\lambda ={\frac{\pi ^2}{2f_4}} ,\qq
\mu ^2={\frac{2f_2}{f_4}}\, \Lambda ^2.\ee
 Note the
presence of the conformal coupling of the scalar to the curvature
scalar, $+{\textstyle\frac{1}{12}} |\varphi |^2R$. From the fluctuation
of the Dirac operator, we have derived the scalar representation, a
complex doublet $\varphi $. Geometrically, it is a connection on the
finite space and as such unified with the Yang-Mills bosons, which
are connections on spacetime. As a consequence, the Higgs self
coupling $\lambda $ is related to the gauge coupling $g_2$ in the
spectral action, $g_2^2=12\,\lambda $. Furthermore the spectral action
contains a negative mass square term for the Higgs
$-{\textstyle\frac{1}{2}} \mu ^2|\varphi |^2$ implying a non-trivial
ground state or vacuum expectation value
$|\varphi |=v=\mu (4\lambda )^{-1/2}$ in flat spacetime. Reshifting
to the inhomogeneous scalar variable $h=\varphi -v$, which
vanishes in the ground state, modifies the cosmological constant by
$V(v)$ and Newton's constant from the term
${\textstyle\frac{1}{12}}v^2R$:
\bb\Lambda_c=
6\left( 3{\textstyle\frac{f_0}{f_2}} -
{\textstyle\frac{f_2}{f_4}}\right) \Lambda^2,\qq
G={\frac{3\pi }{2f_2}}\Lambda ^{-2}.\ee
 Now the cosmological constant can have either sign, in particular it
can be zero. This is welcome because experimentally the cosmological
constant is very close to zero, $\Lambda _c\ <\ 10^{-119}/G$.
 On the other
hand, in spacetimes of large curvature, like for example the ground
state, the positive conformal coupling of the scalar to the curvature
dominates the negative mass square term $-{\textstyle\frac{1}{2}} \mu
^2|\varphi |^2$. Therefore the vacuum expectation value of the Higgs
vanishes, the gauge symmetry is unbroken and all particles are
massless. It is only after the big bang, when spacetime loses its
strong curvature that the gauge symmetry breaks down
spontaneously and particles acquire masses.

The computation of the spectral action is long, let us set some
waypoints. The square of the fluctuating Dirac operator is
$\dd_t^2=-\Delta +E$, where $\Delta $ is the covariant
Laplacian, in coordinates:
\bb \Delta &=&g^{\mu\tilde\nu}\left[\left(
\frac{\partial}{\partial x^\mu} 1_4\ot 1_\hh+
{\textstyle\frac{1}{4}}\omega_{ab\mu}\gamma^{ab}
\ot 1_\hh+1_4\ot [\rho(A_\mu)+J\rho(A_\mu)J^{-1}]\right)
{\delta^\nu}_{\tilde\nu}-{\Gamma^\nu}_{\tilde
\nu\mu}1_4\ot1_{\hh}\right]
\cr &&\qq\qq\qq\times\left[
\frac{\partial}{\partial x^\nu} 1_4\ot 1_\hh+
{\textstyle\frac{1}{4}}\omega_{ab\nu}\gamma^{ab}
\ot 1_\hh+1_4\ot [\rho(A_\nu)+J\rho(A_\nu)J^{-1}]\right],\ee
and where
$E$, for endomorphism, is a zero order operator, that is
a matrix of size
$4\dim\hh$ whose entries are functions
constructed from the bosonic fields and
their first and second derivatives,
\bb E&=&{\textstyle\frac{1}{2}}\left[
\gamma^{\mu}\gamma^\nu
\ot1_\hh\right]\rr_{\mu\nu} \\ \cr &&\,+\,
\pp{
1_4\ot \varphi \varphi ^*&-i\gamma_5\gamma^\mu\ot
\dee_\mu\varphi &0&0\cr
-i\gamma_5\gamma^\mu\ot (\dee_\mu\varphi )^*&
1_4\ot\varphi ^*\varphi &0&0\cr
0&0&1_4\ot \overline{\varphi
\varphi ^*}&-i\gamma_5\gamma^\mu\ot
\overline{\dee_\mu\varphi }\cr 0&0&
-i\gamma_5\gamma^\mu\ot \overline{(\dee_\mu\varphi )^*}&
1_4\ot\overline{\varphi ^*\varphi }
}\label{E}.\eee
$\rr$ is the total curvature, a 2-form with values in
the (Lorentz $\op$ internal) Lie algebra represented
on (spinors $\ot\ \hh$). It contains the curvature
2-form
$R=\de\omega+\omega^2
$ and the field strength 2-form $F=\de
A+A^2$, in components
\bb\rr_{\mu\nu}={\textstyle\frac{1}{4}}
R_{ab\mu\nu}\gamma^{a}\gamma^b\ot 1_\hh+
1_4\ot[\rho(F_{\mu\nu})+J\rho(F_{\mu\nu})J^{-1}].\ee
The first term in
equation (\ref{E}) produces
the curvature scalar, which we also (!)
denote by $R$,
\bb
{\textstyle\frac{1}{2}}\left[{e^{-1\,\mu}}_c\,{e^{-1\,\nu}}_d\,
\gamma^{c}\gamma^d\right]
{\textstyle\frac{1}{4}}R_{ab\mu\nu}\gamma^{a}\gamma ^b
= {\textstyle\frac{1}{4}}R1_4.\ee
We have also used the possibly dangerous notation $\gamma ^\mu
={e^{-1\,\mu }}_a\gamma ^a$.
 Finally $\dee$ is the covariant derivative appropriate
for the representation of the scalars.
The above formula for the square of the Dirac operator is also
known as Lichn\'erowicz formula. The Lichn\'erowicz formula with
arbitrary torsion can be found in \cite{att}.

Let $f:\rr_+\rightarrow\rr_+$ be a positive, smooth
function with finite moments,
\bb f_0={\textstyle \int_0^\infty} uf(u)\,\de u,&
f_2={\textstyle \int_0^\infty} f(u)\,\de u,&
f_4=f(0),\\
f_6=-f'(0),&
f_8=f''(0),&...\ee
Asymptotically, for large $\Lambda$, the
distribution function of the spectrum is given in
terms of the heat kernel expansion \cite{egbv}:
\bb S=\t f(\dd_t^2/\Lambda^2)=
\frac{1}{16\pi^2}\,\int_M[\Lambda^4f_0a_0+
\Lambda^2f_2a_2+f_4a_4+\Lambda^{-2}f_6a_6+...]\,
\de V, \label{master}\ee
where the $a_j$ are the coefficients of the heat kernel
expansion of the Dirac operator squared \cite{heat},
\bb a_0&=&\t (1_4\ot1_\hh),\\
a_2&=&{\textstyle\frac{1}{6}}R\,\t (1_4\ot1_\hh)-\t
E,\\ a_4&=&{\textstyle\frac{1}{72}}R^2\t
(1_4\ot1_\hh)-
{\textstyle\frac{1}{180}}R_{\mu\nu}R^{\mu\nu}
\t (1_4\ot1_\hh)+
{\textstyle\frac{1}{180}}R_{\mu\nu\rho\sigma}
R^{\mu\nu\rho\sigma}\t (1_4\ot1_\hh)\cr &&+
{\textstyle\frac{1}{12}}\t (\rr_{\mu\nu}
\rr^{\mu\nu})
-{\textstyle\frac{1}{6}}R\,\t E+
{\textstyle\frac{1}{2}}\t E^2 + {\rm surface\ terms}.\ee
As already noted, for large
$\Lambda$ the positive function
$f$ is universal, only the first three
moments, $f_0,\ f_2$ and $f_4$
appear with non-negative powers of $\Lambda $.
For the minimax model, we get (more details can be found in
\cite{rom}):
\bb a_0&=&4\dim\hh=4\times 6,\\
\t E&=&\dim\hh\, R + 16 |\varphi|^2,\\
a_2&=&{\textstyle\frac{2}{3}}\dim\hh\, R-\dim\hh\, R
-16|\varphi|^2=
 -{\textstyle\frac{1}{3}}\dim\hh\, R
-16|\varphi|^2,\\
\t \left( {\textstyle\frac{1}{2}}
[\gamma^{a},\gamma^b]
{\textstyle\frac{1}{2}}
[\gamma^{c},\gamma^d]\right)&=&
4\left[\delta ^{ad}\delta ^{bc}-\delta ^{ac}\delta ^{bd}\right],\\
\t \{\rr_{\mu\nu}\rr^{\mu\nu}\}&=&
-{\textstyle\frac{1}{2}}\dim\hh\,
R_{\mu\nu\rho\sigma}R^{\mu\nu\rho\sigma}\cr &&\qq\qq
-4 \,\t
\{[\rho(F_{\mu\nu})+J\rho(F_{\mu\nu})J^{-1}]^*
[\rho(F^{\mu\nu})+J\rho(F^{\mu\nu})J^{-1}]\}\cr
&=&-{\textstyle\frac{1}{2}}\dim\hh\,
R_{\mu\nu\rho\sigma}R^{\mu\nu\rho\sigma}
-8\,\t
\{\rho(F_{\mu\nu})^*
\rho(F^{\mu\nu})\},\\
\t E^2&=&{\textstyle\frac{1}{4}}\dim\hh\, R^2+
4\,\t\{ \rho(F_{\mu\nu})^*\rho(F^{\mu\nu})\}\cr &&+
16|\varphi|^4+
16(\dee_\mu\varphi)^*(\dee^\mu\varphi)
+8|\varphi|^2R,
\ee
Finally we have up to surface terms,
\bb a_4 &=& {\textstyle\frac{1}{360}}\dim \hh\,
(5\,R^2-8\,{\rm Ricci}^2-7\,{\rm
Riemann}^2)+
{\textstyle\frac{4}{3}}
\t \rho(F_{\mu\nu})^*\rho(F^{\mu\nu})\cr &&+
8|\varphi|^4+
8(\dee_\mu\varphi)^*(\dee^\mu\varphi)
+{\textstyle\frac{4}{3}}
|\varphi|^2R.\ee
We arrive at the spectral action with its conventional
normalization, equation (\ref{mmsp}), after a finite
renormalization $|\varphi |^2\rightarrow {\textstyle\frac{\pi
^2}{f_4}} |\varphi |^2$.

Our first timid excursion into gravity on a noncommutative
geometry produced a rather unexpected discovery. We stumbled
over a Yang-Mills-Higgs model, which is precisely the electro-weak
model for one family of leptons but with the $U(1)$ of hypercharge
amputated. The sceptical reader suspecting a sleight of
hand is encouraged to try and find a simpler, noncommutative
finite spectral triple.

\subsection{A central extension}

We will see in the next section the technical reason for the absence
of  $U(1)$s as automorphisms: all automorphisms of finite spectral
triples connected to the identity are inner, i.e. conjugation by
unitaries. But conjugation by central unitaries is trivial. This
explains  that in the minimax example, $\aa=\hhh\op\cc$, the
component of the automorphism group connected to the identity was
$SU(2)/\zz_2\owns (\pm u,1)$. It is the domain of definition of
the lift, equation (\ref{mmlift}),
\bb L(\pm u,1)=\rho (\pm u,1)J\rho (\pm u,1)J^{-1}=
\pp{\pm u&0&0&0\cr 0&1&0&0\cr 0&0&\pm \bar u&0\cr
 0&0&0&1}.\ee
 It is
tempting to centrally extend the lift to all unitaries of the algebra:
\bb \llll(w,v)=\rho (w,v)J\rho(w,v)J^{-1}=
\pp{\bar vw&0&0&0\cr 0&\bar v^2&0&0\cr 0&0& v\bar w &0\cr
0&0&0&v^2},\qq (w,v)\in SU(2)\times U(1).
\label{mmext}\ee
An immediate consequence of this extension is encouraging: the
extended lift is single-valued and after tensorization with the one from
Riemannian geometry, the multi-valuedness will remain two.

Then redoing the fluctuation of the Dirac operator and
recomputing the spectral action yields gravity coupled to the
complete electro-weak model of the electron and its neutrino with a
weak mixing angle of $\sin^2\theta _w=1/4$.

\section{Connes' do-it-yourself kit}

Our first example of gravity on an almost commutative space leaves
us wondering what other examples will look like. To play on the
Yang-Mills-Higgs machine, one must know the classification of all
real, compact Lie groups and their unitary representations. To play
on the new machine, we must know all finite spectral triples. The
first good news is that the list of algebras and their representations
is infinitely shorter than the one for groups. The other good news
is that the rules of
Connes' machine are not made up opportunistically to suit the
phenomenology of electro-weak and strong forces as in the case of the
Yang-Mills-Higgs machine. On the contrary, as developed
in the last section, these rules derive naturally from geometry.

\subsection{Input}

Our first input item is a finite dimensional, real, associative
involution algebra with unit and that admits a finite dimensional
faithful representation. Any such algebra is a direct sum of simple
algebras with the same properties. Every such simple algebra is an
algebra of $n\times n$ matrices with real, complex or
quaternionic entries, $\aa=M_n(\rr)$, $M_n(\cc)$ or
$M_n(\hhh)$. Their unitary groups $U(\aa):=\{u\in\aa,\,
uu^*=u^*u=1\}$ are $O(n)$, $U(n)$ and $USp(n)$. Note that
$USp(1)=SU(2)$. The centre
$Z$ of an algebra $\aa$ is the set of elements $z\in\aa$ that
commute with all elements $a\in\aa$. The central unitaries form
an abelian subgroup of $U(\aa)$. Let us denote this subgroup by
$U^c(\aa):=U(\aa)\cap Z$. We have
$U^c(M_n(\rr))=\zz_2\owns\pm 1_n$,
$U^c(M_n(\cc))=U(1)\owns\exp (i\theta ) 1_n$, $\theta \in
[0,2\pi )$,
$U^c(M_n(\hhh))=\zz_2\owns\pm 1_{2n}$.
All
 automorphisms of the real, complex and quaternionic matrix
algebras are inner with one exception,
$M_n(\cc)$ has one outer automorphism, complex conjugation,
which is disconnected from the identity automorphism. An inner
automorphism $\sigma
$ is of the form $\sigma (a)=uau^{-1}$ for some $u\in U(\aa)$ and
for all $a\in\aa$. We will denote this inner automorphism by
$\sigma =i_u$ and we will write Int($\aa$) for the group of inner
automorphisms. Of course a commutative algebra, e.g. $\aa=\cc$,
has no inner automorphism. We have Int$(\aa)=U(\aa)/U^c(\aa)$,
in particular
Int$(M_n(\rr))= O(n)/\zz_2,\ n=2,3,...,$
Int$(M_n(\cc))= U(n)/U(1)=SU(n)/\zz_n,\ n=2,3,...,$
Int$(M_n(\hhh))= USp(n)/\zz_2,\ n=1,2,..$. Note the apparent
injustice: the commutative algebra $\ccc^\infty (M)$ has the
nonAbelian automorphism group Diff$(M)$ while the
noncommutative algebra $M_2(\rr)$ has the Abelian automorphism
group $O(2)/\zz_2$. All
exceptional groups are missing from our list of groups. Indeed they
are automorphism groups of non-associative algebras, e.g. $G_2$ is
the automorphism group of the octonions.

The second input item is a faithful representation $\rho $ of the
algebra $\aa$ on a finite dimensional, complex Hilbert space
$\hh$. Any such representation is a direct sum of irreducible
representations.
$M_n(\rr)$ has only one irreducible representation, the
fundamental one on $\rr^n$,
$M_n(\cc)$ has two, the fundamental one and its complex
conjugate. Both are defined on $\hh=\cc^n\owns \psi $ by $\rho
(a)\psi =a\psi $ and by $\rho (a)\psi =\bar a\psi $. $M_n(\hhh)$ has
only one irreducible representation, the fundamental one defined
on
$\cc^{2n}$. For example, while $U(1)$ has an infinite number of
inequivalent irreducible representations, characterized by an
integer `charge', its algebra $\cc$ has only two with charge plus
and minus one. While $SU(2)$ has an infinite number of
inequivalent irreducible representations characterized by its spin,
$0,{\textstyle\frac{1}{2}} ,1,...$, its algebra $\hhh$ has only one,
spin ${\textstyle\frac{1}{2}} $. The main reason behind this
multitude of group representation is that the tensor product of two
representations of one group is another representation of this
group, characterized by the sum of charges for $U(1)$ and by the sum
of spins for $SU(2)$. The same is not true for two 
representations of one associative algebra whose tensor product
fails to be linear. (Attention, the tensor product of two
representations of two algebras does define a representation of the
tensor product of the two algebras. We have used this tensor
product of Hilbert spaces to define almost commutative geometries.)

The third input item is the finite Dirac operator $\dd$ or
equivalently the fermionic mass matrix, a matrix of size
dim$\hh_L\,\times\,$dim$\hh_R$.

These three items can however not be chosen freely, they must still
satisfy all axioms of the spectral triple \cite{tkmz}. I do hope you
have convinced yourself of the nontriviality of this requirement for
the case of the minimax example.

The minimax example has taught us something else. If we want
abelian gauge fields from the fluctuating metric, we must centrally
extend the spin lift, an operation, that at the same time may reduce
the multivaluedness of the original lift. Central extensions are by
no means unique, its choice is our last input item \cite{fare}.

To simplify notations, we
concentrate on complex matrix algebras $M_n(\cc)$ in the following
part. Indeed the others, $M_n(\rr)$ and $M_n(\hhh)$,
do not have central unitaries close to the identity. We have already
seen that it is important to separate the commutative and
noncommutative parts of the algebra:
\bb\aa=\cc^M\oplus
\bigoplus_{k=1}^N M_{n_k}(\cc)\ \owns
a=(b_1,...b_M,c_1,...,c_N),\qq n_k\geq 2.
\label{algebra}\ee
Its group of unitaries is
\bb U(\aa)=U(1)^M\times
\matrix{N\cr \times\cr {k=1}} U(n_k)\
\owns\ u=(v_1,...,v_M,w_1,...,w_N)\ee
and its group of central
unitaries
\bb U^c(\aa)=U(1)^{M+N}\
\owns\ u_c=
( v_{c1},...,v_{cM},w_{c1}1_{n_1},...,w_{cN}1_{n_N}).\ee
All automorphisms
connected to the identity are inner,
there are outer automorphisms, the complex conjugation and, if
there are identical summands in $\aa$, their permutations.
In compliance with the minimax principle, we disregard the discrete
automorphisms.  Multiplying a unitary $u$ with
a central unitary
$u_c$ of course does not affect its inner automorphism
$i_{u_cu}=i_u$. This ambiguity distinguishes
between `harmless' central unitaries $v_{c1},...,v_{cM}$ and
the others, $w_{c1},...,w_{cN}$, in the sense that
\bb {\rm Int}(\aa)=U^n(\aa)/U^{nc}(\aa),\label{inntrue}\ee
where we have defined the group of noncommutative unitaries
\bb U^n(\aa):=\matrix{N\cr \times\cr {k=1}} U(n_k)\
\owns\ w\ee
and $U^{nc}(\aa):=U^n(\aa)\cap U^c(\aa) \owns w_c$.
The map
\bb i:U^n(\aa)&\longrightarrow&{\rm Int}(\aa)\cr
w&\longmapsto&i_w\ee
 has kernel Ker$\,i=U^{nc}(\aa)$.

The lift of an inner
automorphism to the Hilbert space has a simple closed
form
\cite{tresch}, $L=\hat L\circ i^{-1}$ with
\bb \hat L(w)=\rho(1,w)J\rho(1,w)J^{-1}.\ee
It satisfies $p(\hat L(w))=i(w)$.
If the kernel of $i$ is contained in the kernel of $\hat L$, then
the lift is well defined, as e.g. for $\aa=\hhh$,
$U^{nc}(\hhh)=\zz_2$.
\begin{eqnarray}
&&{\rm Aut}_\hh(\aa)\nonumber \\
&& \hskip -2mm p\
\parbox{6mm}{\begin{picture}(20,10)
\put(0,15){\vector(0,-1){30}}
\put(15,-15){\vector(-1,4){8}}
\put(15,-15){\vector(0,1){33}}
\end{picture}}
 L
\parbox{8mm}{\begin{picture}(20,10)
\put(30,-15){\vector(-1,2){16}}
\put(32,-10){$\hat{L}$}
\end{picture}}
\parbox{12mm}{\begin{picture}(20,10)
\put(65,-15){\vector(-2,1){67}}
\end{picture}}
\ell
\\[4mm]
&&{\rm Int}(\aa)\stackrel{i}{\longleftarrow}
U^n(\aa)\begin{array}{c}\\[-3mm]
\hookleftarrow \\[-5mm]
\stackrel{\vector(1,0){15}}{\mbox{\footnotesize $\det$}}
\end{array}
U^{nc}(\aa) \nonumber
\end{eqnarray}
For more complicated real or
quaternionic algebras, $U^{nc}(\aa)$ is finite and the lift $L$
is multi-valued with a finite number of values. For
noncommutative, complex algebras, their continuous family of
central unitaries cannot be eliminated except for very special
representations and we face a continuous infinity of values.
The solution of this problem follows an old strategy: {\it `If
you can't beat them, adjoin them'.} Who is {\it them?} The
harmful central unitaries $w_c\in U^{nc}(\aa)$ and adjoining
means central extending.
 The central extension (\ref{mmext}), only concerned a discrete
group and a harmless $U(1)$. Nevertheless it generalizes naturally
 to the present setting:
\bb
 \llll:{\rm Int}(\aa)\times U^{nc}(\aa)&\longrightarrow&
\qq\qq{\rm Aut}_\hh(\aa)\cr
(w_\sigma ,w_c)\qq \qq &\longmapsto &
(\hat L\circ i^{-1})(w_\sigma )\,\ell(w_c)\ee
with
 \bb\ell(w_c)&:=&\rho\!\! \left(
\prod_{j_1=1}^N(w_{cj_1})^{q_{1,j_1}},
...,\prod_{j_M=1}^N(w_{cj_M})^{q_{M,j_M}},\right.\\ &&
\left.\qq
\prod_{j_{M+1}=1}^N(w_{cj_{M+1}})^{q_{{M+1},j_{M+1}}}
1_{n_1},
...,\prod_{j_{M+N}=1}^N(w_{cj_{M+N}})^{q_{{M+N},j_{M+N}}}
1_{n_N}
\right)
 J
\rho (...)
\,J^{-1}
\label{ell}\eee
with the $(M+N)\times N$ matrix of charges $q_{kj}$. The
extension satisfies indeed
$p(\ell(w_c))=1\in{\rm Int}(\aa)$ for all $w_c\in
U^{nc}(\aa)$.

Having adjoined the harmful, continuous central unitaries, we
may now stream line our notations and write the group of
inner automorphisms as
\bb {\rm Int}(\aa)=\left(
\matrix{{N} \cr \times \cr k=1} SU(n_k)\right) /\Gamma
\owns[w_\sigma] = [(w_{\sigma 1},...,w_{\sigma N})]\
{\rm mod}\
\gamma \label{innfake} ,\ee
where $\Gamma $ is the discrete group
\bb\Gamma=\matrix{{N}\cr \times\cr
{k=1}}\zz_{n_k}\ \owns\ (z_11_{n_1},...,z_N1_{n_N}),\qq
z_{k}=\exp[-m_{k}2\pi i/n_k],\ m_k=0,...,n_k-1
\label{discrete}\ee
and the quotient is factor by factor. This way to write inner
automorphisms is convenient for complex matrices, but not
available for real and quaternionic matrices. Equation
(\ref{inntrue}) remains the general characterization of
inner automorphisms.

The lift $L(w_\sigma )=(\hat L\circ i^{-1})(w_\sigma )$, $w_\sigma =
w\ {\rm mod}\ U^{nc}(\aa)$,
is multi-valued with, depending
on the representation, up to $ |\Gamma |=\prod_{j=1}^N n_j$
values. More precisely the multi-valuedness of $L$ is indexed
by the elements of the kernel of the projection $p$ restricted
to the image $L({\rm Int}(\aa))$. Depending on the choice of
the charge matrix
$q$, the central extension $\ell$ may reduce this
multi-valuedness. Extending harmless central unitaries is
useless for any reduction. With
the multi-valued group homomorphism
\bb (h_\sigma ,h_c) : U^n(\aa)&\longrightarrow &
{\rm Int}(\aa)\times U^{nc}(\aa)\cr
(w_j) & \longmapsto &((w_{\sigma j} , w_{cj}))=((w_j(\det
w_j)^{-1/n_j},(\det w_j)^{1/n_j}))\label{isom},\ee
 we can write the two lifts
$L$ and
$\ell$
together in closed form
$\llll:U^n(\aa)\rightarrow
{\rm Aut}_\hh(\aa)$:
\bb\llll(w)&=&L(h_\sigma (w))\,\ell(h_c(w))\cr \cr
&=&
\rho\!\! \left(
\prod_{j_1=1}^N(\det w_{j_1})^{\tilde q_{1,j_1}},
...,\prod_{j_M=1}^N(\det w_{j_M})^{\tilde
q_{M,j_M}},\right.\cr &&\left.\qq
w_1\prod_{j_{M+1}=1}^N(\det w_{j_{M+1}})^{\tilde
q_{{M+1},j_{M+1}}}, ...,w_N\prod_{j_{N+M}=1}^N(\det
w_{j_{N+M}})^{\tilde q_{{N+M},j_{N+M}}}\right)%\cr
\nonumber\\[2mm]
&&\times\,
J \rho (...) J^{-1}.\ee
We have set
\bb\tilde q:=\left( q-\pp{0_{M\times N}\cr\cr  1_{N\times
N}}
\right) \pp{n_1&&\cr &\ddots&\cr &&n_N}^{-1}.\ee
Due to the
phase ambiguities in the roots of the determinants, the
extended lift
 $\llll$ is multi-valued in general. It is single-valued if the
matrix
$\tilde q$ has integer entries, e.g.
$q=\pp{0\cr 1_N}$, then $\tilde q=0$ and
$\llll(w)=\hat L(w)$. On the other hand, $q=0$ gives
$\llll(w)=\hat L(i^{-1}(h_\sigma (w)))$, not always well
defined as already noted. Unlike the extension
(\ref{mmext}), and unlike the map $i$, the extended lift $\llll$ is
not necessarily even. We do impose this symmetry
$\llll (-w)=\llll (w)$, which translates into
conditions on the charges, conditions that depend on the
details of the representation $\rho $.

Let us note that the lift
$\llll$ is simply a representation up to a phase and as such it is
not the most general lift. We could have added harmless central
unitaries if any present, and, if the representation
$\rho $ is reducible, we could have chosen different charge
matrices in different irreducible components. If you are not
happy with central extensions, then this is a sign of good taste.
Indeed commutative algebras like the calibrating example have no
inner automorphisms and a huge centre. Truly noncommutative
algebras have few outer automorphism and a small centre. We
believe that almost commutative geometries with their central
extensions are only low energy approximations of a truly
noncommutative geometry where central extensions are not an
issue.

\subsection{Output}

From the input data of a finite spectral triple, the central charges
and the three moments of the spectral function, noncommutative
geometry produces a  Yang-Mills-Higgs model coupled
to gravity. Its entire Higgs sector is computed from the input
data, Figure 6. The Higgs representation derives from the
fluctuating metric and the Higgs potential from the spectral
action.

\begin{figure}[h]
\label{slot3}
\epsfxsize=8cm
\hspace{4.1cm}
\epsfbox{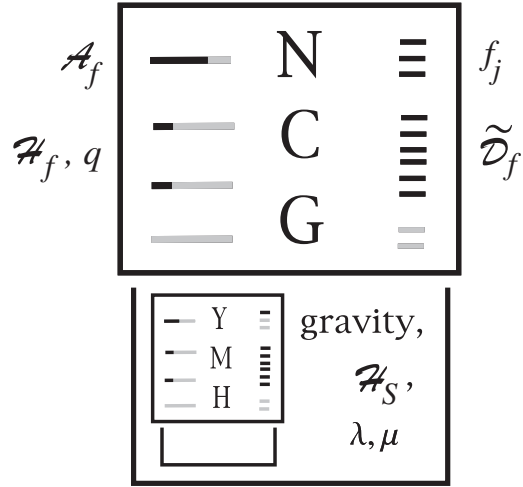}
\caption{Connes' slot machine}
\end{figure}

To see how the Higgs representation derives in general from the
fluctuating Dirac operator $\dd$, we must write it as `flat' Dirac
operator
$\tilde \dd$ plus internal 1-form $H$ like we have done in
equation (\ref{mmcov}) for the minimax example without
extension. Take the extended lift $\llll (w)=\rho (w) J\rho
(w)J^{-1}$ with the unitary
\bb w&=&\prod_{j_1=1}^N(\det w_{j_1})^{\tilde q_{1j_1}},
...,\prod_{j_M=1}^N(\det w_{j_M})^{\tilde
q_{Mj_M}},\cr
&&w_1\prod_{j_{M+1}=1}^N(\det w_{j_{M+1}})^{\tilde
q_{{M+1},j_{M+1}}}, ...,w_N\prod_{j_{N+M}=1}^N(\det
w_{j_{N+M}})^{\tilde q_{{N+M},j_{N+M}}}.\ee
Then
\bb\dd&&= \llll\tilde \dd\llll^{-1}\cr &&=
\left(\rho(w)\,J\rho(w)J^{-1}\right)\,
\tilde \dd\,
\left(\rho(w)\,J\rho(w)J^{-1}\right)^{-1} =
\rho(w)\,J\rho(w)J^{-1}
\tilde \dd\,
\rho(w^{-1})\,J\rho(w^{-1})J^{-1} \cr
&&=
\rho(w)J\rho(w)J^{-1}(\rho(w^{-1})\tilde \dd+
[\tilde \dd,\rho(w^{-1})])J\rho(w^{-1})J^{-1}\cr
&&=
J\rho(w)J^{-1}\tilde \dd J\rho(w^{-1})J^{-1}
+\rho(w)[\tilde \dd,\rho(w^{-1})]=
J\rho(w)\tilde \dd \rho(w^{-1})J^{-1}
+\rho(w)[\tilde \dd,\rho(w^{-1})]\cr
&&=
J(\rho(w)[\tilde \dd ,\rho(w^{-1})]+\tilde \dd)J^{-1}
+\rho(w)[\tilde \dd,\rho(w^{-1})]\cr
&&=\tilde \dd\,+\,H\,+\,JHJ^{-1},
\label{fluct}\ee
with the internal 1-form, the Higgs scalar, $H=\rho (w)[\tilde
\dd,\rho (w^{-1})]$. In the chain (\ref{fluct}) we have used
successively the following three axioms of spectral
triples, $[\rho(a),J\rho(\tilde a)J^{-1}]=0$, the first
order condition $[[\tilde \dd,\rho(a)],J\rho(\tilde a)J^{-1}]=0$
and $[\tilde \dd,J]=0$. Note that the unitaries, whose
representation commutes with the internal Dirac operator, drop
out from the Higgs, it transforms as a singlet under their
subgroup.

\begin{figure}[h]
\label{versus}
\epsfxsize=11cm
\hspace{2.1cm}
\epsfbox{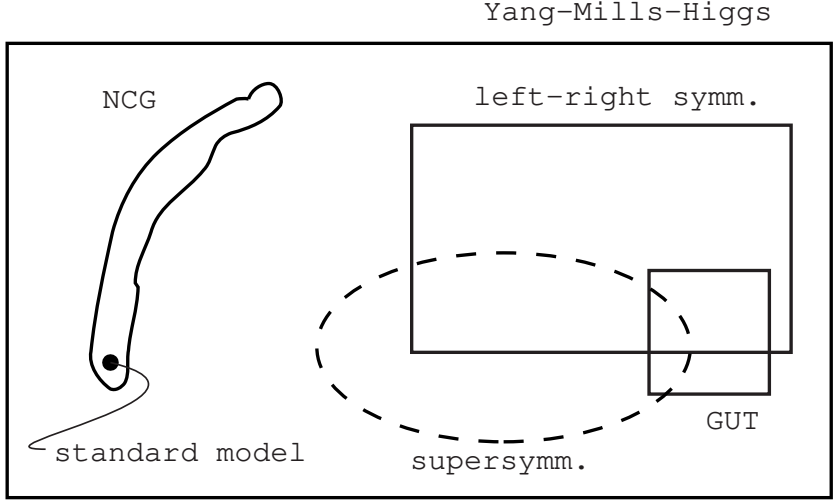}
\caption{Pseudo forces from noncommutative geometry}
\end{figure}

The constraints from the axioms of noncommutative geometry are
so tight that only very few Yang-Mills-Higgs models can be derived
from noncommutative geometry as pseudo forces. No left-right
symmetric model can \cite{florian}, no Grand Unified Theory can
\cite{fedele}, for instance the $SU(5)$ model needs 10-dimensional
fermion representations, $SO(10)$  16-dimensional ones, $E_6$ is
not the group of an associative algebra. Moreover the last two
models are left-right symmetric. Much effort has gone into the
construction of a supersymmetric model from noncommutative
geometry, in vain \cite{kw}. The standard model on the other
hand fits perfectly into Connes' picture, Figure 7.

\subsection{The standard model}

The first noncommutative formulation of the standard model was
published by Connes \& Lott \cite{ccl} in 1990. Since then it
has evolved into its present form \cite{book, tresch, grav, cc}
and triggered quite an amount of literature \cite{stand}.

\subsubsection{Spectral triple}

The internal algebra $\aa$ is chosen as to reproduce
$SU(2)\times U(1)\times SU(3)$ as subgroup of
$U(\aa)$,
\bb \aa=\hhh\op\cc\op
M_3(\cc)\,\owns\,(a,b,c).\ee
The internal Hilbert
space is copied from the Particle Physics Booklet
\cite{data},
\bb \hh_L&=&
\left(\cc^2\ot\cc^N\ot\cc^3\right)\ \op\
\left(\cc^2\ot\cc^N\ot\cc\right), \\
\hh_R&=&\left(\cc\ot\cc^N\ot\cc^3\right)\
\op\ \left(\cc\ot\cc^N\ot\cc^3\right)\
\op\ \left(\cc\ot\cc^N\ot\cc\right).\ee
 In each summand, the first factor
denotes weak isospin doublets or singlets, the second
denotes
$N$ generations, $N=3$, and the third denotes colour
triplets or singlets.
Let us choose the following basis
of the internal Hilbert space, counting fermions and
antifermions (indicated by the superscript $\cdot^c$ for `charge
conjugated') independently,
$\hh=\hh_L\op\hh_R\op\hh^c_L\op\hh^c_R
=\cc^{90}$:
\bb
& \pp{u\cr d}_L,\ \pp{c\cr s}_L,\ \pp{t\cr b}_L,\
\pp{\nu_e\cr e}_L,\ \pp{\nu_\mu\cr\mu}_L,\
\pp{\nu_\tau\cr\tau}_L;&\cr \cr
&\matrix{u_R,\cr d_R,}\qq \matrix{c_R,\cr s_R,}\qq
\matrix{t_R,\cr b_R,}\qq  e_R,\qq \mu_R,\qq
\tau_R;&\cr  \cr
& \pp{u\cr d}^c_L,\ \pp{c\cr s}_L^c,\
\pp{t\cr b}_L^c,\
\pp{\nu_e\cr e}_L^c,\ \pp{\nu_\mu\cr\mu}_L^c,\
\pp{\nu_\tau\cr\tau}_L^c;&\cr\cr
&\matrix{u_R^c,\cr d_R^c,}\qq
\matrix{c_R^c,\cr s_R^c,}\qq
\matrix{t_R^c,\cr b_R^c,}\qq  e_R^c,\qq \mu_R^c,\qq
\tau_R^c.&\eee
This is the current eigenstate basis, the representation
$\rho$ acting on
$\hh$ by 
\bb \rho(a,b,c):=
\pp{\rho_{L}&0&0&0\cr
0&\rho_{R}&0&0\cr
0&0&{\bar\rho^c_{L}}&0\cr
0&0&0&{\bar\rho^c_{R}}}\ee
with
\bb\rho_{L}(a):=\pp{
a\ot 1_N\ot 1_3&0\cr
0&a\ot 1_N},\qq
\rho_{R}(b):= \pp{
b 1_N\ot 1_3&0&0\cr 0&\bar b 1_N\ot 1_3&0\cr
0&0&\bar
b1_N}, \label{repr1}
\ee\bb
  \rho^c_{L}(b,c):=\pp{
1_2\ot 1_N\ot c&0\cr
0&\bar b1_2\ot 1_N},\qq
\rho^c_{R}(b,c) := \pp{
1_N\ot c&0&0\cr 0&1_N\ot c&0\cr
0&0&\bar b1_N}.  \label{repr2}
\ee
 The
apparent asymmetry between particles and
antiparticles -- the former are subject to weak, the
latter to strong interactions --  will disappear after
application of the lift $\llll$ with
\bb J=\pp{0&1_{15N}\cr 1_{15N}&0}\circ
\ {\rm complex\ conjugation}.\ee
 For the sake of
completeness, we record the chirality as matrix
\bb \chi=\pp{-1_{8N}&0&0&0\cr 0&1_{7N}&0&0\cr
 0&0&-1_{8N}&0\cr 0&0&0&1_{7N} }.\ee
The
internal Dirac operator
\bb \tilde \dd=\pp{0&\mm&0&0\cr
\mm^*&0&0&0\cr
0&0&0&\bar\mm\cr
0&0&\bar\mm^*&0}\ee
is made of the fermionic mass matrix of the standard
model,
\bb\mm=\pp{
\pp{1&0\cr 0&0}\ot M_u\ot 1_3\,+\,
\pp{0&0\cr 0&1}\ot M_d\ot 1_3
&0\cr
0&\pp{0\cr 1}\ot M_e},\ee
with
\bb M_u:=\pp{
m_u&0&0\cr
0&m_c&0\cr
0&0&m_t},&& M_d:= C_{KM}\pp{
m_d&0&0\cr
0&m_s&0\cr
0&0&m_b},\\[2mm] && M_e:=\pp{
m_e&0&0\cr
0&m_\mu&0\cr
0&0&m_\tau}.\ee
From the booklet we know that all indicated fermion
masses are different from each other and that the
Cabibbo-Kobayashi-Maskawa matrix  $C_{KM}$ is
non-degenerate in the sense that  no quark is
simultaneously mass and weak current eigenstate.

We must acknowledge the fact -- and this is far from trivial --
that the finite spectral triple of the standard model satisfies all of
Connes' axioms:\\
$\bullet$ It is orientable, $\chi = \rho (-1_2,1,1_3)J\rho
(-1_2,1,1_3)J^{-1}.$\\
$\bullet$ Poincar\'e duality holds. The standard model
has  three minimal
projectors,
\bb p_1=(1_2,0,0),\qq p_2=(0,1,0), \qq p_3=
\left(0,0,\pp{1&0&0\cr 0&0&0\cr 0&0&0}\right)\ee
and the intersection form
\bb \cap=-2N\pp{0&1&1\cr 1&-1&-1\cr 1&-1&0},\ee
is non-degenerate. We note that Majorana masses are forbidden
because
of the axiom $\tilde \dd\chi=-\chi\tilde \dd.$ On the other hand if
we wanted to give Dirac masses to all three neutrinos we would
have to add three right-handed neutrinos to the standard model.
Then the intersection form,
\bb \cap=-2N\pp{0&1&1\cr 1&-2&-1\cr 1&-1&0},\ee
would become degenerate and Poincar\'e duality would fail.\\
$\bullet$ The first order axiom is satisfied precisely because of the
first two of the six ad hoc properties of the standard model
recalled in subsection \ref{winner}, colour couples vectorially and
commutes with the fermionic mass matrix, $[\dd,\rho
(1_2,1,c)]=0$. As an immediate consequence the Higgs scalar =
internal 1-form will be a colour singlet and the gluons will remain
massless, the third ad hoc property of the standard model in its
conventional formulation.\\
$\bullet$
There seems to be some arbitrariness in the choice of the
representation under $\cc\owns b$. In fact this is not true, any
choice different from the one in equations
(\ref{repr1},\ref{repr2}) is either incompatible with the axioms of
spectral triples or it leads to charged massless particles
incompatible with the Lorentz force or to a symmetry breaking
with equal top and bottom masses. Therefore, the only flexibility
in the fermionic charges is from the choice of the central charges
\cite{fare}.

\subsubsection{Central charges}

The standard model has the following groups,
\bb U(\aa)\ =&\ SU(2)\times U(1)\times U(3)\qq& \owns
u=(u_0,v, w),\\
U^c(\aa)\ =&\ \zz_2\times U(1)\times U(1)\qq&\owns
u_c=(u_{c0},v_{c}, w_{c}1_3),\\
U^n(\aa)\ =&SU(2)\qq\times\qq U(3)\qq&\owns\qq\qq\
(u_0,w),\\
U^{nc}(\aa)\ =&\zz_2\qq\times\qq U(1)\qq&\owns\qq\qq\
(u_{c0},w_{c}1_3),\\
{\rm Int}(\aa)\ =&\ [SU(2)\qq\times\qq SU(3)]/\Gamma
&\owns u_\sigma =(u_{\sigma 0}, w_{\sigma }),\\
\Gamma \ =&\zz_2\qq\times\qq \zz_3\qq&\owns\gamma =
(\exp[-m_02\pi i/2],\exp[-m_22\pi i/3]),\label{disstan}
\ee
with
$m_0=0,1$ and $m_2=0,1,2$.
Let us compute the receptacle
of the lifted automorphisms,
\bb {\rm Aut}_{\hh}(\aa)=
[U(2)_L\times U(3)_c&&\!\!\!\!\!\!\!\times U(N)_{qL}
\times U(N)_{\ell L}\times U(N)_{uR}
\times U(N)_{dR}]/[U(1)\times U(1)]\cr &&
\times U(N)_{eR}.\label{smrecept}\ee
The subscripts indicate on which
multiplet the
$U(N)$s act. The kernel of the projection down to the
automorphism group Aut($\aa)$ is
\bb{\rm ker}\,p=
[U(1)\times U(1)&&\!\!\!\!\!\!\!\times U(N)_{qL}
\times U(N)_{\ell L}\times U(N)_{uR}
\times U(N)_{dR}]/[U(1)\times U(1)]\cr &&
\times U(N)_{eR},\ee
and its restrictions to the images of the lifts are
\bb {\rm ker}\,p \cap L({\rm Int}(\aa))=\zz_2\times\zz_3,\qq
{\rm ker}\,p \cap \llll(U^n(\aa))=\zz_2\times U(1).\ee
The kernel of $i$ is $\zz_2\times U(1)$ in sharp contrast to
the kernel of $\hat L$, which is trivial.
The isospin $SU(2)_L$ and the
colour $SU(3)_c$ are the image of the lift $\hat L$.
If $q\not=0,$ the image of
$\ell$ consists of one
$U(1)\owns w_c=\exp [i\theta ]$ contained in the five
flavour
$U(N)$s. Its embedding depends on $q$:
\bb \llll(1_2,1,
w_c1_3)=\ell(w_c)=&&\\
{\rm diag}\,(&\!\!\!\!u_{qL}1_2\ot1_N\ot1_3,
u_{\ell L}1_2\ot1_N,u_{u R}1_N\ot1_3,u_{d
R}1_N\ot1_3,u_{e R}1_N;&\cr  &\!\!\!\!\bar u_{qL}1_2\ot1_N\ot1_3,
\bar u_{\ell L}1_2\ot1_N,\bar
u_{u R}1_N\ot1_3,\bar u_{d R}1_N\ot1_3,\bar u_{e R}1_N)&\eee
with $u_j=\exp[iy_j\theta ]$ and
\bb
y_{qL} = q_{2},\qquad
y_{\ell L}= -q_{1},\qquad
y_{u R} = q_{1}+q_{2} ,\qquad
y_{d R} = -q_{1}+q_{2}, \qquad
y_{e R} = -2q_{1}. \label{y}
\ee
Independently of the embedding, we have indeed {\it
derived} the three fermionic conditions of the hypercharge fine
tuning (\ref{4cond}). In other words, in noncommutative
geometry the massless electroweak gauge boson necessarily
couples vectorially.

Our goal is now to find the minimal extension $\ell$ that
renders the extended lift symmetric,
$\llll(-u_0,-w)=\llll(u_0,w)$, and that renders
 $\llll(1_2,w)$ single-valued. The first requirement means \{
$ \tilde q_{1}=1$ and
$\tilde q_{2}=0$ \} modulo 2,
with
\bb \pp{\tilde q_1\cr \tilde q_2}
=\,{\textstyle\frac{1}{3}} \left( \pp{ q_1\cr  q_2}
-\pp{0\cr 1}\right).\ee
 The second requirement means that
$\tilde q$ has integer coefficients.

 The first extension which comes to mind has
$q=0$, $\tilde q=\pp{0\cr -1/3}$. With respect to the
interpretation (\ref{innfake}) of the inner
automorphisms, one might object that this is not an
extension at all. With respect to the {\it generic}
characterization (\ref{inntrue}), it certainly is a
non-trivial extension. Anyhow it  fails both tests. The most
general extension that passes both tests has the form
\bb \tilde q=\pp{2z_1+1\cr 2z_2},\qq
q=\pp{6z_1+3\cr 6z_2+1},\qq
z_1,z_2\in\zz.\ee
Consequently, $y_{\ell_L}=-q_1$  cannot vanish, the neutrino
comes out electrically neutral in compliance with the Lorentz
force. As common practise, we normalize the hypercharges to
$y_{\ell_L}=-1/2$ and compute the last remaining hypercharge
$y_{q_L}$,
 \bb
y_{q_L}=\,\frac{q_{2}}{2q_{1}}\,
={\frac{{\textstyle\frac{1}{6}}+z_2 }{
1+2z_1  }}.\ee
We can change the sign of $y_{q_L}$ by
permuting
$u$ with $d^c$ and $d$ with $u^c$. Therefore it is
sufficient to take $z_1=0,1,2,...$
The minimal such extension, $z_1=z_2=0$,
recovers nature's choice $y_{q_L}={\textstyle\frac{1}{6}}$.
Its lift,
\bb \llll(u_0,w)=\rho (u_0,\det w,w)J
\rho (u_0,\det w,w)J^{-1},\ee
is the anomaly free fermionic
representation of the standard model considered as
$SU(2)\times U(3)$ Yang-Mills-Higgs model. The
double-valuedness of
$\llll$ comes from the discrete group $\zz_2$ of central
quaternionic unitaries
$(\pm 1_2,1_3)\ \in\  \zz_2\ \subset\ \Gamma\ \subset\
U^{nc}(\aa) $.  On the other hand, O'Raifeartaigh's \cite{or}
$\zz_2$ in the group of the standard model (\ref{smgr}),
$\pm (1_2, 1_3)\ \in\ \zz_2\ \subset\ U^{nc}(\aa)$, is not a
subgroup of
$\Gamma $. It reflects the symmetry of
$\llll$.

\subsubsection{Fluctuating metric}

The stage is set now for fluctuating the metric by means of the
extended lift. This algorithm answers en passant a long standing
question in Yang-Mills theories: To gauge or not to gauge? Given a
fermionic Lagrangian, e.g. the one of the standard model, our first
reflex is to compute its symmetry group. In noncommutative
geometry, this group is simply the internal receptacle
(\ref{smrecept}). The painful question in Yang-Mills theory is
what subgroup of this symmetry group should be gauged? For us,
this question is answered by the choices of the spectral triple and of
the spin lift. Indeed the image of the extended lift is the gauge
group. The fluctuating metric promotes its generators to gauge
bosons, the $W^\pm$, the $Z$, the photon and the gluons. At the
same time, the Higgs representation is derived, equation
(\ref{fluct}):
\bb H=\rho (u_0, \det w, w)[\tilde\dd,\rho (u_0, \det w,
w)^{-1}]=\pp{0&\hat H&0&0\cr \hat H^*&0&0&0\cr 0&0&0&0\cr
0&0&0&0}\ee
with
\bb \hat H=
\pp{\pp{h_1M_u&-\bar h_2M_d\cr h_2 M_u&\bar h_1M_d}\ot
1_3 &0\cr
 0&\pp{-\bar h_2M_e\cr \bar h_1M_e}\cr
}\ee
and
\bb \pp{h_1&-\bar h_2\cr h_2&\bar h_1}=\pm u_0
\pp{\det w&0\cr 0&\det\bar w}-1_2.\ee
The Higgs is characterized by one complex doublet, $(h_1,h_2)^T$.
Again it will be convenient to pass to the homogeneous Higgs
variable,
\bb \dd&=&\llll\tilde \dd\llll^{-1}=\tilde \dd +H+JHJ^{-1}\cr\cr
&=&\Phi +J\Phi J^{-1}=\pp{0&\hat\Phi &0&0\cr \hat \Phi ^*&0&0&0
\cr 0&0&0&\bar{\hat\Phi} \cr 0&0&\bar{\hat\Phi} ^*&0}\ee
with
\bb \hat \Phi =
\pp{\pp{\varphi _1M_u&-\bar \varphi _2M_d\cr \varphi _2
M_u&\bar \varphi _1M_d}\ot 1_3 &0\cr
 0&\pp{-\bar \varphi _2M_e\cr \bar \varphi _1M_e}\cr
}=\rho _L(\phi )\mm\ee
and
\bb\phi = \pp{\varphi _1&-\bar \varphi _2\cr \varphi _2&\bar
\varphi _1}=\pm u_0
\pp{\det w&0\cr 0&\det\bar w}.\ee

In order to satisfy the first order condition,
the representation of
$M_3(\cc)\owns c$ had to commute with the Dirac operator.
Therefore the Higgs is a colour singlet and the gluons will remain
massless. The first two of the six intriguing properties of the
standard model listed in subsection \ref{winner} have a geometric
{\it raison d'\^etre}, the first order condition.
In turn, they imply the third
property: we have just shown that the Higgs $\varphi
=(\varphi _1,\varphi _2)^T$ is a colour singlet. At the same time
the fifth property follows from the fourth: the Higgs
 of the standard model is an isospin doublet
 because of the parity violating
couplings of the quaternions $\hhh$.
 Furthermore, this Higgs has
hypercharge
$y_\varphi =-{\textstyle\frac{1}{2}} $ and the last fine tuning of
the sixth property (\ref{4cond})  also derives from Connes'
algorithm: the Higgs has a component with vanishing electric
charge, the physical Higgs, and the photon will remain massless.

In conclusion, in Connes version of the standard model there is
only one intriguing input property, the fourth: explicit parity
violation in the algebra representation $\hh_L\oplus\hh_R$, the
five others are mathematical consequences.

\subsubsection{Spectral action}

Computing the spectral action $S_{CC}=f(\dd_t^2/\Lambda ^2)$ in
the standard model is not more difficult than in the minimax
example, only the matrices are a little bigger,
\bb \dd_t=\llll_t\tilde \dd_t \llll_t^{-1}=
\pp{\ddd_L&\gamma _5\hat\Phi    &0&0\cr
\gamma _5\hat\Phi ^*&\ddd_R&0&0\cr
0&0&C\ddd_LC^{-1}&\gamma _5\bar{\hat\Phi }   \cr
0&0&\gamma _5\bar{\hat\Phi }^*&C\ddd_RC^{-1}}.\ee
The trace of the powers of $\hat\Phi $ are computed from the
identities $\hat\Phi = \rho_L (\phi )\mm$ and $\phi ^*\phi =\phi
\phi ^*= (|\varphi_1|^2+|\varphi _2|^2)1_2=|\varphi |^21_2$ by
using that $\rho _L$ as a representation respects multiplication
and involution.

The spectral action produces the complete action of the standard
model coupled to gravity with the following relations for coupling
constants:
\bb g_3^2=g_2^2={\textstyle\frac{9}{N}}  \lambda\label{smconstr}
 .\ee
Our choice of central charges, $\tilde q=(1,0)^T$, entails a further
relation, $g_1^2={\textstyle\frac{3}{5}} g_2^2$, i.e. $\sin^2\theta
_w=3/8$. However only products of the Abelian gauge coupling
$g_1$ and the hypercharges $y_j$ appear in the Lagrangian. By
rescaling the central charges, we can rescale the hypercharges
and consequently the Abelian coupling $g_1$. It seems quite moral
that noncommutative geometry has nothing to say about Abelian
gauge couplings.

\begin{figure}[h]
\label{merge}
\epsfxsize=11cm
\hspace{2.2cm}
\epsfbox{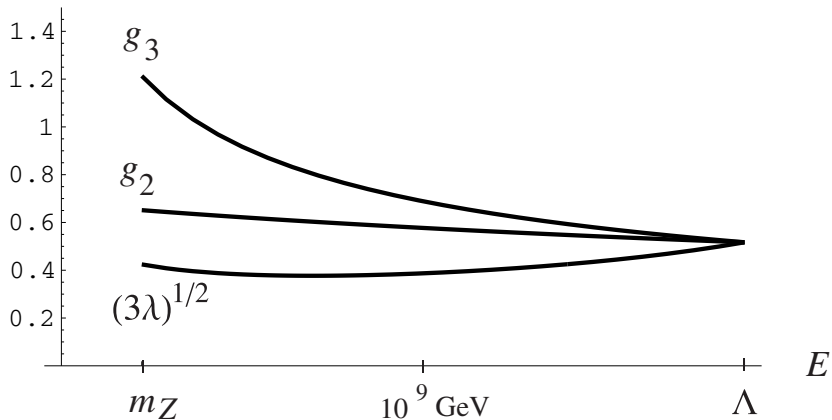}
\caption{Running coupling constants}
\end{figure}

Experiment tells us that the weak and strong couplings are
unequal, equation (\ref{gaugecoup}) at  energies
corresponding to the $Z$ mass, $g_2=0.6518\pm
0.0003,\ g_3=1.218\pm 0.01.$ Experiment also tells us that the
coupling constants are not constant, but that they evolve with energy.
This evolution can be understood theoretically in terms of
renormalization: one can get rid of short distance divergencies in
perturbative quantum field theory by allowing energy depending
 gauge, Higgs, and Yukawa couplings where the theoretical evolution
depends on the particle content of the model. In the standard model,
$g_2$ and $g_3$ come together with increasing energy, see Figure
8. They would become equal at astronomical energies,
$\Lambda =10^{17}$ GeV, if one believed that
between presently explored energies,
$10^2$ GeV, and the `unification scale' $\Lambda $, no new
particles exist. This hypothesis has become popular under the name
`big desert'
 since Grand Unified Theories.
It was believed that new gauge bosons, `lepto-quarks' with masses of
order $\Lambda $ existed. The lepto-quarks together with the $W^\pm$,
the $Z$, the photon and the gluons generate the simple group $SU(5)$,
with only one gauge coupling,
$g_5^2:=g_3^2=g_2^2={\textstyle\frac{5}{3}} g_1^2$ at
$\Lambda $. In the minimal $SU(5)$ model, these lepto-quarks
would mediate proton decay with a half life that today is excluded
experimentally.

If we believe in the big desert, we can imagine that -- while almost
commutative at present energies -- our geometry becomes truly
noncommutative at time scales of
$\hbar /\Lambda \sim 10^{-41}$ s. Since in such a geometry smaller
time intervals cannot be resolved, we expect the coupling constants to
become energy independent at the corresponding energy scale
$\Lambda $. We remark that the first motivation for
noncommutative geometry in spacetime goes back to
Heisenberg and was precisely the
regularization of short distance divergencies in quantum field
theory, see e.g. \cite{jackiw}. The big desert is an opportunistic
hypothesis and remains so in the context of noncommutative
geometry. But in this context, it has at least the merit of being
consistent with three other physical ideas:
\begin{description}\item[Planck time:]
There is an old hand waving argument
combining Heisenberg's uncertainty relation of phase
space with the Schwarzschild horizon to find an
uncertainty relation in spacetime with a scale
$\Lambda$ smaller than the Planck energy $(\hbar
c^5/G)^{1/2}\sim 10^{19}$ GeV: To measure a position with a
precision
$\Delta x$ we need, following Heisenberg, at least a momentum
$\hbar/\Delta x$ or, by special relativity, an energy
$\hbar c/\Delta x$.  According to general relativity,
such an energy creates an horizon of size $G\hbar
c^{-3}/\Delta x$. If this horizon exceeds $\Delta x$ all
information on the position is lost. We can only
 resolve positions with $\Delta x$ larger than the Planck length,
$\Delta x> (\hbar  G /c^{3})^{1/2}\sim 10^{-35}$ m. Or we can only
resolve time with  $\Delta t$ larger than the Planck time,
$\Delta t> (\hbar  G /c^{5})^{1/2}\sim 10^{-43}$ s. This is
compatible with the above time uncertainty of $\hbar /\Lambda \sim
10^{-41}$ s.
\item[Stability:]
We want the Higgs self coupling $\lambda $ to remain positive
\cite{cmpp} during its perturbative evolution for all energies up to
$\Lambda $. A negative Higgs self coupling would mean that no
ground state exists, the Higgs potential is unstable. This requirement
is met for the self coupling given by the constraint (\ref{smconstr})
at energy
$\Lambda $, see Figure 8.
\item[Triviality:]
We want the Higgs self coupling $\lambda $ to remain
perturbatively small  \cite{cmpp} during its evolution for
all energies up to
$\Lambda $ because its evolution is computed from a
perturbative expansion. This requirement as well is met for
the self coupling given by the constraint (\ref{smconstr}), see
Figure 8. If the top mass was larger than 231 GeV or if there were
$N=8$ or more generations this criterion would fail.
\end{description}
Since the big desert gives a minimal and consistent picture we are
curious to know its numerical implication. If we accept the constraint
(\ref{smconstr})  with $g_2=0.5170$ at the energy
$\Lambda=0.968\ 10^{17}$ GeV
and evolve it down to lower energies using the perturbative
renormalization flow of the standard model, see Figure 8, we
retrieve
the experimental nonAbelian gauge couplings $g_2$
and $g_3$  at
the $Z$ mass  by construction of
$\Lambda $. For the Higgs coupling, we obtain
\bb \lambda =0.06050  \ \pm\  0.0037\qq  {\rm at}\qq  E=m_Z.\ee
The indicated error comes from the experimental error in the top
mass,
$m_t= 174.3\pm 5.1\ {\rm GeV}$, which affects the evolution of the
Higgs coupling. From the Higgs coupling at low energies we compute
the Higgs mass,
\bb m_H=4\sqrt 2 \,{\frac{\sqrt\lambda }{ g_2  }}\,m_W=171.6\
\pm\ 5\ \rm GeV.\ee
For details of this calculation see \cite{bridge}.

\subsection{Beyond the standard model}

A social reason, that made the Yang-Mills-Higgs machine
popular, is that it is an inexhaustible source of employment. Even
after the standard model, physicists continue to play on the machine
and try out extensions of the standard model by adding new
particles, `let the desert bloom'. These particles can be
gauge bosons coupling only to  right-handed fermions in order to
restore left-right symmetry. The added particles can be lepto-quarks for
grand unification or supersymmetric particles. These models are
carefully tuned not to upset the phenomenological success of the
standard model. This means in practice to choose Higgs representations
and potentials that give masses to the added particles, large enough
to make them undetectable in present day experiments, but not too
large so that experimentalists can propose bigger
machines to test these models. Independently there are always short
lived deviations from the standard model predictions in new
experiments. They never miss to trigger new, short lived models with
new particles to fit the `anomalies'. For instance, the literature
contains hundreds of superstring inspired Yang-Mills-Higgs models,
each of them with hundreds of parameters, coins, waiting for the
standard model to fail.

Of course, we are trying the same game in Connes' do-it-yourself kit.
So far, we have not been able to find one single consistent extension of
the standard model \cite{florian, fedele, kw,beyond}. The reason is
clear, we have no handle on the Higgs representation and potential,
which are on the output side, and, in general, we meet two problems:
light physical scalars and degenerate fermion masses in irreducible
multiplets. The extended standard model with arbitrary numbers of
quark generations, $N_q\geq 0$, of lepton generations, $N_\ell\geq
1$, and of colours $N_c$, somehow manages  to avoid both problems
and we are trying to prove that it is unique as such. The minimax
model has
$N_q=0,\ N_\ell=1,\ N_c=0$. The standard model has
$N_q=N_\ell=:N$ and $N_c=3$ to avoid Yang-Mills anomalies
\cite{zinn}. It also has 
$N=3$ generations. So far, the only realistic extension of the standard
model that we know of in
noncommutative geometry, is the addition
of right-handed neutrinos  and of  Dirac masses in one or two
generations. These might be necessary to account for observed neutrino
oscillations \cite{data}.

\section{Outlook and conclusion}

Noncommutative geometry reconciles Riemannian geometry and
uncertainty and we expect it to reconcile general relativity with
quantum field theory. We also expect it to improve our still incomplete
understanding of quantum field theory. On the perturbative level
such an improvement is happening right now: Connes,
Moscovici, and  Kreimer discovered a subtle
link between a noncommutative generalization of the index
theorem and perturbative quantum field theory. This link is a
Hopf algebra relevant to both theories \cite{cmk}.

In general, Hopf
algebras play the same role in noncommutative geometry as Lie
groups play in Riemannian geometry and we expect new examples of
noncommutative geometry from its merging with the theory of Hopf
algebras. Reference \cite{shahn} contains a simple example where
quantum group techniques can be applied to noncommutative particle
models.

The running of coupling constants from perturbative quantum field
theory must be taken into account in order to perform the high
precision test of the standard model at present day energies.
We have invoked an extrapolation of this running to astronomical
energies to make
the constraint $g_2=g_3$ from the spectral action compatible with
experiment. This extrapolation is still based on quantum loops  in
flat Minkowski space. While
 acceptable at energies below the scale
$\Lambda $ where gravity and the noncommutativity of space seem
negligible, this approximation is unsatisfactory from a conceptual
point of view and one would like to see  quantum fields
constructed on a noncommutative space. At the end of the nineties
first examples of  quantum fields
on the (flat) noncommutative torus or its non-compact
 version, the Moyal plane, were
published \cite{ncqf}. These examples came straight from the
spectral action. The noncommutative torus is motivated from
quantum mechanical phase space and was the
first example of a noncommutative spectral triple \cite{nct}.
Bellissard \cite{belliss} has shown that the noncommutative torus is
relevant in solid state physics: one can understand the quantum Hall
effect by taking the Brillouin zone to be noncommutative. Only
recently other examples of noncommutative spaces like
noncommutative spheres where uncovered \cite{coladu}. Since 1999,
quantum fields on the noncommutative torus are being studied
 extensively including
the fields of the standard model \cite{msm}. So far, its internal part
is not treated as a noncommutative geometry and Higgs bosons and
potentials are added opportunistically.
This problem is avoided naturally by considering the tensor
product of the noncommutative torus with a finite spectral triple,
but I am sure that the
axioms of noncommutative geometry can be rediscovered by playing
long enough with model building.

In quantum mechanics and in general relativity, time and space play
radically different roles. Spatial position is an observable in
quantum mechanics, time is not. In general relativity, spacial
position loses all meaning and only proper time can be measured.
Distances are then measured by a particular observer as (his proper)
time of flight of photons going back and forth multiplied by the
speed of light, which is supposed to be universal. This definition of
distances is operational thanks to the high precision of present day
atomic clocks, for example in the GPS. The `Riemannian' definition of
the meter, the forty millionth part of a complete geodesic on
earth, had to be abandoned in favour of  a quantum mechanical
definition of the second via the spectrum of an atom. Connes'
definition of geometry via the spectrum of the
Dirac operator is the precise counter part of today's experimental
situation. Note that the meter stick is an extended (rigid ?) object. On the
other hand an atomic clock is a pointlike object and experiment tells us
that the atom is sensitive to the potentials at the location of the clock,
the potentials of all forces, gravitational, electro-magnetic, ... The
special role of time remains to be understood in noncommutative
geometry
\cite{rov} as well as the notion of spectral triples with Lorentzian
signature and their 1+3 split \cite{kal}.

Let us come back to our initial claim: Connes derives the standard model
of electro-magnetic, weak and strong forces from noncommutative
geometry and, at the same time, unifies them with gravity.  If we say
that the Balmer-Rydberg formula is derived from quantum mechanics,
then this claim has three levels:\\
{\bf Explain the nature of the variables:} The choice of the discrete
variables
$n_j$,
 contains already a -- at the time revolutionary -- piece of
physics, energy quantization. Where does it come from?\\
{\bf Explain the ansatz:} Why should one take the power law
(\ref{ansatz})?\\
{\bf Explain the experimental fit:} The ansatz comes with discrete
parameters,
 the `bills' $q_j$, and continuous parameters, the `coins' $g_j$, which are
determined by an experimental fit. Where do the fitted values, `the
winner', come from?

 How about deriving gravity from Riemannian geometry?
Riemannian geometry has only one possible variable, the metric $g$.
The minimax principle dictates the Lagrangian ansatz:
\bb S[g]=\int_M [\Lambda ^c-{\textstyle\frac{1}{16\pi G}} R^q
]\,\de V.\ee
Experiment rules on the parameters: $q=1$, $G=6.670\cdot 10^{-11}\ \rm
m^3s^{-2}kg$, Newton's constant, and $\Lambda ^c\sim 0$. Riemannian
geometry remains silent on the third level. Nevertheless, there is
general agreement, gravity derives from Riemannian geometry.

Noncommutative geometry has only one possible variable, the Dirac
operator, which in the commutative case coincides with the metric. Its
fluctuations explain the variables of the additional forces, gauge and
Higgs bosons. The minimax principle dictates the Lagrangian ansatz:
the spectral action. It reproduces the Einstein-Hilbert action and the
ansatz of Yang, Mills and Higgs, see Table 3. On the third level,
noncommutative geometry is not silent, it produces lots of constraints,
all compatible with the experimental fit. And their exploration is not
finished yet.

\begin{table}[h]
\begin{center}
\setlength{\unitlength}{1.0cm}
\begin{picture}(10,4.8)(0,1)
\put(0,5){\parbox{2cm}{\rm Riemannian geometry}}
\put(3.5,5){\vector(1,0){2,5}}
\put(4,5.5){\parbox[b]{2cm}{\rm Einstein}}
\put(7,5){\parbox{2cm}{\rm gravity}}
\put(1.1,4){\oval(0.2,0.6)[t]}
\put(1,4){\vector(0,-1){2}}
\put(0,1){\parbox{2cm}{\rm noncommutative geometry}}
\put(3.5,1){\vector(1,0){2.5}}
\put(4,1.5){\parbox[b]{2cm}{\rm Connes}}
\put(7,1){\parbox{5cm}{\rm gravity +
Yang-Mills-Higgs}}
\put(-0.7,3){\parbox{2cm}{\rm Connes}}
\end{picture}
\end{center}
\caption{Deriving some
YMH forces from gravity}
\end{table}

 I hope to have convinced one or the other reader that
noncommutative geometry contains elegant solutions of long
standing problems in fundamental physics and that it proposes
concrete strategies to tackle the remaining ones.
I would like to conclude our outlook with a sentence by Planck who tells
us how important the opinion of our young, unbiased colleagues is.
Planck said, a new theory is accepted, not because the others are
convinced, because they die.

\vspace{1cm}\noindent
It is a pleasure to thank Eike Bick and Frank Steffen for the
organization of a splendid School. I thank the participants for their
unbiased criticism and Kurusch Ebrahimi-Fard, Volker Schatz, and
Frank Steffen for a careful reading of the manuscript.

\section{Appendix}

\subsection{Groups}

Groups are an extremely powerful tool in physics. Most symmetry
transformations form a group. Invariance under continuous
transformation groups entails conserved quantities, like energy,
angular momentum or electric charge.

A group G is a set equipped with an associative, not necessarily
commutative (or `Abelian') multiplication law that has a neutral
element 1. Every group element $g$ is supposed to have an inverse
$g^{-1}.$

We denote by $\zz_n$ the {\it cyclic group} of $n$ elements. You can
either think of $\zz_n$ as the set $\{0,1,...,n-1\}$ with
multiplication law being addition modulo $n$ and neutral element
0. Or equivalently, you can take the set $\{1,\exp(2\pi i/n),\exp(4\pi
i/n),..., \exp((n-1)2\pi i/n)\}$ with multiplication and neutral element
1.
$\zz_n$ is an Abelian subgroup of the permutation group on $n $
objects.

Other immediate examples are matrix groups: The {\it general linear
groups}
$GL(n,\cc)$ and
$GL(n,\rr)$ are the sets of complex (real), invertible $n\times n$
matrices. The multiplication law is matrix multiplication and the
neutral element is the $n\times n$ unit matrix $1_n$. There are
many important subgroups of the general linear groups:
$SL(n,\cdot)$,
$\cdot=\rr$ or
$\cc$, consist only of matrices with unit determinant. $S$ stands for
special and will always indicate that we add the condition of unit
determinant. The {\it orthogonal group} $O(n)$ is the group of real
$n\times n$ matrices $g$ satisfying $gg^T=1_n$. The {\it special
orthogonal group}
$SO(n)$ describes the rotations in the Euclidean space $\rr^n$. The
{\it Lorentz group} $O(1,3)$ is the set of real $4\times 4$ matrices $g$
satisfying $g\eta g^T=\eta $, with $\eta =$diag$\{1,-1,-1,-1\}$. The
{\it unitary group} $U(n)$ is the set of complex $n\times n$ matrices
$g$ satisfying $gg^*=1_n$. The {\it unitary symplectic group}
$USp(n)$ is the group of complex $2n\times 2n$ matrices $g$
satisfying
$gg^*=1_{2n}$ and $g {\cal I} g^T={\cal I}$ with
\bb {\cal I}:=
\pp{\pp{0&1\cr -1&0}&\cdots&0\cr
\vdots&\ddots&\vdots\cr 0&\cdots&\pp{0&1\cr -1&0}}.\ee

The {\bf center} $Z(G)$ of a group $G$ consists of those elements in
$G$ that commute  with all elements in $G$, $Z(G)=\{z\in G, zg=gz\
{\rm for\ all}\ g\in G\}$. For example, $Z(U(n))=U(1)\owns
\exp (i\theta )\,1_n, $ $Z(SU(n))=\zz_n\owns \exp (2\pi i k/n)\, 1_n$.

All matrix groups are subsets of $\rr^{2n^2}$ and therefore we can
talk about {\bf compactness} of these groups. Recall that a subset of
$\rr^N$ is compact if and only if it is closed and bounded. For
instance,
$U(1)$ is a circle in
$\rr^2$ and therefore compact. The Lorentz group on the other
hand is unbounded because of the boosts.

The matrix groups are {\it Lie groups} which means that they contain
infinitesimal elements $X$ close to the neutral element:
$\exp X=1+X+O(X^2)\in G.$ For instance,
\bb X=\pp{0&\epsilon &0\cr -\epsilon &0&0\cr 0&0&0},
\qq\epsilon
\ \rm small,\ee
describes an infinitesimal rotation around the $z$-axis by an
infinitesimal angle
$\epsilon $. Indeed
\bb \exp X=\pp{\cos\epsilon &\sin\epsilon &0\cr
-\sin\epsilon &\cos\epsilon &0\cr 0&0&1}\ \in SO(3),\qq
0\le\epsilon <2\pi ,\ee
is a rotation around the $z$-axis by an
arbitrary angle $\epsilon $. The infinitesimal transformations
$X$ of a Lie group $G$ form its {\bf Lie algebra} $\gg$. It is closed
under
 the commutator $[X,Y]=XY-YX$.
For the above
matrix groups the Lie algebras are denoted by lower case letters.
For example, the Lie algebra of the special unitary group $SU(n)$ is
written as
$su(n)$. It is the set of complex $n\times n$ matrices $X$
satisfying $X+X^*=0$ and $\t X=0$. Indeed,
$1_n=(1_n+X+...)(1_n+X+...)^*=1_n+X+X^*+O(X^2)$ and
$1=\det\exp X =\exp\t X$. Attention, although defined in terms of
complex matrices, $su(n)$ is a {\bf real} vector space. Indeed, if a
matrix $X$ is anti-Hermitean, $X+X^*=0$, then in general, its complex
scalar multiple
$iX$ is no longer anti-Hermitean.

 However, in real vector spaces,
eigenvectors do not always exist and we will have to {\bf complexify}
the real vector space $\gg$: Take a basis of $\gg$. Then $\gg$ consists
of linear combinations of these basis vectors with real coefficients.
The {\bf complexification} $\gg^\cc$ of $\gg$ consists of linear
combinations with complex coefficients.

The {\it translation group} of $\rr^n$ is $\rr^n$ itself. The
multiplication law now is vector addition and the neutral element is
the zero vector. As the vector addition is commutative, the
translation group is Abelian.

The {\it diffeomorphism group} Diff$(M)$ of an open subset $M$ of
$\rr^n$ (or of a manifold) is  the set of differentiable maps
$\sigma $ from
$M$ into itself that are invertible (for the composition $\circ$) and
such that its inverse is differentiable. (Attention, the last condition is
not automatic, as you see by taking $M=\rr\owns x$ and $\sigma
(x)=x^3$.) By virtue of the chain rule we can take the composition as
multiplication law. The neutral element is the identity map on $M$,
$\sigma =1_M$ with
$1_M(x)=x$ for all $x\in M$.

\subsection{Group representations}

We said that $SO(3)$ is the rotation group. This needs a little
explanation. A rotation is given by an axis, that is a unit eigenvector
 with unit eigenvalue, and an angle. Two rotations can be
carried out one after the other, we say `composed'. Note that the
order is important, we say that the 3-dimensional rotation group is
nonAbelian. If we say that the rotations form a group, we mean that
the composition of two rotations is a third rotation. However, it is not
easy to compute the multiplication law, i.e., compute the axis and
angle of the third rotation as a function of the axes and angles of the
two initial rotations. The equivalent `representation' of the rotation
group as
$3\times 3$ matrices is much more convenient because the
multiplication law is simply matrix multiplication. There are several
`representations' of the
$3$-dimensional rotation group in terms of matrices of different
sizes, say $N\times N$. It is sometimes useful to know all these
representations. The $N\times N$ matrices are linear maps,
`endomorphisms', of the
$N$-dimensional vector space $\rr^N$ into itself. Let us denote by
End$(\rr^N)$ the set of all these matrices. By definition, a
representation of the group $G$ on the vector space
$\rr^N$ is a map $\rho :G\rightarrow {\rm End}(\rr^N)$
reproducing the multiplication law as matrix multiplication or in
nobler terms as composition of endomorphisms. This means
 $\rho (g_1g_2)=\rho (g_1)\,\rho (g_2)$ and $\rho (1)=1_N$.
The representation is called {\bf faithful} if the map $\rho $ is
injective. By the minimax principle we are interested in the faithful
representations of lowest dimension. Although not always unique,
physicists call them  fundamental representations. The
fundamental representation of the $3$-dimensional rotation group
is defined on the vector space $\rr^3$. Two $N$-dimensional
representations
$\rho _1$ and $\rho_2$ of a group $G$ are {\bf equivalent} if
there is an invertible $N\times N$ matrix $C$ such that $\rho
_2(g)=C\rho _1(g)C^{-1}$ for all $g\in G$. $C$ is interpreted as
describing a change of basis in $\rr^N$. A representation is called
{\bf irreducible} if its vector space has no proper invariant subspace,
i.e. a subspace
$W\subset\rr^N$, with $W\not=\rr^N, \{0\}$ and $\rho (g)
W\subset W$ for all $g\in G$.

Representations can be defined in the same manner on complex
vector spaces, $\cc^N$. Then every $\rho (g)$ is a complex,
invertible matrix. It is often useful, e.g. in quantum mechanics, to
represent a group on a Hilbert space, we put a scalar product on the
vector space, e.g. the standard scalar product on $\cc^N\owns
v,w$, $(v,w):=v^*w$. A {\bf unitary} representation is a
representation whose matrices $\rho (g)$ all respect the scalar
product, which means that they are all unitary. In quantum
mechanics, unitary representations are important because they
preserve probability. For example, take the {\bf adjoint
representation} of
$SU(n)\owns g$. Its Hilbert space is the complexification of
its Lie algebra $su(n)^\cc\owns X,Y$ with scalar product $(X,Y):=\t
(X^*Y)$. The representation is defined by conjugation, $\rho
(g)X:=gXg^{-1}$, and it is unitary, $(\rho (g)X,\rho (g)Y)=(X,Y)$. In
Yang-Mills theories, the gauge bosons live in the adjoint
representation. In the Abelian case, $G=U(1)$, this representation is
1-dimensional, there is one gauge boson, the photon, $A\in
u(1)^\cc=\cc.$ The photon has no electric charge, which means that
it transforms trivially, $\rho (g)A=A$ for all $g\in U(1)$.

 {\bf Unitary equivalence} of representations is
defined by change of orthonormal bases. Then $C$ is a unitary matrix.
A key theorem for particle physics states that all irreducible unitary
representations of any compact group are finite dimensional. If we
accept the definition of elementary particles as orthonormal basis
vectors of unitary representations, then we understand why Yang and
Mills only take compact groups. They only want a finite number of
elementary particles. Unitary equivalence expresses the quantum
mechanical superposition principle observed for instance in the
$K^0-\bar K^0$ system. The unitary matrix $C$ is sometimes
referred to as mixing matrix.

Bound states of elementary particles are described by tensor
products: the tensor product of two unitary representations $\rho
_1$ and
$\rho _2$ of one group defined on two Hilbert spaces $\hh_1$
and $\hh_2$ is the unitary representation $\rho _1\ot\rho _2$
defined on
$\hh_1\ot\hh_2\owns \psi _1\ot \psi _2$ by $(\rho _1\ot\rho
_2)(g)\,(\psi _1\ot \psi _2):=\rho _1(g)\,\psi _1\ot \rho
_2(g)\,\psi _2$.  In the case of electro-magnetism,
$G=U(1)\owns\exp(i\theta )$ we know that all irreducible unitary
representations are 1-dimensional,
$\hh=\cc\owns\psi $ and characterized by the electric charge $q$,
$\rho (\exp(i\theta ))\psi =\exp(i q\psi )\psi $. Under tensorization
the electric charges are added. For $G=SU(2)$, the irreducible
unitary representations are characterized by the spin,
$\ell=0,{\textstyle\frac{1}{2}} ,1,...$ The addition of spin from
quantum mechanics is precisely tensorization of these
representations.

Let $\rho $ be a representation of a Lie group $G$ on a vector
space and let $\gg$ be the Lie algebra of $G$. We
denote by
$\tilde\rho $ the Lie algebra representation of the group
representation $\rho $. It is defined on the same vector space by
$\rho (\exp X)=\exp(\tilde \rho (X))$. The $\tilde \rho (X)$s are
not necessarily invertible endomorphisms. They satisfy
$\tilde\rho ([X,Y])=[\tilde \rho (X),\tilde \rho (Y)]:=\tilde \rho
(X)\tilde \rho (Y)-\tilde \rho (Y)\tilde \rho (X).$

An {\bf affine} representation is the same construction as above, but
we allow the
$\rho(g)$s to be invertible affine maps, i.e. linear maps plus
constants.

\subsection{Semi-direct product and Poincar\'e group}

The direct product $G\times H$ of two groups $G$ and
$H$ is again a group with multiplication law: $(g_1,
h_1)(g_2,h_2):=(g_1g_2,h_1h_2).$ In the direct product, all
elements of the first factor commute with all elements of the second
factor: $(g,1^H)(1^G,h)=(1^G,h)(g,1^H).$ We write $1^H$ for the
neutral element of $H$. Warning, you sometimes see the misleading
notation
$G\ot H$ for the direct product.

To be able to define the semi-direct product $G\ltimes H$ we
must have an action of $G$ on $H$, that is a map $\rho:
G\rightarrow {\rm Diff}(H)$ satisfying $\rho _g(h_1h_2)=
\rho _g(h_1)\,\rho _g(h_2)$, $\rho _g(1^H)=1^H$,
$\rho _{g_1g_2}=\rho_{g_1}\circ\rho _{g_2}$ and $\rho
_{1^G}=1_H$. If $H$ is a vector space carrying a representation or
an affine representation $\rho $ of the group $G$, we can view
$\rho $ as an action by considering $H$ as translation group.
Indeed, invertible linear maps and affine maps are diffeomorphisms
on
$H$.  As a set, the semi-direct product
$G\ltimes H$ is the direct product, but the multiplication law
is modified  by help of the action:
\bb(g_1,h_1)(g_2,h_2):=(g_1g_2,h_1\,\rho _{g_1}(h_2)).\ee
We retrieve the direct product if the action is trivial, $\rho
_g=1_H$ for all $g\in G$.
Our first example is the invariance group of electro-magnetism
coupled to gravity ${\rm Diff}(M)\ltimes
\,^MU(1).$ A diffeomorphism $\sigma(x)$ acts on a gauge function
$g(x) $ by $\rho _\sigma (g):=g\circ\sigma ^{-1}$ or more
explicitly
$(\rho _\sigma (g))(x):=g(\sigma ^{-1}(x))$. Other examples come
with other gauge groups like $SU(n)$ or spin groups.

Our second example is
the Poincar\'e group, $O(1,3)\ltimes \rr^4$, which is the isometry
group of Minkowski space. The semi-direct product is important
because Lorentz transformations do not commute with translations.
Since we are talking about the Poincar\'e group, let us mention the
theorem behind the definition of particles as orthonormal basis
vectors of unitary representations: The irreducible, unitary
representations of the Poincar\'e group are characterized by
mass and spin. For fixed mass $M\geq 0$ and spin $\ell$, an
orthonormal basis is labelled by the momentum $\vec p$ with
$E^2/c^2-\vec p^2=c^2M^2$,
$\psi =\exp(i(Et-\vec p\cdot\vec x)/\hbar)$ and the
$z$-component $m$ of the spin with $|m|\leq\ell$, $\psi
=Y_{\ell,m}(\theta ,\varphi )$.

\subsection{Algebras}

Observables can be added, multiplied and multiplied by scalars. They
form naturally an associative algebra $\aa$, i.e. a vector space
equipped with an associative product and neutral elements 0 and 1.
Note that the multiplication does not always admit inverses, $a^{-1}$,
e.g. the neutral element of addition, 0, is not invertible. In quantum
mechanics, observables are self adjoint. Therefore, we need an
involution
$\cdot^*$ in our algebra. This is an anti-linear map from the algebra
into itself,
$(\lambda a+b)^* =\bar\lambda a^*+b^*, \ \lambda \in\cc,\
a,b\in\aa,$ that reverses the product,
$(ab)^*=b^*a^*$, respects the unit,
$1^*=1$, and is such that $a^{**}=a$. The set of $n\times n$ matrices
with complex coefficients, $M_n(\cc)$, is an example of such an
algebra, and more generally, the set of endomorphisms or operators
on a given Hilbert space $\hh$. The multiplication is matrix
multiplication or more generally composition of operators, the
involution is Hermitean conjugation or more generally the adjoint of
operators.

A representation $\rho $ of an abstract algebra $\aa$ on a Hilbert
space
$\hh$ is a way to write $\aa$ concretely as operators as in the last
example,
$\rho :\aa\rightarrow {\rm End}(\hh)$. In the group case, the
representation had to reproduce the multiplication law. Now it has to
reproduce, the linear structure:
$\rho (\lambda a+b)=\lambda \rho (a)+\rho (b),\ \rho
(0)=0,$ the multiplication: $\rho (ab)=\rho (a)\rho (b),\ \rho (1)=1,$
and the involution:  $ \rho (a^*)=\rho
(a)^*.$ Therefore the tensor product of two representations $\rho
_1$ and $\rho _2$ of $\aa$ on Hilbert spaces $\hh_1\owns\psi _1$
and
$\hh_2\owns\psi _2$ is not a representation: $((\rho _1\ot\rho
_2)(\lambda a))\,(\psi _1\ot\psi _2)=(\rho _1(\lambda a)\,\psi_1) \ot
(\rho _2(\lambda a)\,\psi _2)=\lambda ^2 (\rho _1\ot\rho
_2)(a)\,(\psi _1\ot\psi _2)$.

The group of unitaries $U(\aa):=\{u\in\aa,\,
uu^*=u^*u=1\}$ is a subset of the algebra $\aa$. Every algebra
representation induces a unitary representation of its group of
unitaries. On the other hand, only few unitary representations of
the group of unitaries extend to an algebra representation. These
representations describe elementary particles. Composite particles
are described by tensor products, which are not algebra
representations.

An anti-linear operator $J$ on a Hilbert space $\hh\owns\psi,\tilde
\psi $ is a map from $\hh$ into itself satisfying $J(\lambda \psi
+\tilde \psi )=\bar\lambda J(\psi )+J(\tilde \psi ).$ An anti-linear
operator $J$ is anti-unitary if it is invertible and preserves the
scalar product, $(J\psi ,J\tilde \psi )= (\tilde \psi ,\psi )$. For example,
on $\hh=\cc^n\owns \psi $ we can define an anti-unitary operator $J$
in the following way. The image of the column vector $\psi $ under
$J$ is obtained by taking the complex conjugate of $\psi $ and then
multiplying it with a unitary $n\times n$ matrix $U$, $J\psi
=U\bar\psi $ or $J=\, U\,\circ$ complex conjugation. In fact, on a
finite dimensional Hilbert space, every anti-unitary operator is of this
form.


\begin{thebibliography}{47}

\bibitem{cls}
A. Connes, A. Lichn\'erowicz \& M. P. Sch\"utzenberger, {\it
Triangle de Pens\'ees}, O. Jacob (2000), English version: {\it Triangle
of Thoughts}, AMS (2001)
\bibitem{ac}
G. Amelino-Camelia, {\it Are we at the dawn of
quantum gravity phenomenology?}  Lectures given at
35th Winter School of Theoretical Physics: From
Cosmology to Quantum Gravity, Polanica,
Poland, 1999, gr-qc/9910089
\bibitem{gr}
S. Weinberg, {\it Gravitation and Cosmology}, Wiley (1972)\\
R. Wald, {\it  General Relativity}, The University of
Chicago Press (1984)
\bibitem{bd}
J. D. Bj\o rken \& S. D. Drell, {\it
Relativistic Quantum Mechanics}, McGraw--Hill (1964)
\bibitem{or}
L. O'Raifeartaigh, {\it Group Structure of
Gauge Theories}, Cambridge University Press (1986)
\bibitem{gs}
M. G\"ockeler \& T. Sch\"ucker, {\it Differential
Geometry, Gauge Theories, and Gravity}, Cambridge
University Press (1987)
\bibitem{group}
R. Gilmore, {\it Lie Groups, Lie Algebras and Some of Their
Applications}, Wiley (1974)\\
H. Bacry, {\it Lectures Notes in Group Theory and Particle Theory},
Gordon and Breach (1977)
\bibitem{algebra}
N. Jacobson, {\it Basic Algebra I, II}, Freeman (1974,
1980)
\bibitem{jogi}
J. Madore, {\it An Introduction to Noncommutative
Differential Geometry and Its Physical Applications},
Cambridge University Press (1995)\\
G. Landi, {\it An Introduction to Noncommutative
Spaces and Their Geometry}, hep-th/9701078,
Springer (1997)
\bibitem{costarica}
J. M. Gracia-Bond\'\i a, J. C. V\'arilly \& H. Figueroa,
{\it Elements of Noncommutative Geometry},
Birkh\"auser (2000)
\bibitem{van}
J. W. van Holten, {\it Aspects of BRST quantization}, hep-th/0201124,
in this volume
\bibitem{zinn}
J. Zinn-Justin, {\it Chiral anomalies and topology}, hep-th/0201220,
in this volume
\bibitem{data}
The Particle Data Group, {\it Particle Physics Booklet}
and { http://pdg.lbl.gov}
\bibitem{renorm}
G. 't Hooft, {\it Renormalizable Lagrangians for massive
Yang-Mills fields}, Nucl. Phys. B35 (1971) 167\\
G. 't Hooft \& M. Veltman, {\it Regularization and renormalization
of gauge fields}, Nucl. Phys. B44 (1972) 189\\
G. 't Hooft \& M. Veltman, {\it Combinatorics of gauge fields}, Nucl.
Phys. B50 (1972) 318\\
B. W. Lee \& J. Zinn-Justin, {\it Spontaneously broken gauge
symmetries I, II, III and IV}, Phys. Rev. D5 (1972) 3121, 3137, 3155,
Phys. Rev. D7 (1973) 1049
\bibitem{gsw}
S. Glashow, {\it Partial-symmetries of weak interactions}, Nucl.
Phys. 22 (1961) 579\\
A. Salam in Elementary Particle Physics: Relativistic Groups and
Analyticity, Nobel Symposium no. 8, page 367, eds.: N.  Svartholm,
Almqvist \& Wiksell, Stockholm 1968\\
S. Weinberg, {\it A model of leptons}, Phys. Rev. Lett. 19 (1967) 1264
\bibitem{joke}
J. Iliopoulos, {\it An introduction to gauge theories}, Yellow Report,
CERN (1976)
\bibitem{gilles}
G. Esposito-Far\`ese, {\it Th\'eorie de
Kaluza-Klein et gravitation quantique}, Th\'ese de
Doctorat, Universit\'e d'Aix-Marseille II, 1989
\bibitem{book}
 A. Connes, {\it Noncommutative Geometry}, Academic
Press (1994)
\bibitem{tresch}
A. Connes, {\it Noncommutative
geometry and reality},  J. Math. Phys. 36 (1995) 6194
\bibitem{grav}
A. Connes, {\it Gravity coupled with
matter and the foundation of noncommutative
geometry}, hep-th/9603053, Comm. Math. Phys. 182
(1996) 155
\bibitem{neu}
H. Rauch, A. Zeilinger, G. Badurek, A. Wilfing, W. Bauspiess \&
U. Bonse, {\it Verification of coherent spinor rotations of
fermions}, Phys. Lett. 54A (1975) 425
\bibitem{cartan}
E. Cartan, {\it Le\c cons sur la
th\'eorie des spineurs}, Hermann (1938)
\bibitem{bris}
A. Connes, {\it Brisure de sym\'etrie spontan\'ee et g\'eom\'etrie du
point de vue spectral}, S\'eminaire Bourbaki, 48\`eme ann\'ee, 816 (1996)
313\\
A. Connes, {\it Noncommutative differential geometry and
the structure of space time}, Operator Algebras and Quantum Field
Theory, eds.: S. Doplicher et al., International Press, 1997
\bibitem{lift}
T. Sch\"ucker, {\it Spin group and almost
commutative geometry}, hep-th/0007047
\bibitem{bourg} J.-P. Bourguignon \& P. Gauduchon,
{\it Spineurs, op\'erateurs de Dirac et variations de
m\'etriques}, Comm. Math. Phys. 144
(1992) 581
\bibitem{bonse}
U. Bonse \& T. Wroblewski,
{\it Measurement of neutron quantum interference in
noninertial frames}, Phys. Rev. Lett. 1 (1983) 1401
\bibitem{cow}
R. Colella, A. W. Overhauser \& S. A. Warner, {\it Observation of
gravitationally induced quantum interference}, Phys. Rev. Lett.
34 (1975) 1472
\bibitem{cc}
 A. Chamseddine \& A. Connes, {\it The
spectral action principle}, hep-th/9606001,
Comm. Math. Phys.186 (1997) 731
\bibitem{lr}
G. Landi \& C. Rovelli, {\it Gravity from Dirac
eigenvalues}, gr-qc/9708041,
Mod. Phys. Lett. A13 (1998) 479
\bibitem{heat}
P. B. Gilkey, {\it Invariance Theory, the Heat Equation,
and the Atiyah-Singer Index Theorem}, Publish or
Perish (1984)\\
S. A. Fulling, {\it Aspects of Quantum Field Theory in
Curved Space-Time}, Cambridge University Press
(1989)
\bibitem{ikm}
B. Iochum, T. Krajewski \& P. Martinetti, {\it Distances
in finite spaces from noncommutative geometry},
hep-th/9912217, J. Geom. Phys. 37 (2001) 100
\bibitem{dkm}
M. Dubois-Violette, R. Kerner \& J.
Madore, {\it Gauge bosons in a noncommutative
geometry}, Phys. Lett. 217B (1989) 485
\bibitem{ccl}
A. Connes, {\it Essay on physics and noncommutative
geometry}, in {\it The Interface of Mathematics and
Particle Physics}, eds.: D. G. Quillen et al., Clarendon
Press (1990)\\
A. Connes \& J. Lott, {\it Particle models and noncommutative
geometry}, Nucl. Phys. B 18B (1990) 29\\
 A. Connes \& J. Lott, {\it The metric
aspect of noncommutative geometry}, in the
proceedings of the 1991 Carg\`ese Summer Conference,
eds.: J. Fr\"ohlich et al., Plenum Press (1992)
\bibitem{kk}
J. Madore, {\it Modification of Kaluza Klein theory},
Phys. Rev. D 41 (1990) 3709
\bibitem{mw}
P. Martinetti \& R. Wulkenhaar, {\it Discrete Kaluza-Klein from
scalar fluctuations in noncommutative geometry}, hep-th/0104108,
J. Math. Phys. 43 (2002) 182
\bibitem{att}
 T. Ackermann \& J. Tolksdorf, {\it A generalized
Lichnerowicz formula, the Wodzicki residue and
gravity}, hep-th/9503152, J. Geom. Phys. 19 (1996) 143
\\
 T. Ackermann \& J. Tolksdorf, {\it The generalized
Lichnerowicz formula and analysis of Dirac
operators}, hep-th/9503153, J. reine angew. Math. 471
(1996) 23
\bibitem{egbv}
R. Estrada, J. M. Gracia-Bond\'\i a \& J. C. V\'arilly,
{\it On summability of distributions and spectral
geometry}, funct-an/9702001, Comm. Math. Phys. 191 (1998) 219
\bibitem{rom}
B. Iochum, D. Kastler \& T. Sch\"ucker, {\it
On the universal Chamseddine-Connes action:
 details of the action computation}, hep-th/9607158,
J. Math. Phys. 38 (1997) 4929\\
L. Carminati, B. Iochum, D. Kastler \& T.
Sch\"ucker, {\it On Connes' new principle of general
relativity: can spinors hear the forces of space-time?},
hep-th/9612228, Operator Algebras and Quantum Field
Theory, eds.: S. Doplicher et al., International Press,
1997
\bibitem{tkmz}
M. Paschke \& A. Sitarz, {\it Discrete spectral triples
and their symmetries}, q-alg/9612029,
J. Math. Phys. 39 (1998) 6191 \\
T. Krajewski, {\it
Classification of finite spectral triples}, hep-th/9701081, J. Geom.
Phys. 28 (1998) 1
\bibitem{fare}
S. Lazzarini \& T. Sch\"ucker,
{\it A farewell to unimodularity}, hep-th/0104038,  Phys.Lett. B 510
(2001) 277
\bibitem{florian}
B. Iochum \& T. Sch\"ucker, {\it A left-right symmetric
model \`a la Connes-Lott}, hep-th/9401048, Lett. Math.
Phys. 32 (1994)  153\\
 F. Girelli, {\it Left-right symmetric models
in noncommutative geometry? } hep-th/0011123, Lett. Math. Phys. 57
(2001) 7
\bibitem{fedele}
F. Lizzi, G. Mangano, G. Miele \& G. Sparano, {\it Constraints on
unified gauge theories from noncommutative geometry},
hep-th/9603095, Mod. Phys. Lett. A11 (1996) 2561
\bibitem{kw}
 W. Kalau \& M. Walze, {\it Supersymmetry and
noncommutative geometry}, hep-th/9604146,
J. Geom. Phys. 22 (1997) 77
\bibitem{stand}
D. Kastler, {\it Introduction to noncommutative geometry
and Yang-Mills model building}, Differential geometric
 methods in
theoretical physics, Rapallo (1990), 25\\
${}$   --- , {\it A detailed account of Alain
Connes' version of the standard model in
non-commutative geometry, I, II and III}, Rev. Math. Phys.
5 (1993) 477, Rev. Math.
Phys. 8 (1996) 103  \hfil\break
 D. Kastler
\& T. Sch\"ucker, {\it Remarks on Alain Connes'
approach to the standard  model in non-commutative
geometry}, Theor. Math. Phys. 92 (1992) 522, English
version, 92 (1993) 1075, hep-th/0111234\\
${}$   --- , {\it A detailed account of Alain Connes'
version of the standard model in non-commutative
geometry, IV}, Rev. Math. Phys. 8 (1996) 205\\
${}$   --- , {\it The standard model \`a la
Connes-Lott}, hep-th/9412185, J. Geom. Phys. 388
(1996) 1 \\
J. C. V\'arilly \&  J. M. Gracia-Bond\'\i a,
{\it Connes' noncommutative differential geometry and
the standard model}, J. Geom. Phys. 12 (1993) 223
\hfil\break
T. Sch\"ucker \& J.-M. Zylinski, {\it Connes' model
building kit}, hep-th/9312186, J. Geom. Phys. 16 (1994)
1 \\
E. Alvarez, J. M. Gracia-Bond\'\i a \& C. P. Mart\'\i n,
 {\it Anomaly cancellation and the gauge group of the
Standard Model in Non-Commutative Geometry},
hep-th/9506115, Phys. Lett. B364 (1995) 33\\
R. Asquith, {\it Non-commutative geometry and the
strong force}, hep-th/9509163, Phys. Lett. B 366 (1996)
220
\\
C. P. Mart\'\i n, J. M. Gracia-Bond\'\i a \& J. C. V\'arilly,
{\it The standard model as a noncommutative geometry:
the low mass regime},
 hep-th/9605001, Phys. Rep. 294 (1998) 363 \\
L. Carminati, B. Iochum \& T. Sch\"ucker, {\it
The noncommutative constraints on the standard
model \`a la Connes}, hep-th/9604169, J. Math. Phys. 38
(1997) 1269\\
R. Brout, {\it Notes on Connes' construction of the standard model},
hep-th/9706200, Nucl. Phys. Proc. Suppl. 65 (1998) 3
\\
J. C. V\'arilly, {\it Introduction to noncommutative
geometry}, physics/9709045, EMS
Summer School on Noncommutative Geometry and
Applications, Portugal, september 1997, ed.: P.
Almeida \\
T. Sch\"ucker, {\it Geometries and forces},
hep-th/9712095, EMS
Summer School on Noncommutative Geometry and
Applications, Portugal, september 1997, ed.: P.
Almeida\\
J. M. Gracia-Bond\'\i a, B. Iochum \& T. Sch\"ucker,
{\it The Standard Model in Noncommutative Geometry
and Fermion Doubling}, hep-th/9709145, Phys. Lett. B
414 (1998) 123\\
D. Kastler, {\it Noncommutative geometry and basic
physics},
 Lect. Notes Phys. 543 (2000) 131\\
${}$   --- , {\it Noncommutative geometry and
fundamental physical interactions: the Lagrangian
level},  J. Math. Phys. 41 (2000) 3867\\
K. Elsner, {\it Noncommutative geometry: calculation of the standard
model Lagrangian}, hep-th/0108222, Mod. Phys. Lett. A16 (2001) 241\\
 F. Scheck, W. Werner \& H. Upmeier (eds.), {\it
Noncommutative Geometry and the Standard Model of Elementary
Particle Physics},  Lecture
notes in physics 596, Springer (2002)\\
\bibitem{jackiw}
R. Jackiw, {Physical instances of noncommuting coordinates},
hep-th/0110057
\bibitem{cmpp}
N. Cabibbo, L. Maiani, G. Parisi \& R. Petronzio, {\it
Bounds on the fermions and Higgs boson masses in
grand unified theories}, Nucl. Phys. B158 (1979) 295
\bibitem{bridge}
L. Carminati, B. Iochum \& T. Sch\"ucker, {\it
Noncommutative Yang-Mills and noncommutative
relativity: A bridge over troubled water},
hep-th/9706105, Eur. Phys. J. C8 (1999) 697
\bibitem{beyond}
B. Iochum \& T. Sch\"ucker, {\it
Yang-Mills-Higgs versus Connes-Lott},
hep-th/9501142, Comm. Math. Phys. 178 (1996) 1\\
I. Pris \& T. Sch\"ucker, {\it Non-commutative
geometry beyond the standard model}, hep-th/9604115,
J. Math. Phys. 38 (1997) 2255\\
I. Pris \& T. Krajewski, {\it Towards a $Z'$ gauge boson
in noncommutative geometry}, hep-th/9607005, Lett.
Math. Phys. 39 (1997) 187\\
M. Paschke, F. Scheck \& A. Sitarz, {\it Can (noncommutative)
geometry accommodate leptoquarks?} hep-th/9709009, Phys . Rev. D59
(1999) 035003\\
 T. Sch\"ucker \& S. ZouZou, {\it Spectral action beyond the
standard model}, hep-th/0109124
\bibitem{cmk}
A. Connes \& H. Moscovici, {\it Hopf Algebra, cyclic
cohomology and the transverse index theorem}, Comm.
Math. Phys. 198 (1998) 199\\
D. Kreimer, {\it On the Hopf algebra structure of
perturbative quantum field theories}, q-alg/9707029,
Adv. Theor. Math. Phys. 2 (1998) 303\\
A. Connes \& D. Kreimer, {\it Renormalization in
quantum field theory and the Riemann-Hilbert
problem. 1. The Hopf algebra structure of graphs and
the main theorem}, hep-th/9912092,
Comm. Math. Phys. 210 (2000) 249\\
A. Connes \& D. Kreimer, {\it Renormalization in
quantum field theory and the Riemann-Hilbert
problem. 2. the beta function, diffeomorphisms and
the renormalization group}, hep-th/0003188,
Comm. Math. Phys. 216 (2001) 215 \\
for a recent review, see
J. C. V\'arilly, {\it Hopf algebras in noncommutative geometry},
hep-th/010977
\bibitem{shahn}
S. Majid \& T. Sch\"ucker, {$\zz_2\times\zz_2$ \it Lattice as
Connes-Lott-quantum group model}, hep-th/0101217, J. Geom. Phys. 43
 (2002) 1
\bibitem{ncqf}
J. C. V\'arilly \& J. M. Gracia-Bond\'\i a,
{\it On the ultraviolet behaviour of quantum fields over
noncommutative manifolds}, hep-th/9804001, Int. J. Mod. Phys. A14
(1999) 1305\\
T. Krajewski, {\it G\'eom\'etrie non commutative et interactions
fondamentales}, Th\'ese de
Doctorat, Universit\'e de Provence, 1998, math-ph/9903047\\
C. P. Mart\'\i n \& D. Sanchez-Ruiz, {\it The one-loop UV divergent
structure of U(1) Yang-Mills theory on noncommutative $\rr^4$},
hep-th/9903077, Phys. Rev. Lett. 83 (1999) 476 \\
M. M. Sheikh-Jabbari, {\it Renormalizability of the supersymmetric
Yang-Mills theories on the noncommutative torus}, hep-th/9903107,
JHEP 9906 (1999) 15 \\
T. Krajewski \& R. Wulkenhaar, {\it Perturbative quantum gauge
fields on the noncommutative torus}, hep-th/9903187, Int. J. Mod.
Phys. A15 (2000) 1011\\
S. Cho, R. Hinterding, J. Madore \&  H. Steinacker, {\it Finite field
theory on noncommutative geometries}, hep-th/9903239,
Int. J. Mod. Phys. D9 (2000) 161
\bibitem{nct}
 M. Rieffel, {\it  Irrational Rotation $C^*$-Algebras}, Short Comm.
I.C.M. 1978\\
 A. Connes, {\it $C^*$ alg\`ebres et g\'eom\'etrie
diff\'erentielle}, C.R. Acad. Sci. Paris, Ser. A-B (1980) 290, English
version hep-th/0101093\\
 A. Connes \& M. Rieffel, {\it Yang-Mills for non-commutative
two-tori,} Contemp. Math. 105 (1987) 191
\bibitem{belliss}
J. Bellissard, {\it $K-$theory of $C^*-$algebras in solid state physics},
in: Statistical Mechanics and Field Theory: Mathematical Aspects,
eds.: T. C. Dorlas et al., Springer (1986)\\
J. Bellissard, A. van Elst \& H. Schulz-Baldes, {\it The
noncommutative geometry of the quantum Hall effect}, J. Math.
Phys. 35 (1994) 5373
\bibitem{coladu}
A. Connes \& G. Landi, {\it Noncommutative manifolds, the instanton
algebra and isospectral deformations}, math.QA/0011194, Comm.
Math. Phys. 216 (2001) 215\\
A. Connes \& M. Dubois-Violette, {\it Noncommutative
finite-dimensional manifolds I. Spherical manifolds and related
examples}, math.QA/0107070
\bibitem{msm}
M. Chaichian, P. Pre\v snajder, M. M. Sheikh-Jabbari \& A. Tureanu,
{\it Noncommutative standard model: model building},
hep-th/0107055\\
X. Calmet, B. Jur\v co, P. Schupp, J. Wess \& M. Wohlgenannt,
{\it The standard model on non-commutative space-time},
hep-ph/0111115
\bibitem{rov}
A. Connes \& C. Rovelli, {\it Von Neumann algebra
automorphisms and time-thermodynamics relation in
general covariant quantum theories}, gr-qc/9406019,
Class. Quant. Grav. 11 (1994) 1899\\
 C. Rovelli, {\it Spectral noncommutative geometry
and quantization: a simple example}, gr-qc/9904029, Phys.
Rev. Lett. 83 (1999) 1079\\
M. Reisenberger \& C. Rovelli, {\it Spacetime states and covariant
quantum theory}, gr-qc/0111016
\bibitem{kal}
W. Kalau, {\it Hamiltonian formalism in
non-commutative geometry}, hep-th/9409193,
J. Geom. Phys. 18 (1996) 349\\
E. Hawkins, {\it Hamiltonian gravity and
noncommutative geometry}, gr-qc/9605068, Comm. Math. Phys. 187
(1997) 471\\
 T. Kopf \& M. Paschke, {\it A spectral quadruple for the De Sitter
space}, math-ph/0012012\\
A. Strohmaier, {\it On noncommutative and semi-Riemannian
geometry}, math-ph/0110001

\end{thebibliography}
 \end{document}